\documentclass[9pt,twocolumn,twoside]{pnas-new}

\setboolean{displaywatermark}{false}

\templatetype{pnasresearcharticle} 

\usepackage{balance}
\usepackage{mathrsfs}
\usepackage{amsmath}
\usepackage{amsfonts}
\usepackage{mathtools}
\usepackage{tikz}
\usepackage{graphicx}
\usetikzlibrary{positioning}
\usepackage[export]{adjustbox}
\usepackage{rotate}
\usepackage{dsfont}
\usepackage{relsize}
\usepackage{epstopdf}
\DeclareGraphicsExtensions{.jpg,.png,.pdf,.eps}
\usepackage{pstricks}
\usepackage{multirow}
\usepackage{color}
\usepackage{import}

\usepackage{dsfont}
\usepackage{relsize}

\usepackage{filecontents}
\usepackage{multibib}

\usepackage{epstopdf}

\newcommand{\mcal}[1]{\mathscr{#1}}

\newcommand{\bfv}[1]{\mathbf{#1}}

\newcommand{\hp}{\hat{p}}
\newcommand{\hdelta}{\hat{\delta}}
\newcommand{\hX}{\hat{x}}
\newcommand{\hx}{\hat{x}}
\newcommand{\hz}{\hat{z}}

\newcommand{\hA}{\hat{A}}
\newcommand{\ha}{\hat{a}}
\newcommand{\hV}{\hat{V}}
\newcommand{\hu}{\hat{u}}

\newcommand{\hh}{\hat{h}}
\newcommand{\hm}{\hat{m}}

\newcommand{\bl}{\big\langle}
\newcommand{\br}{\big\rangle}
\newcommand{\hti}{\hat{t\hspace{0.01cm}}}

\title{Mechanisms for bacterial gliding motility on soft substrates}

\author[a]{Jo\"el Tchoufag}
\author[b]{Pushpita Ghosh} 
\author[c]{Connor B. Pogue}
\author[c]{Beiyan Nan}
\author[a,d,1]{Kranthi K. Mandadapu}

\affil[a]{Department of Chemical and Biomolecular Engineering, University of California, Berkeley, CA 94720}
\affil[b]{Tata Institute of Fundamental Research Hyderabad, 500107, India}
\affil[c]{Department of Biology, Texas A\&M University, College Station, TX 77843}
\affil[d]{Chemical Sciences Division, Lawrence Berkeley National Laboratory, Berkeley, CA}
\leadauthor{Tchoufag} 
\authorcontributions{J.T., P.G., B.N. and K.K.M. designed research; J.T. and K.K.M. performed theoretical and numerical analysis; C.P. and B.N. carried out experimental research; all authors wrote the paper.}
\correspondingauthor{\textsuperscript{1}To whom correspondence should be addressed. E-mail: kranthi@berkeley.edu}


\begin{abstract}
The motility mechanism of certain rod-shaped bacteria has long been a mystery, since no external appendages (pili, flagella or cilia) are involved in their motion which is known as "gliding". The physical principles behind gliding motility still remain poorly understood. As a canonical example of such organisms, myxobacteria exhibit a gliding motility where the gliding speed depends on the substrate stiffness \cite{Shi93, Fontes99}: an effect known as mechanosensitivity. While there exist some physical models for the mechanosensitivity of eukaryotic cells in tissues due to adhesion \cite{Ladoux12}, the mechanism of myxobacterial gliding motility remains unclear mainly due to the existence of a thin slime layer secreted between the cell and the substrate \cite{Ducret12}. Here we identify the physical principles behind gliding motility, and develop a theoretical model that predicts a two-regime behavior of the gliding speed as a function of the substrate stiffness. Our theory describes the elastic, viscous, and capillary interactions between the bacterial membrane carrying a traveling wave, the slime layer acting as a lubricating viscous film, and the substrate which we model as a soft solid. Defining the gliding motility as the horizontal translation under zero net force, we find the two-regime behavior is due to two different mechanisms of motility thrust. On stiff substrates, the gliding thrust arises from the elasto-hydrodynamic interactions between the bacteria, the slime and the substrate, whereby the bacterial shape deformations create a flow of slime exerting a pressure along the bacterial length. This pressure in conjunction with the bacteria shape provides the necessary thrust for propulsion. However, we show that such a mechanism cannot lead to gliding on very soft substrates. Instead, we show that including capillary effects along with the elasto-hydrodynamic interactions leads to formation of a ridge at the slime-substrate-air interface, which creates a thrust in the form of a localized pressure gradient at the tip of the bacteria. To test our theory, we performed experiments with isolated \textit{M. xanthus} cells on agar substrates of varying stiffness. Over the whole range of substrate stiffness here investigated, the measured gliding speeds are found to be in good agreement with the predictions from our elasto-capillary-hydrodynamic model. The physical mechanisms of mechanosensitivity we propose serve as an important step towards an accurate theory of friction and substrate-mediated interaction between bacteria in a swarm of cells proliferating in soft media \cite{Gallegos06}. 
\end{abstract}

\begin{document}
\maketitle
\ifthenelse{\boolean{shortarticle}}{\ifthenelse{\boolean{singlecolumn}}{\abscontentformatted}{\abscontent}}{}
The inspection of a swarm of myxobacteria, such as \textit{Myxococcus xanthus}, reveals two types of motility: Social (S)-motility or "twitching" involving type IV pili and Adventurous (A)-motility or "gliding" occuring without any external appendage. Here we seek to shed light on the nature of the interaction between a gliding A-motile cell and its substrate. The physical approaches that have been undertaken before to explain myxobacteria gliding are either on the molecular scale or continuum models on the scale of the whole cell \cite{Wolgemuth02}. The former is primarily concerned with identifying genes, proteins and molecular motors and their role in empowering a cell to glide \cite{Hodgkin79, McBride01, Nan11b, Faure16}. Here our focus is not to elucidate the internal molecular mechanisms \cite{Faure16, Xing06, Nan13, Mandadapu15, Fu18} but rather to identify the physical principles governing the gliding motion of myxobacteria and how they interact with their environment. As such, our theory belongs to the class of models analyzing the bacteria at the cellular level. 

Our model is built upon two essential features established through previous experiments on myxobacteria. The first feature is the geometry of the cell's basal surface that interacts with the substrate. As recently revealed through total internal reflection fluorescence (TIRF) microscopy experiments of \textit{M. xanthus} cells \cite{Nan11}, this basal surface possesses an oscillatory structure which we approximate by a sinusoidal shape $b(x,t)$ (see Fig.~\ref{fig:configMain}). The TIFR experiments were carried out on cells expressing green fluorescence protein (GFP) in the cytoplasm and yet showed a modulation of intensity with a period of $L \sim 1\mu m$ \cite{Nan11} (similar to the periodicity of localization of MreB filaments \cite{Mauriello10, Fu18}). Given that GFP was evenly distributed in the cytoplasm, the TIRF images are thus evidence that the distance from the cytoplasm to the microscope glass is modulated. Images from Atomic Force Measurement also revealed that surfaces of single \textit{M. xanthus} cells display a helical pattern \cite{Pelling05}. Such shape deformation may therefore be a necessary condition for gliding. The second feature is that myxobacteria gliding is always accompanied by a trail of slime in the wake of the motile cells \cite{Yu07,Ducret12}. Using the so-called wet-surface enhanced ellipsometry contrast, a thin layer of slime was observed underneath the basal cell surface of even nonmotile mutants \cite{Ducret12}. Slime deposition was thus suggested to be a general means for gliding organisms to adhere to and move over surfaces. Recently, the stick-slip dynamics of twitching \textit{M. xanthus} cells was also well explained by understanding the slime to function both as a glue and as a lubricant \cite{Gibiansky13}. Here, we corroborate these findings and demonstrate that the slime acts as a crucial agent which not only lubricates the motion of myxobacteria, but also creates a coupling between the cell and the deformable substrate. 
  
\begin{figure}[t!]
\centering
 \begin{tikzpicture}
  \node (img1)  {\includegraphics[scale=0.36]{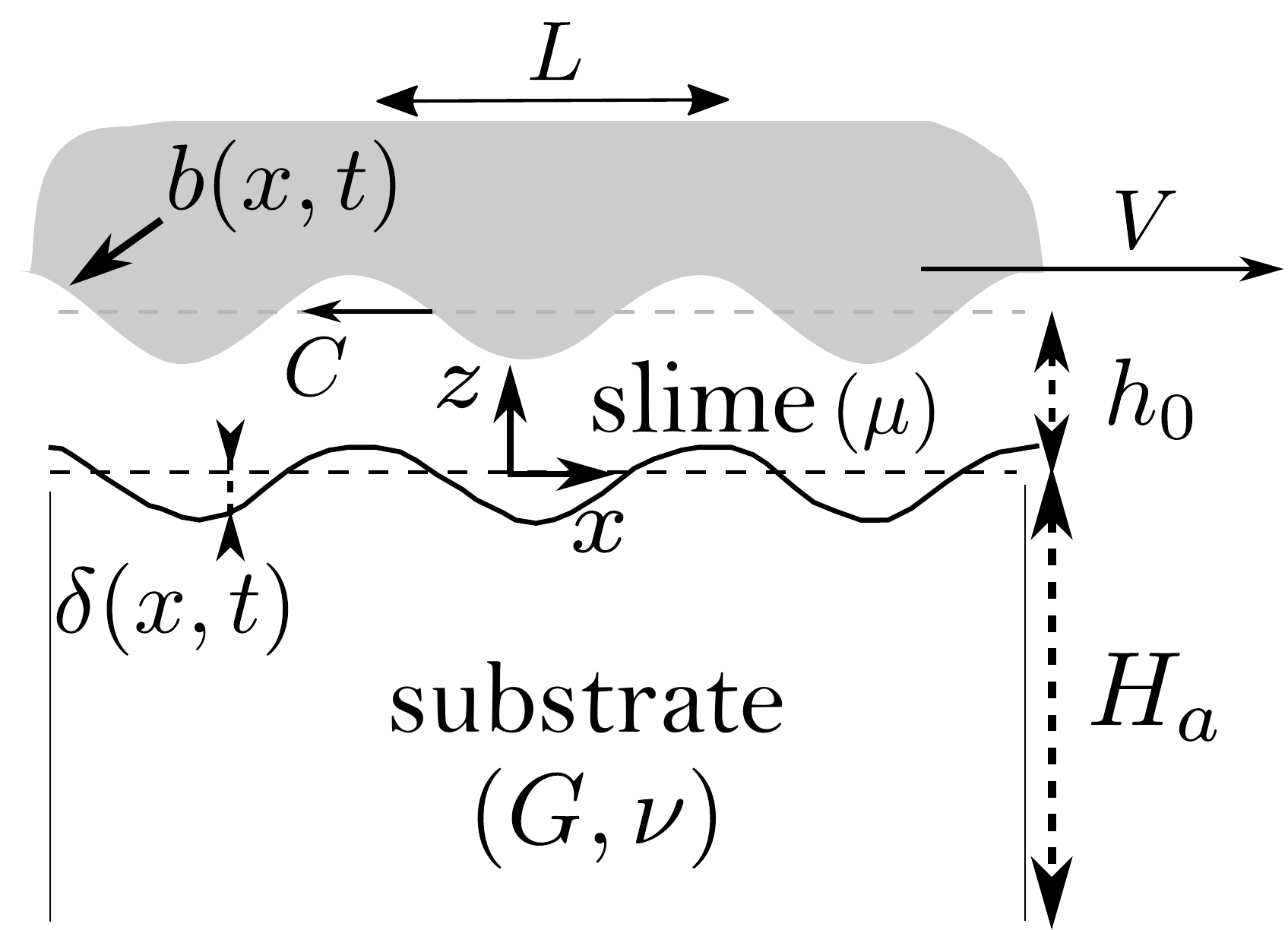}};
  \end{tikzpicture}
 \caption{Schematic description: a gliding bacterium (in grey) with a sinusoidal basal shape. The contact with the soft substrate is lubricated by a thin film of slime. See the text for a description of the variables.}
 \label{fig:configMain}
 \end{figure} 

To model the slime-mediated interaction between myxobacteria and their substrate, we make the following assumptions. We consider small shear rates of the exopolysaccharide (EPS) slime and hence treat it as a Newtonian viscous fluid, in spite of its polymeric constitution \cite{Ducret12}. The good comparison of our model with experiments will justify \textit{a posteriori} that the non-Newtonian rheology of the slime has a second order influence on the myxobacteria-substrate interaction. Furthermore, we neglect inertial effects given that for the characteristic speed ($V\sim 3 \, \mu$m/min \cite{Nan16}), mean height of the interstitial gap between the bacteria and the substrate $h_0\sim 10$ nm \cite{Ducret12}, and large viscosity of bacterial EPS $\mu \sim 5-20$ Pa.s \cite{Choi91, Isobe92}, the typical Reynolds number is $Re\ll 1$ as in the locomotion of most microorganisms \cite{Purcell77}. Hence, we model the dynamics of the slime using the Stokes equations \cite{Happel83}. Given the geometric ratio of the interstitial gap $\epsilon\coloneqq\dfrac{h_0}{L}\sim 10^{-2} \ll 1$, we use the classical lubrication approximation \cite{Reynolds86, Leal07} for the slime film and further simplify the Stokes equations. Moreover, ignoring the rotation of the myxobacteria along its long axis, we confine the problem to two dimensions. Lastly, we account for the deformation $\delta(x,t)$ of the soft substrate, caused by the pressure in the lubricating slime (see Fig.~\ref{fig:configMain}). The thickness of the slime layer is thus $\left[h(x,t) - \delta(x,t)\right]$, where $h(x,t) = h_0 + b(x,t)$. With these simplifications and the lubrication approximation formulated in the reference frame of the bacteria, the velocity component in the gliding direction, $u_x(x,z,t)$, reads (see Supplementary Information (SI) section 1.2)
 \begin{equation}
u_x(x,z,t)=\frac{1}{2\mu}\frac{\partial p}{\partial x}\left(z-h\right)\left(z-\delta\right)+V\frac{z-h}{h-\delta},
\label{eq:Velocity}
\end{equation}
where $p(x,t)$ is the pressure in the slime and $V$ is the gliding velocity of the bacteria which remains to be solved for. Thus, the flow in the slime layer is a superposition of a Poiseuille flow driven by the pressure gradient $\dfrac{\partial p}{\partial x}$ and a Couette flow due to a boundary moving at speed $V$ \cite{Leal07}. We then use Eq.~(\ref{eq:Velocity}) to integrate the mass balance equation across the film thickness, and obtain the modified Reynolds equation (see SI section 1.2)
 \begin{equation}
-\frac{\partial}{\partial t}(h-\delta)+\frac{\partial}{\partial x}\left[ \frac{1}{12\mu}\frac{\partial p}{\partial x}(h-\delta)^3+\frac{1}{2} V(h-\delta) \right] =0.
\label{eq:ReynoldsMain}
\end{equation}

In the frame of reference translating with the cell, we consider the propagation of a traveling wave along the cell surface \cite{Nan11,Sun11} as several recent experiments report gliding is strongly correlated with molecular motor complexes moving with helical trajectories on a scaffold provided by MreB actin \cite{Nan11, Sun11, Nan13, Faure16, Fu18}. Some of these motor complexes, namely AglR molecules, were observed to slow down once they arrive at the basal cell boundary, in contact with the substrate. Due to their low velocity, they appear almost stationary in the fixed laboratory frame \cite{Nan16}. Given that these motor complexes appear almost stationary in the laboratory frame, they correspond to a traveling disturbance on the cell surface in the reference frame translating with the bacteria. In accord with this observation, it was affirmed in \cite{Nan11} that when \emph{viewed externally, the motors driving the rotation of the helical rotor generate transverse waves on the ventral surface}. Certainly, the Reynolds equation (\ref{eq:ReynoldsMain}) admits such traveling wave solutions as we now establish.

Let us consider a membrane carrying a unidirectional traveling wave of speed ${C}$, such that the shape $h(x,t)$ satisfies
\begin{equation}\label{eq:wave}
\frac{\partial h}{\partial t}=C\frac{\partial h}{\partial x}.
\end{equation}
Assuming the $x-$axis is oriented positively to the right as in Fig.~\ref{fig:configMain}, positive and negative values of $C$ correspond to left and right traveling waves, respectively. To obtain traveling wave solutions of Eq.~(\ref{eq:ReynoldsMain}), we search for a substrate deformation field that also satisfies $\dfrac{\partial \delta}{\partial t}=C\dfrac{\partial \delta}{\partial x}$, where we assumed that the substrate deformation exhibits traveling waves of the same speed as that of the bacterial surface. Using this ansatz, the Reynolds equation (\ref{eq:ReynoldsMain}) can be rewritten as (see SI section 1.3). 
\begin{equation}
\frac{\partial}{\partial x}\left[ \frac{\partial p}{\partial x}(h-\delta)^3+6\mu (V-2C) (h-\delta) \right] =0.
\label{eq:Reynolds}
\end{equation}
 
Equation (\ref{eq:ReynoldsMain}) shows that the pressure depends (non-linearly) on the substrate deformation $\delta(x,t)$ which we now set out to determine. In many situations, the horizontal and vertical length scales of the substrate are on the order of centimeters and millimeters respectively \cite{Chen90,Croze11}. Since both dimensions are much larger than the typical length of myxobacteria, we can represent the substrate as a semi-infinite medium. Moreover, gel substrates are generally viscoelastic, where the relative importance of viscous to elastic effects depend on the frequency of excitation of the material. Here, the characteristic frequency is that of the traveling disturbance $f= C/L\approx 3\mu\mathrm{m.s}^{-1}/1\mu\mathrm{m} = 3 \hspace{0.1cm} \mathrm{Hz}$, where the estimated wave speed corresponds to the experimental speed of the AlgR molecules \cite{Nan13}. In the case of agar gels at moderate to high concentrations ($\geq 1\%$) and at such frequencies, the loss modulus is expected to remain smaller than the storage modulus by one or more orders of magnitude \cite{Nayar12}. Therefore, we first treat the substrate as a purely elastic half-space. Furthermore, we formulate the problem in the context of linear elasticity (small deformations) and make the quasi-static assumption that the time-dependent solution can be found by solving a static problem at every instant. In this framework, the deformation of the substrate surface under the action of the film pressure is given by (see SI section 2.1-2.5)

\begin{equation}
\delta (x,t)= \frac{(1-\nu)}{\pi G}\int_{-\ell/2}^{\ell/2}  p(x',t)\ln\frac{\left| x-x'\right |}{x_0} \mathrm{d}x'.
\label{eq:thickdeform}
\end{equation}
Here, $\ell$ is the length of the entire bacteria. In Eq.~(\ref{eq:thickdeform}), $x_0$ is a constant that sets the zero displacement point on the substrate \cite{Johnson85} and $G=E/2(1+\nu)$ is the shear modulus for Young's modulus $E$ and Poisson's ratio $\nu$. Note that we restrict our analysis to incompressible substrates, for which $\nu=1/2$, in order to ensure that the deformation of the substrate surface occurs in the vertical direction only (see SI section 2.5). A motion in the horizontal direction would imply a slip velocity at the slime-substrate interface, in contradiction with the no-slip condition assumed earlier to solve for the velocity field in the thin film. 

In order to compare the viscous and elastic forces at play in the problem, we rewrite the equations in their dimensionless form. To that end, we first note that the characteristic deformation scale of the substrate emerges from Eq.~(\ref{eq:thickdeform}) as $\Delta = (1-\nu)PL/G$, where $P = \mu CL/h_0^2$ is the characteristic pressure scale (See SI section 3). This can then be used to non-dimensionalize the thickness of the slime layer to yield $\left(h -\delta\right)=h_0\left(\hat{h} - \eta \hat{\delta}\right)$, where $\hat{h}=h/h_0$, $\hat{\delta}=\delta/\Delta$. Here, $\eta = \Delta/h_0$ is a dimensionless number that compares the characteristic deformation of the substrate due to the slime pressure to the mean thickness of the film. 

For a given cell shape, $\eta$ essentially captures the influence of the substrate deformability on variations of the lubrication gap during gliding. Therefore, it is also known as the softness parameter and be recast as \cite{Maha04,Skotheim05}
\begin{equation}
\eta=\frac{(1-\nu)PL}{Gh_0},
\label{eq:softness}
\end{equation}
such that large values of $\eta$ denote soft substrates. Using the above dimensionless parameters and defining $\hat{p} = p/P$, $\hat{x} = {x}/L$, $\hti=t/(L/C)$ and $\hat{V} = {V}/C$, the governing equations for the slime layer (\ref{eq:ReynoldsMain}) and the substrate (\ref{eq:thickdeform}) can be non-dimensionalized to become
\begin{subequations}
\begin{align}
 \frac{\partial}{\partial\hx}\left[\frac{\partial\hp}{\partial\hx}(\hh-\eta \hdelta)^3+6(\hV-2)(\hh-\eta \hdelta)\right] & =0, \label{eq:DEMaina} \\
\hdelta(\hx,\hti)- \frac{1}{\pi }\int_{-n/2}^{n/2}  \hp(\hx',\hti) \ln\frac{\left| \hx-\hx'\right |}{\hx_0} \mathrm{d}\hx' &=0,
\end{align}
\label{eq:DEMain}%
\end{subequations} 
where $n=\ell/L$ is a measure of the bacterial length, for a fixed value of the wavelength $L$. 

In order to solve the system of equations (\ref{eq:DEMain}) for different values of $\eta$, we specify two boundary conditions on the pressure so the problem admits a unique solution. We first set $\hp(n/2,\hti)=0$. In other words, we assume that the slime pressure at leading edge of the cell drops to the atmospheric (zero) pressure. Second, given that myxobacteria move without inertia in the Stokes regime and that the forces for gliding result from the internal motors, the sum of the external forces on the cell must vanish according to Newton's second law. In the $z$-direction, this implies that the lift force from the film pressure must vanish. In the lubrication approximation, this lift-free condition reads (see SI section 4)
\begin{equation}
\int_{-n/2}^{n/2}  \hp(\hx,\hti)\hspace{0.03cm} \mathrm{d}\hx=0,
\label{eq:Lift-free}
\end{equation}
and consequently, the second boundary condition for Eq.~(\ref{eq:DEMaina}) is a global condition. In the $x$-direction, the drag-free condition provides an equation, which must be solved inversely to obtain the gliding speed $V$ (see SI section 4)
\begin{equation}
- \int_{-n/2}^{n/2}  \left(\hp\frac{\partial\hat{b}}{\partial \hx}+\frac{1}{2}\frac{\partial\hp}{\partial \hx}\left(\hh-\eta\hdelta\right)+\frac{\hat{V}}{\hh-\eta\hdelta}\right) \mathrm{d}\hx=0.
\label{eq:Drag-free}
\end{equation}
The lift-free and drag-free constraints, (\ref{eq:Lift-free}) and (\ref{eq:Drag-free}), together define the gliding motion of myxobacteria at the cellular level.

The system of coupled equations (\ref{eq:DEMain}), (\ref{eq:Lift-free}), and (\ref{eq:Drag-free}) with the condition $\hp(n/2,\hti)=0$ admits a unique solution $\bfv{q}(\hx,\hti)\coloneqq[\hp(\hx,\hti),\hdelta(\hx,\hti),\hV(\hti)]^T$ for a given softness number $\eta$ and a given geometry of the basal cell shape. Throughout this study, we consider bacterial shapes of the form $b(x,t)=A\sin\left[\frac{2\pi}{L}(x+Ct)\right]$, corresponding to the dimensionless height $\hh(\hx,\hti)=1+\hA\sin\left[2\pi (\hat{x}+\hti)\right]$ where the amplitude is non-dimensionalized by $h_0$. As such, it must verify $\hA\in [0,1]$ for a thin film to exist between the cell and the substrate. 

We solve the problem numerically for a given $\eta$ and $\hh(\hx,\hti)$ and obtain $\bfv{q}(\hx,\hti)$ at different instants. Due to the time-periodicity of the bacterial membrane, we expect $\bfv{q}(\hx,\hti)$ to be periodic, hence the gliding speed as well (see SI section 5.4). A net gliding motion only then exists when the mean velocity, averaged over a wave period, is non-zero. Therefore, the results hereafter will be presented in their time-averaged form, expressed by the notation $\left<\cdot\right>$.

Fig.~\ref{fig:speedamp} shows the average gliding velocity as a function of the softness parameter for different amplitudes of the basal cell shape.
\begin{figure}[t]
\centering
 \begin{tikzpicture}
  \node(img1)  {\includegraphics[scale=0.8]{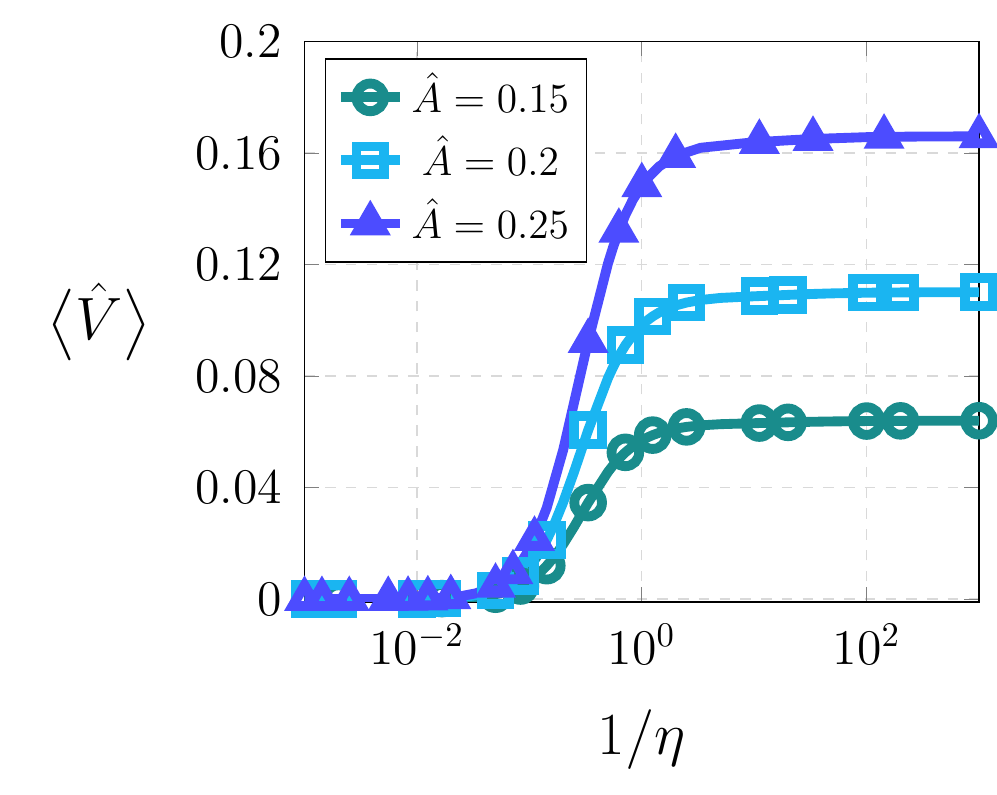}};
 \end{tikzpicture}%
 \caption{Time-averaged gliding speed as function of the softness parameter for different amplitudes of the bacteria shape. Here, we consider a bacteria shape with five wavelenths, i.e. $n=5$. The gliding speed increases with the substrate stiffness and with the amplitude of the bacterial shape.}
 \label{fig:speedamp}
 \end{figure}
Clearly, the gliding velocity decreases with the softness parameter $\eta$. For very small values of $\eta$, the substrate is a very stiff solid and we recover the prediction of the speed of a periodic wavy sheet (Taylor's swimmer \cite{Taylor51}) near a rigid wall. As one could expect, the precision of this agreement improves with $n$, since the bacterial length increases ($L$ being fixed) and hence its approximation by a periodic wavy sheet (see SI section 5.5). Indeed, it is known that in the lubrication regime near a rigid wall ($1/\eta\rightarrow\infty$), small amplitude wavy membranes have a swimming velocity given by \cite{Katz74, Pak16}
\begin{equation}
\left<\hat{V}_{\eta=0}\right> = \frac{3}{2+1/\hA^2},
\label{eq:RigidSpeed}
\end{equation}
which corresponds to the asymptotic values of the small-$\eta$ plateaus in Fig.~\ref{fig:speedamp}. Since $\hA=A/h_0$, increasing the dimensionless amplitude is equivalent to decrease the mean gap $h_0$. Therefore, we recover the well-known result that the locomotion speed of a wavy sheet increases as it approaches a rigid boundary \cite{Katz74}. In our context, this conclusion is equivalent to state that on stiff but elastically deformable substrates, the gliding motility speed increases with the deformation amplitude of myxobacteria membrane. 

\begin{figure*}[t!]
 \begin{tikzpicture}
 \node[xshift=-3cm, yshift=0cm] (img1)  {\includegraphics[scale=0.65]{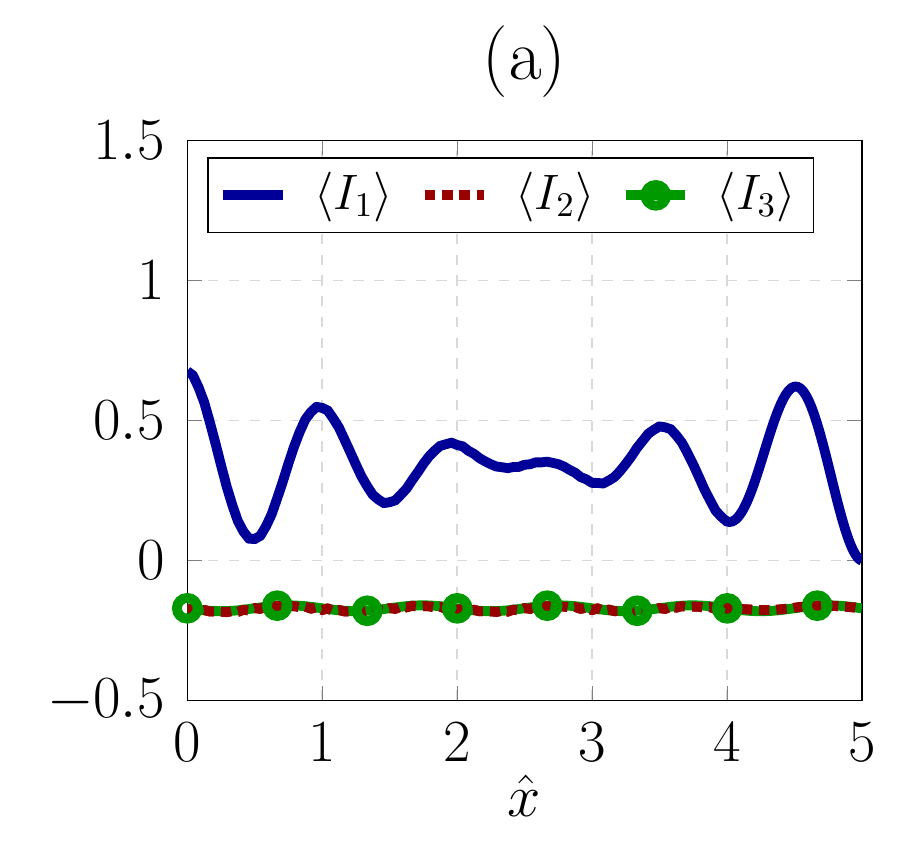}};
 \node[right=of img1, xshift=-1.4cm, yshift=0cm] (img2)  {\includegraphics[scale=0.65]{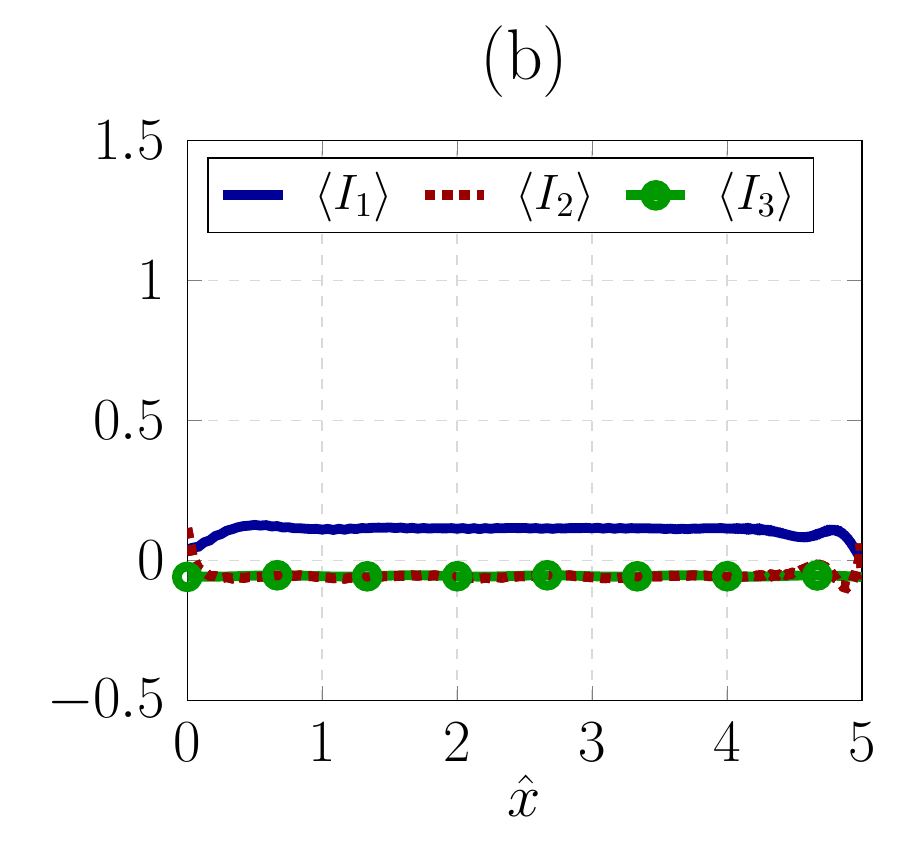}};
 \node[right=of img2, xshift=-1.4cm, yshift=0cm] (img3)  {\includegraphics[scale=0.65]{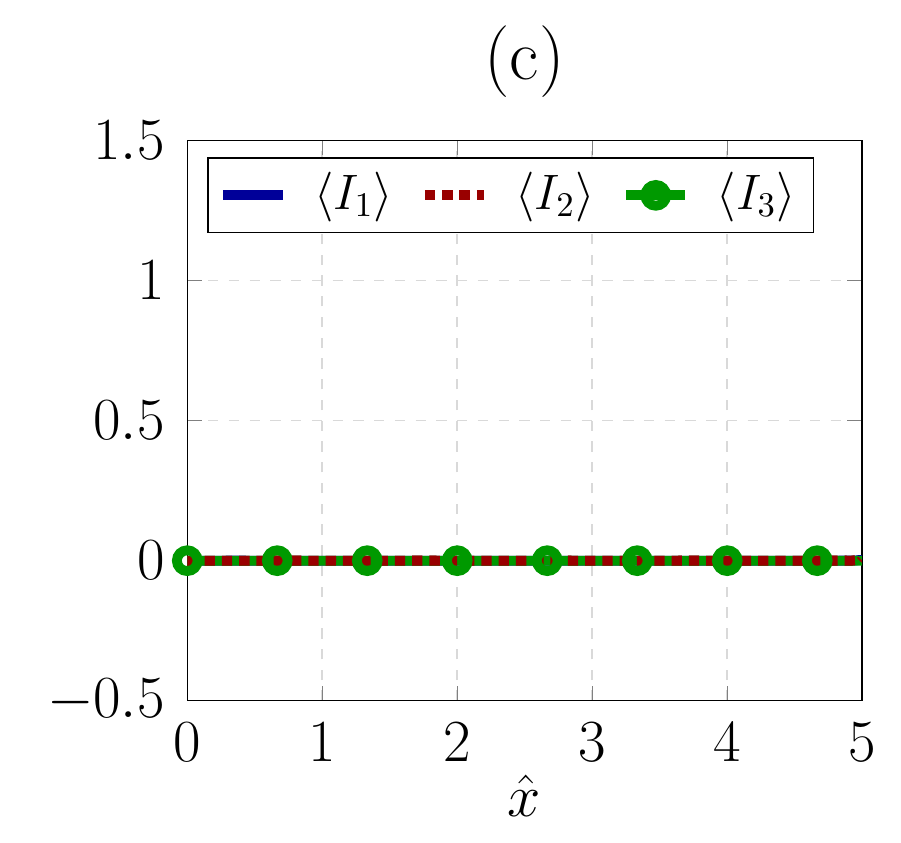}};
 \end{tikzpicture}
 \caption{Time-averaged distribution of the contributions $I_1$, $I_2$ and  $I_3$ to the horizontal force balance for different values of $\eta$. The softness parameter is (a) $\eta=0.001$ (very stiff substrate), (b) $\eta=5$ (midly soft substrate), (c) $\eta=1000$ (very soft substrate). Here, the input parameters are $A=0.25$ and $n=5$.}
 \label{fig:Iforce}
 \end{figure*}

However, as we increase the softness parameter, the gliding velocity transitions from a non-zero value for very stiff substrates to zero in the limit of very soft substrates. For all amplitudes $\hA$, Fig.~\ref{fig:speedamp} shows that substantial deviations from the rigid wall solution occur when the substrate number $\eta\approx 1$, which corresponds to a substrate whose deformation is comparable to that of the slime thickness. 
The vanishing nature of the gliding speed in the limit $\eta\rightarrow\infty$ can be understood by analyzing the horizontal force balance that must be satisfied for gliding to occur. Defined by Eq.~(\ref{eq:Drag-free}), this force balance involves three contributions $I_1(\hx,\hti)$, $I_2(\hx,\hti)$ and $I_3(\hx,\hti)$ given by

\begin{center}
$I_1= -\hp\dfrac{\partial\hat{b}}{\partial \hx}$, $I_2=-\dfrac{1}{2}\dfrac{\partial\hp}{\partial \hx}\left(\hh-\eta\hdelta\right)$, $I_3=\dfrac{-\hV}{\hh-\eta\hdelta}$.
\end{center}
Firstly, $I_1$ represents the action of the slime pressure on the bacteria induced by variations of the bacterial geometry. Hence, it vanishes for bacteria with spatially uniform shapes. Secondly, $I_2$ denotes the effect of pressure variations along the bacteria and would vanish in absence of pressure gradients. Lastly, $I_3$ constitutes the resistance experienced by the bacteria as it must shear the slime to achieve gliding.

We show in Fig.~\ref{fig:Iforce} the time-averaged spatial distribution of the components $I_1$, $I_2$ and $I_3$ along the length of the bacteria, for three values of the softness parameter $\eta\in \{0.001, 5, 1000\}$. As expected, $\left<I_3\right>$ is always negative and contributes to the friction experienced by the bacteria. Next, regarding the component $\left<I_2\right>$, its integral over the bacterial basal shape is negative and thus constitutes an additional contribution to friction. Lastly, Fig.~\ref{fig:Iforce} shows the term $\left<I_1\right>$ is positive over the length of the bacteria and provides the necessary thrust balancing the Poiseuille and Couette frictions from $\left<I_2\right>$ and $\left<I_3\right>$, respectively. 

\begin{figure}[b!]
\centering
 \begin{tikzpicture}
  \node (img1)  {\includegraphics[scale=0.4]{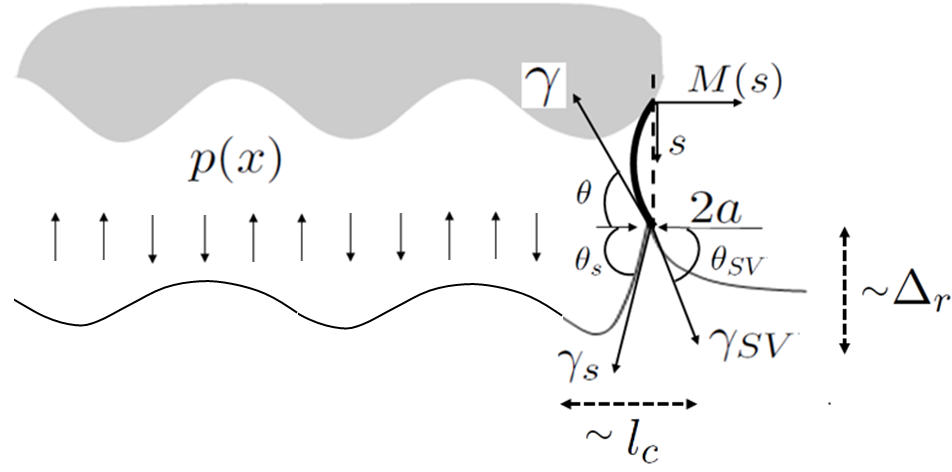}};
 \end{tikzpicture}
 \caption{\footnotesize{Schematic description of the elasto-capillary-hydrodynamics problem: a gliding bacterium (in grey) with a sinusoidal ventral shape. A ridge is formed and balanced by the surface tensions at the slime-air, substrate-slime and air-substrate interfaces. See the text for a description of the variables.}}
 \label{fig:setupecMain}
 \end{figure}

Fig.~\ref{fig:Iforce} therefore reveals the decrease of the gliding speed with the softness number $\eta$ is correlated to that of the local thrust generation along the bacteria. This can be explained in the following manner. For small values of $\eta$, the substrate is very stiff and remains almost unperturbed by the oscillations of the bacterial surface. This implies the thickness of the gap between the bacterial surface and the substrate oscillates and induce a lubricating flow of slime exerting on the cell pressure oscillations in phase with the bacterial shape (see SI-section 5.6). The resulting flow of slime exerts the thrust $\left<I_1 (\hx)\right>$ along the bacteria which leads to a non-zero gliding speed $\left<\hV\right>$. Therefore, this mechanism of thrust generation, whereby the shape oscillations are converted into a lubrication flow of slime, require little to no deformation of the substrate. As such, it cannot be sustained for very soft substrates, when $\eta\rightarrow\infty$. In fact, in this limit, the substrate is very compliant and instantaneously conforms to the oscillating bacterial surface so that $\eta\hat{\delta}\approx \hat{b}$, as supported by our computations (see SI-section 5.6). Given that $\hp \sim\hdelta\approx \hat{b}/\eta$, the pressure distribution along the bacteria thus decays to zero as $\eta\rightarrow\infty$, thereby leading to a vanishing pressure gradient as well. Therefore, the pressure-dependent terms $\left<I_1 (\hx)\right>$ and $\left<I_2 (\hx)\right>$ tend to zero (as $\sim 1/\eta$) in the limit of very soft substrates. As a result, the gliding speed $\left<\hV\right>$ of the force-free bacteria tends to zero in the limit $\eta\rightarrow\infty$.
\begin{figure*}[ht!]
 \begin{tikzpicture}
   \node[xshift=-3cm, yshift=0cm] (img1)  {\includegraphics[scale=0.65]{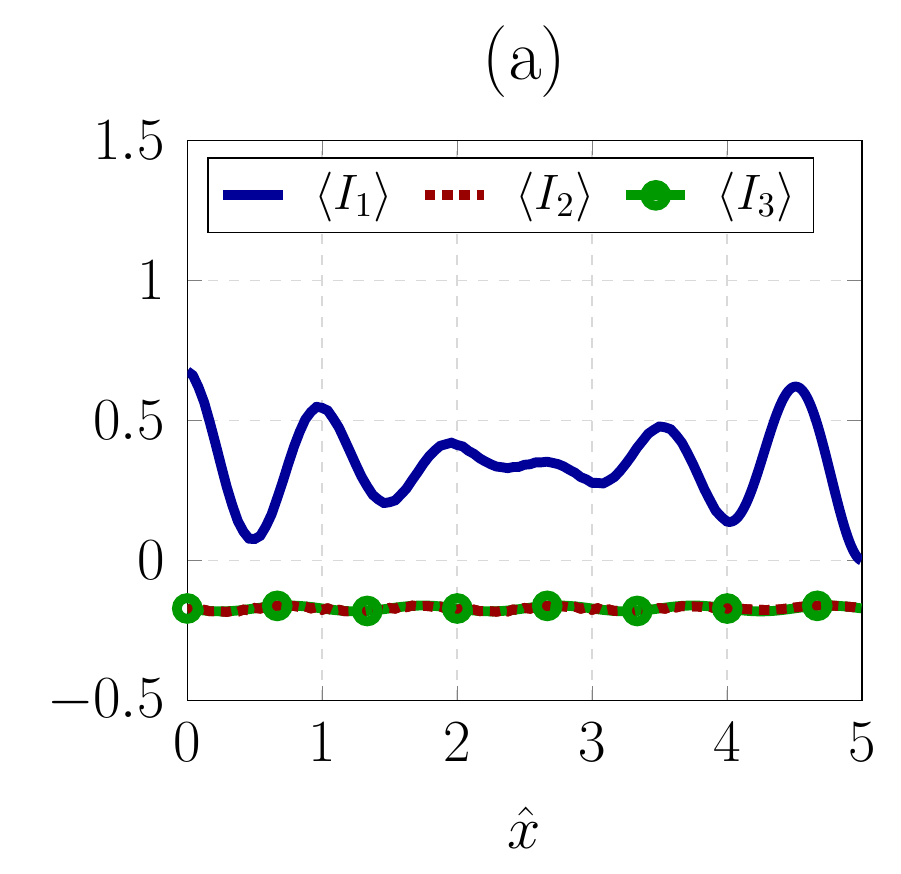}};
 \node[right=of img1, xshift=-1.2cm, yshift=0cm] (img2)  {\includegraphics[scale=0.65]{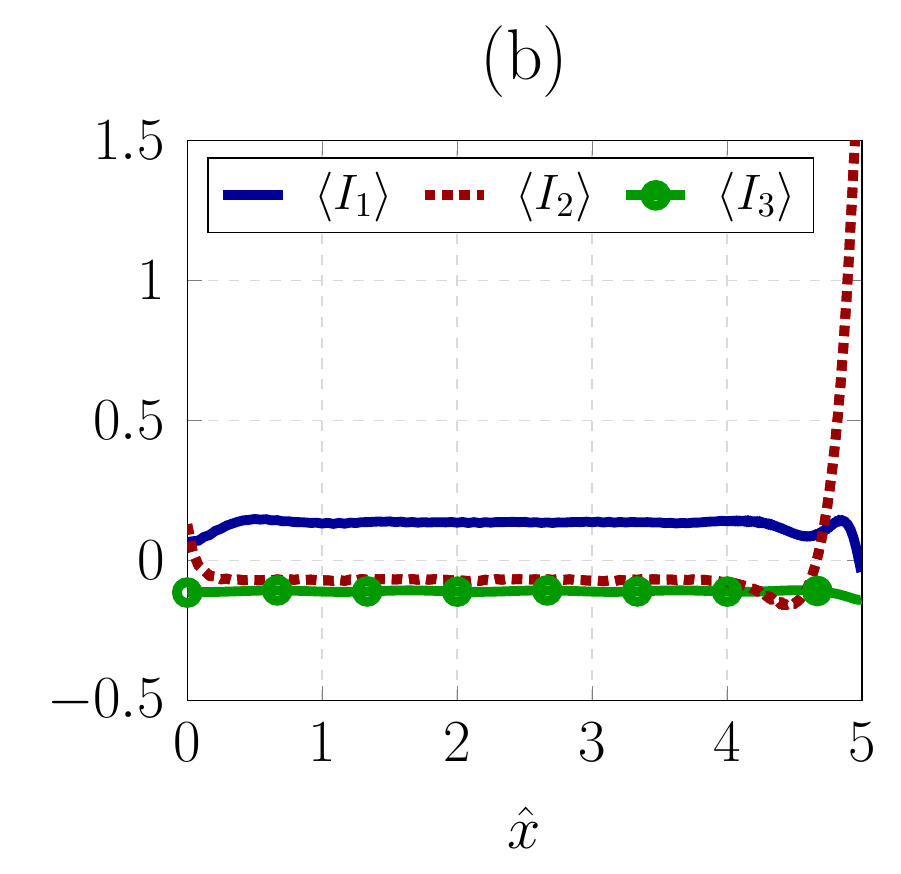}};
 \node[right=of img2, xshift=-1.2cm, yshift=0cm] (img3)  {\includegraphics[scale=0.65]{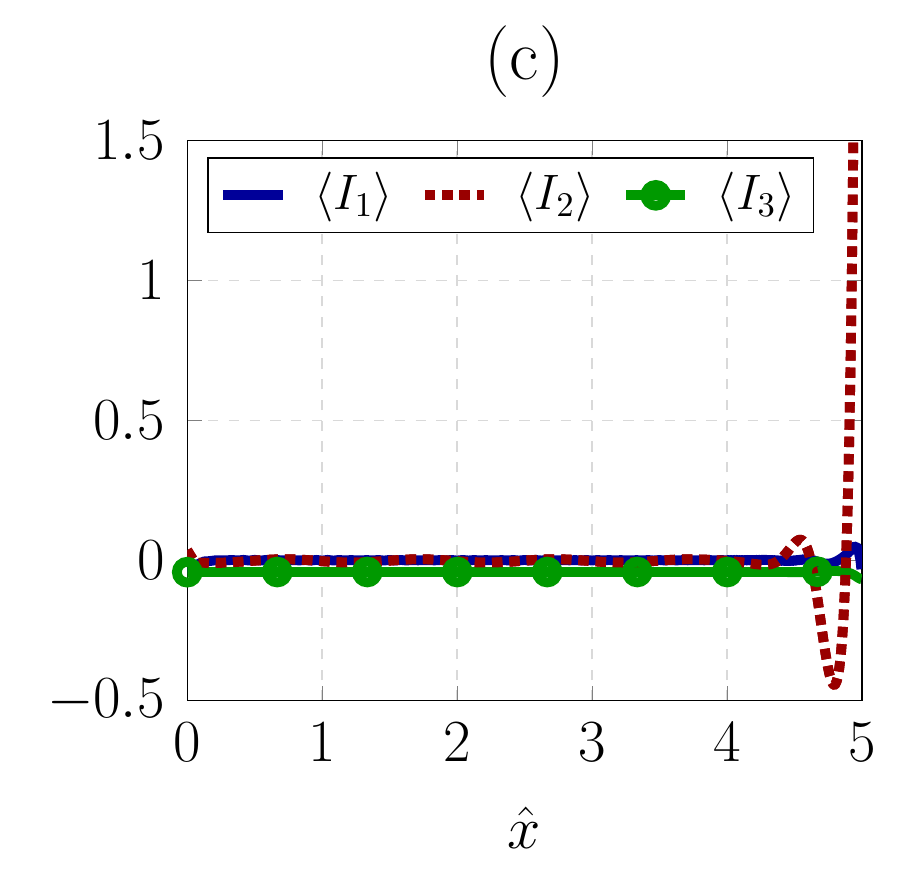}};
 \end{tikzpicture}
 \caption{Time-averaged distribution of the contributions $I_1$, $I_2$ and  $I_3$ to the horizontal force balance for different values of $\eta$. The softness parameter is (a) $\eta=0.001$ (very stiff substrate), (b) $\eta=5$ (midly soft substrate), (c) $\eta=1000$ (very soft substrate). Here, the input parameters are $A=0.25$, $n=5$, $\ha\approx3.14\times 10^{-3}$, $Ca\approx1.67\times 10^{-3}$.}
 \label{fig:Iforce2}
 \end{figure*}

However, this large-$\eta$ behavior is not corroborated by our experiments which instead show the gliding speed to be quasi-constant as the softness increases, so that \textit{M. xanthus} cells glide even on extremely soft substrates (see Fig.~\ref{fig:ExpSpeed}). Such a remarkable feature shows that modeling the substrate as a pure elastic half-space breaks down for large values of the softness parameter. Certainly, for very soft substrates, surface tension effects at the surface can no longer be totally ignored in creating the substrate deformation $\hdelta(\hx,\hti)$. The elasto-capillary balance of the substrate is particularly critical at the leading edge of the bacteria. In fact, the slime-air interfacial tension can generate, from the soft substrate, a "ridge" (see Fig.~\ref{fig:setupecMain}) which takes a shape determined by the balance of tensions at the substrate-slime-air triple line (see SI section 6.3). 

Capillary ridges are well-documented in the literature of soft solids \cite{Shanahan86, Shanahan88, Marchand12,Lubbers14, Style17}. Here, we consider that the growth of such a ridge creates a curvature of the slime-air interface, inducing a pressure difference at the leading edge of the cell. If the capillary effects are important, then the zero pressure condition at the leading edge, $\hp(n/2,\hti)=0$, no longer holds. This pressure must be obtained from an elasto-capillary-hydrodynamic problem, which we now set out to describe.  
Following the work by Limat \cite{Limat12,Dervaux15}, we show that when accounting for the capillary corrections, the substrate deformation now takes the dimensionless form (see SI sections 6.1-6.4)
\begin{equation}
\hdelta (\hx,\hti)\approx -\frac{1}{\pi}\int_{-n/2}^{n/2}\hp(\hx',\hti)\ln \left(|\hx-\hx'|+\frac{\xi}{\pi}\right) \mathrm{d}\hx',
\label{eq:thickdeformCapillary}
\end{equation}
In Eq.~(\ref{eq:thickdeformCapillary}), we introduced the elastocapillarity number $\xi$ defined by
\begin{center}
$\xi=\dfrac{2\gamma_s(1-\nu)}{GL}$,
\end{center}
where  $\gamma_s$ is the slime-substrate interfacial tension, which is around $\sim 56$mN/m for Agar gels at all concentrations \cite{Yoshitake08}. This dimensionless parameter compares capillary stresses at the slime-substrate interface to the elastic stresses in the bulk of the substrate. Therefore, for a given soft elastic substrate, the elasto-capillary length $\xi L$ provides the scale below (resp. above) which capillary (resp. elastic) effects dominate in the solid.  

Furthermore, we can write the balance of interfacial tensions (Young-Dupr\'e and Neumann's relationships \cite{Dervaux15}) at the capillary ridge and use Laplace's law across the slime-air interface to find the leading edge pressure. In the limit of small deformations and curvatures, and assuming a parabolic slime-air interface, we obtain the dimensionless pressure at the leading edge as (see SI sections 6.3-6.4)
\begin{align}
\hp(n/2,\hti) & = -\frac{\epsilon/Ca}{\hat{h}(n/2)-\eta\hdelta(n/2)} \times \nonumber \\ 
& \left(2\sqrt{1+\left[\mcal{R}\dfrac{\xi}{2\ha}\ln\left(1+\frac{2\ha}{\xi}\right)\right]^2}-2\right)^{1/2}.
\label{eq:pressureBC}
\end{align}
In Eq.~(\ref{eq:pressureBC}), $\mcal{R}=\gamma/\gamma_s$ is the ratio of the slime-air and slime-substrate interfacial tensions, $\epsilon$ is the lubrication parameter and $\ha$ is a measure of the the slime-air interfacial thickness which removes mathematical singularities of the displacement and slope at the triple line \cite{deGennes85,Lubbers14}. Lastly, \textit{Ca} is the wave-based capillary number comparing viscous to interfacial forces at the tip of the bacteria and is defined as
\begin{center}
$\textit{Ca}=\dfrac{\mu C}{\gamma}$.
\end{center}

Eq.~(\ref{eq:pressureBC}) accounts for both elastic and capillary effects for all values of the substrate softness. For very stiff substrates, the elasto-capillarity number $\xi\rightarrow 0$ and we recover $\hp(n/2,\hti) \approx 0$. Likewise, the substrate deformation given by Eq.~(\ref{eq:thickdeformCapillary}) converges to the equation of the deformation of a purely elastic half-space. 

For extremely soft substrates, the substrate deformation is dominated by the slime-substrate interfacial tension so that the elasto-capillary length is much larger than the bacterial length, i.e. $\xi L\gg nL$. We thus expand of the tip pressure and deformation fields in the limit $n \ll \xi\rightarrow\infty$ and also use the lift-free condition to reduce Eqs.~(\ref{eq:thickdeformCapillary}) and (\ref{eq:pressureBC}) to (see SI section 6.5)
\begin{align}
 \eta\hdelta(\hX,\hti)=\hdelta_{\infty}(\hX,\hti) & \simeq \frac{\mcal{R}Ca}{2\epsilon^3}\int_{-n/2}^{n/2}\hp(\hx',\hti)|\hx-\hx'|\mathrm{d}\hx' , \label{eq:BCCa1} \\
\hp(n/2,\hti) & \simeq  -\frac{\epsilon}{Ca}\frac{\left(2\sqrt{1+\mcal{R}^2}-2\right)^{1/2}}{\hat{h}(n/2,\hti)-\hdelta_{\infty}(n/2,\hti)}. \label{eq:BCCa2}
\end{align}
Eq.~(\ref{eq:BCCa2}) shows that the pressure at the leading edge scales as $\textit{Ca}^{-1}$. In other words, when \textit{Ca} is small, there exists at the leading edge a strong localized pressure sink, as hypothesized. This pressure pulls a ridge with a characteristic height ($\Delta_r$) and extent ($l_c$) as indicated in Fig.~\ref{fig:setupecMain}. The length scale $l_c$ is similar to a boundary layer thickness below (resp. above) which the localized capillary pressure is significant (resp. negligible). Assuming $l_c\ll L$, the drag-free gliding condition, Eq.~\ref{eq:Drag-free}, reduces to the balance between $I_2$ (driving pressure gradient) and $I_3$ (friction from gliding) for very soft substrates. Such a conclusion is supported by the simulations of the full elasto-capillary-hydrodynamics equations as shown in Fig.~\ref{fig:Iforce2} which shows the contributions of $I_1$, $I_2$ and $I_3$ to the horizontal force balance.

\begin{figure}[t!]
 \begin{tikzpicture}
   \node[xshift=0cm, yshift=0cm] (img1)  {\includegraphics[scale=0.7]{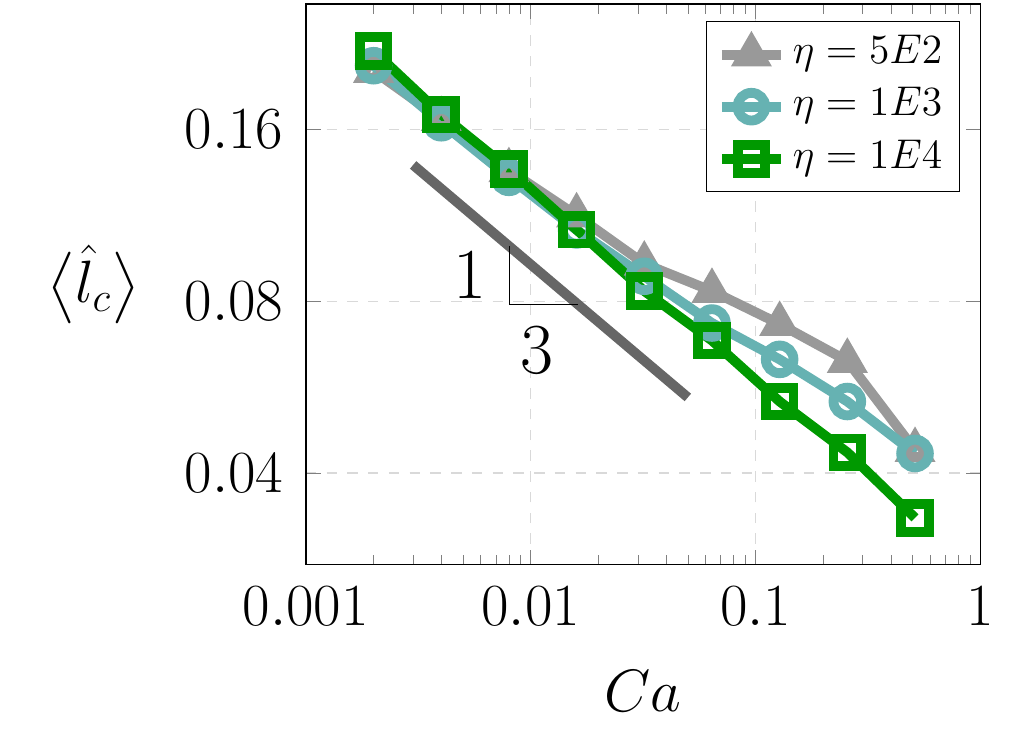}};
\node[ xshift=-1cm,yshift=-0.6cm] {$\mathlarger{\mathlarger{(a)}}$};
 \node[below=of img1,xshift=-0.2cm,yshift=1.2cm] (img2)  {\includegraphics[scale=0.7]{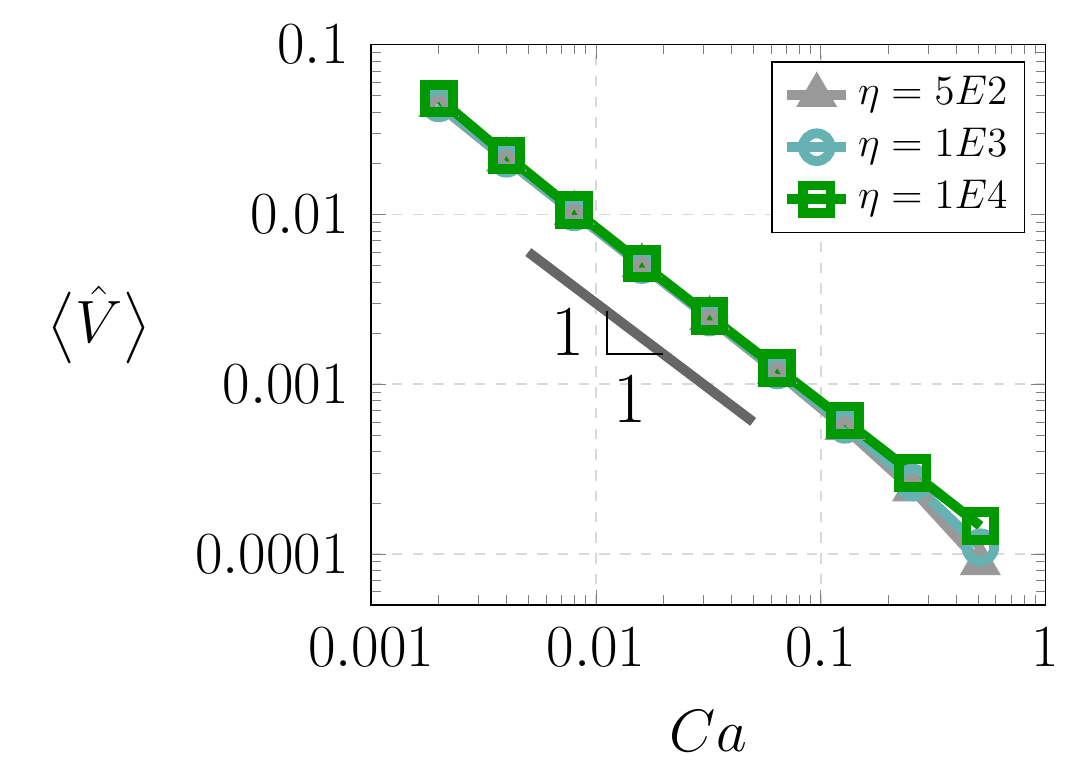}};
\node[ xshift=-1cm,yshift=-6cm] {$\mathlarger{\mathlarger{(b)}}$};
 \end{tikzpicture}
 \caption{Scaling with the capillary number of (a) the horizontal lengthscale of the ridge and (b) the gliding speed, for different values of the softness parameter. Here $l_c$ is arbitrarily defined as the distance from the leading edge where the slope of the substrate first vanishes, i.e. $\dfrac{d\left<\hdelta\right>}{d\hx}\left(\hx=\dfrac{n}{2}-\hat{l}_c\right)=0$. Here, the input parameters are $\epsilon=0.008$, $\hA=0.25$, $\ha\approx3.14\times 10^{-3}$ and $\mcal{R}\simeq0.1$.}
 \label{fig:scaling}
 \end{figure}
The comparison between Fig.~\ref{fig:Iforce} (a) and Fig.~\ref{fig:Iforce2} (a) shows that for very stiff substrates, the full solution reduces to that obtained by considering only the elasto-hydrodynamic interactions. However, as the softness increases, Fig.~\ref{fig:Iforce2} (b)-(c) show the growing effect of the leading edge pressure. While $I_3$ still contributes to a Couette-like friction, the term $I_2$ is now responsible for a non-zero thrust, largely due to the localized strong pressure gradient at the tip of the myxobacteria. In the capillarity-dominated regime ($n\ll \xi \rightarrow \infty$), we derive an asymptotic expression for the gliding speed on very soft substrates $\hV_{\eta=\infty}$, as (see SI section 6.5)
\begin{equation}
\hV_{\eta=\infty}(\hti)\approx 2-\Delta\hp\frac{(1-\alpha)}{\beta}+\alpha,
\label{eq:gspeedMain}
\end{equation}
where $$\Delta\hp=\hp(n/2,\hti)-\hp(-n/2,\hti),$$
\begin{center}
$\alpha=\left(1-\dfrac{3\zeta_2^2}{2\zeta_1\zeta_3}\right)^{-1}$,  $\beta=6\zeta_2$, and $\zeta_j=\mathlarger{\mathlarger{\int_{-n/2}^{n/2}}}\dfrac{\mathrm{d}\hx}{(\hh-\hdelta_{\infty})^j}$.
\end{center}

Since the leading edge pressure is proportional to the inverse capillary number, Eq.~(\ref{eq:gspeedMain}) also predicts that $\hV_{\infty}\sim Ca^{-1}$. In fact, a more refined scaling analysis of the governing equations in the limit of very soft substrates predicts (see SI section 7):
\begin{subequations}
\begin{align}
\dfrac{l_c}{h_0}  \sim & \mcal{R}^{1/3}Ca^{-1/3}, \\
 \dfrac{V}{C}  \sim & \dfrac{\epsilon \left(\sqrt{1+\mcal{R}^2}-1\right)^{1/2}}{n}Ca^{-1}.
\end{align}
\end{subequations}
In other words, while the gliding speed is independent of the bacterial length in an elasticity-dominated regime, the gliding speed $\left<\hV\right>$ scales with the bacterial length as $1/n$ when capillary effects dominate. This scaling law is obtained by balancing the localized capillary force at the slime-air interface and the viscous friction over the entire bacteria. Although the thrust results from the capillary-induced pressure at the leading edge and is thus essentially independent of bacterial length, the friction coefficient, given by $I_3/\hV$, depends on the cell geometry and increases with $n$. Hence, the resulting velocity decreases with $n$. The scaling laws given above are confirmed by computations of the full elasto-capillary-hydrodynamics equations, as shown in Fig.~\ref{fig:scaling} and Fig.~\ref{fig:SpeedCa}. As expected, we find $l_c\sim Ca^{-1/3}$ and $\left<V\right>\sim Ca^{-1}$ when the softness number is very large. 
The contrast between the elasticity-dominated and the capillary-dominated regimes is emphasized in Fig.~\ref{fig:SpeedCa}, which shows that $\left<\hV\right>$ is independent on the wavenumber for very stiff (small $\eta$) substrates, while $\left<\hV\right>\sim 1/n$ for very soft (large $\eta$) substrates. 

\begin{figure}[t!]
\centering
 \begin{tikzpicture}
   \node[xshift=0cm, yshift=0cm] (img1)  {\includegraphics[scale=0.7]{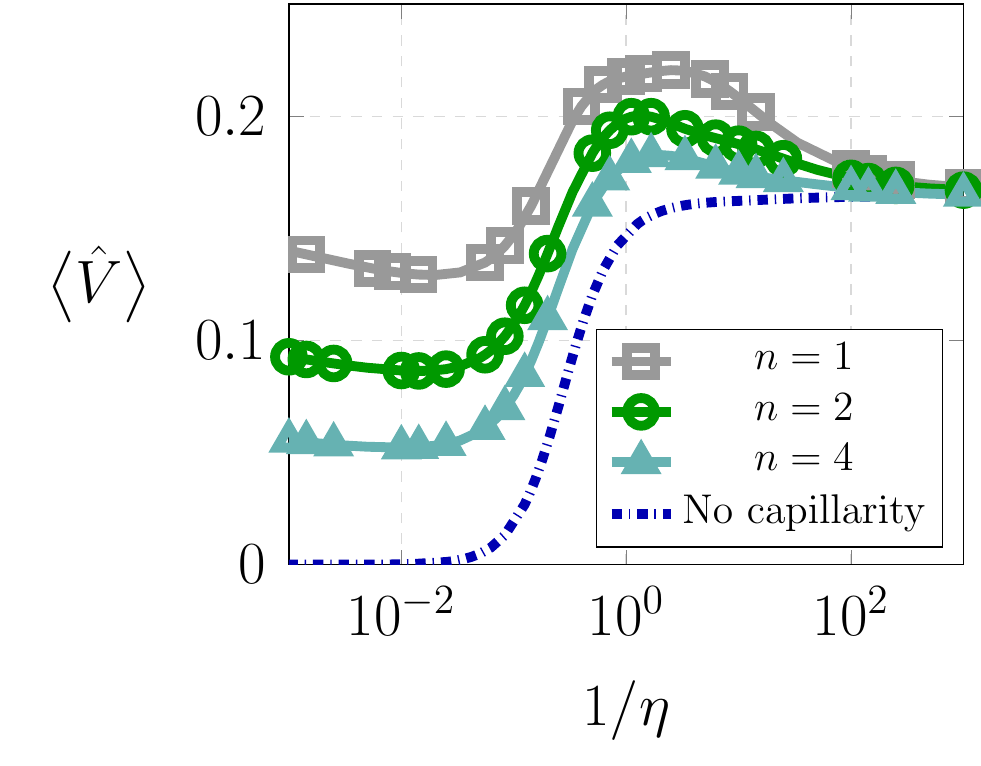}};
 \end{tikzpicture}
 \caption{Time-averaged gliding speed as a function of the softness parameter for different wavenumbers. Here, the input parameters are $\epsilon=0.008$, $\hA=0.25$, $\mcal{R}\approx 0.1$ and $Ca\approx1.67\times 10^{-3}$.}
 \label{fig:SpeedCa}
 \end{figure}

To test the validity of our theory, we cultured \textit{M. xanthus} cells in a rich Charcoal Yeast Extract (CYE) medium, spotted cell suspensions on agar gel pads, and measured the average gliding speed of A-motile cells on gels of different concentrations of agar, corresponding to different stiffnesses. To avoid the potential interference from S-motility, we used a $pilA^{-}$ strain, unable to produce functional type IV pilus. A good fit of collected data obtained by various authors \cite{Strange12, Oyen14} show that the shear modulus ($G$) of agar gels increases with concentration ($\mathcal{C}_{\mathrm{Agar}}$) according to the empirical law: $G\approx 20(\mathcal{C}_{\mathrm{Agar}}-0.1)^2$ kPa. Such power laws are more rigorously derived in the percolation theory for gels \cite{deGennes79}. After recording and post-processing the myxobacteria motion for a sample of $m=40$ cells, we obtained the mean and standard error of the gliding speed as shown in Fig.~\ref{fig:ExpSpeed}. The experimental data show a good agreement with the prediction our model. We obtain a good fit with our numerical solution, using the following parameters
\begin{flushleft}
$h_0=10$ nm, $\ell=5 \, \mu$m, $n=5$, $A\simeq3.33$ nm, $a\simeq 1$ nm, \\
$\mu=10$ Pa.s , $C=3 \, \mu$m/s, $\gamma_s=60$ mN/m, $\mcal{R}\simeq 0.16$.
\end{flushleft}  

\begin{figure}[t!]
\centering
\includegraphics[scale=0.8]{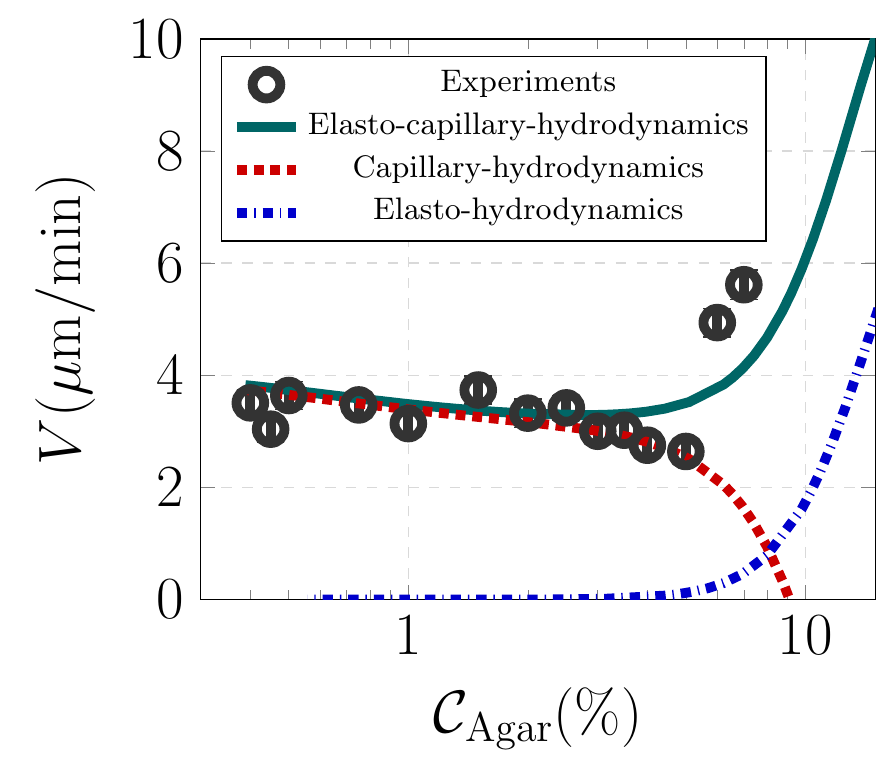}
 \caption{\footnotesize{Mean velocity of A-motile \textit{Myxococcus xanthus} cells as a function of the concentration of agar in the substrate. The experimental data (symbols) are reported in terms of the mean values with an error bar corresponding to the standard error, i.e the uncertainty on the estimate of the mean. The dark green line corresponds to our simulations (see text for parameters) of the elasto-capillary-hydrodynamic problem. The blue line dash-dot corresponds to the case of a pure elastic substrate while the red dash line is the asymptotic solution (\ref{eq:gspeedMain}) for a very soft substrate.}}
 \label{fig:ExpSpeed}
 \end{figure}
Fig.~\ref{fig:ExpSpeed} shows that our model (lines) agrees well with the experimental values of the mean gliding speed. Experimental data on Fig.~\ref{fig:ExpSpeed} confirm the existence of a two-regime behavior of the average gliding speed as a function of the substrate stiffness. At low agar concentrations, the gliding motion is due to capillary effects localized at the tip of the bacteria whose velocity follows the prediction of the asymptotic speed given by Eq.~(\ref{eq:gspeedMain}). As the concentration increases, the substrate gets stiffer and capillary effects at its surface decrease in favor of elastic ones in the bulk. The substrate being stiffer, it becomes harder to deform and causes the oscillations of the bacterial shape to be rather converted into variations of the slime pressure along the bacteria. Thus, there is a gradual switch in the nature of the gliding thrust, from a localized pressure gradient towards the slime-air interface to a distributed slime pressure over the bacteria. As the substrate becomes stiffer with the concentration, the slime pressure based thrust increases, leading to higher gliding speeds. 

As a final calculation, we estimated the thrust $\mathcal{T}$ required for the propulsion of a single myxobacteria cell. On substrates with $\mathcal{C}_{\mathrm{Agar}}\leq 3\%$, the motility thrust is balanced by the friction $I_3$ exerted by the slime on the cell, and consequently, it was calculated as (in dimensional form)
\begin{center}
$\mathcal{T}\approx 2R\mu\mathlarger{\mathlarger{\int_0^\ell}} \dfrac{V}{h-\delta}\mathrm{d}x \sim 2R\ell\dfrac{\mu V}{h_0},$
\end{center}
where $R\simeq 250$ nm is the radius of the rod-shaped bacteria. Using the parameters given above, along with $V\approx 3 \, \mu$m/min, we obtain $\mathcal{T}\approx 125$ pN. This value is in good agreement with other experimental and computational estimates ($\sim 50-150$ pN) of the propulsive force of single A-motile cells \cite{Wolgemuth05, Balagam14} and S-motile cells \cite{Clausen09, Sabass17} (which move at approximately the same speed).

We have presented a model for the gliding of single myxobacteria cells and their underlying substrate. It appears that the motor complexes slow dynamics can be modeled as a traveling wave, while the secreted slime serves as a lubricating film mediating the cell-substrate coupling. Our analysis shows that the mechanosensitivity of myxobacteria to their substrate results from their need to glide under drag-free and lift-free constraints. We find that satisfying these constraints can lead to different balances between the viscous, capillary and elastic forces depending on the substrate stiffness. This leads to two-regime behavior of the gliding velocity. On very soft substrates, the motility thrust is due to the existence of a localized capillary-induced pressure gradient towards the slime-air interface. But for much stiffer substrates, it originates from oscillations of the slime pressure in phase with the shape deformations over the bacteria length. Lastly, we estimated a single myxobacteria cell's thrust to be $\approx 125 \hspace{0.05cm} \mathrm{pN}$, in agreement with single-cell experiments. The speed-stiffness relationship investigated here improves our understanding of friction and substrate-mediated interaction between bacteria in a swarm of cells proliferating in soft media \cite{Gallegos06}. A crucial next step would be to consider the actual shape of the cell and its modification under the torque exerted by the slime pressure. Tackling such problems will require to determine stable three-dimensional deformations of the bacterial curved surface using membrane mechanics \cite{Sahu17} and could give insight in the shape-motility coupling of other rod-shaped cells. 

\acknow{We thank Amaresh Sahu and Yannick Omar for clarifying discussions and helpful comments on the manuscript. J.T. thanks Dr. Camille Duprat for helpful comments on elasto-capillarity aspects of the problem. J.T. and K.K.M. are supported by the NIH grant R01-GM110066.}

\showacknow 

\balance

\onecolumn
\begingroup
\fontsize{10pt}{12pt}\selectfont
\setcounter{figure}{0} 
\setcounter{equation}{0}
\renewcommand{\thefigure}{S\arabic{figure}}
\renewcommand{\theequation}{S\arabic{equation}}

\setcounter{subsection}{1}

\title{\center{\textbf{Supplementary Information \\ ``Mechanisms for bacterial gliding motility on soft substrates''} \\ \vspace{0.1in} \centerline{J. Tchoufag, P. Ghosh, C. B. Pogue, B. Nan and K. K. Mandadapu}}
}

\tableofcontents
\maketitle


\newpage

\section{Theoretical model of the slime flow}

Myxobacteria are rod-shaped cells that are $\sim 5 \, \mu$m in length and $\sim 0.5 \, \mu$m in diameter \cite{S-Nan11}. Experiments reveal that their basal shape $b(x,t)$ can be approximated as a sinusoid of characteristic wavelength $L\sim 1\, \mu$m \cite{S-Nan11, S-Pelling05} as shown in Fig.~\ref{fig:config}. These bacteria possess an ability to translocate themselves on surfaces by a motility mechanism also known as gliding. Myxobacteria gliding typically occurs at a velocity $V \sim 2-4 \, \mu$m/min and is always accompanied by a secretion of thin film of exopolysaccharide (EPS) slime underneath the cell \cite{S-Nan11b, S-Ducret12}. Assuming that the polysaccharide slime secreted by myxobacteria is a high viscosity fluid, its viscosity can be estimated as $\mu \sim 5-20\,$Pa.s \cite{S-Choi91, S-Isobe92}. Therefore, the typical Reynolds number $Re=\dfrac{\rho VL}{\mu}\sim 10^{-11} \approx 0$ as for the locomotion of most microorganisms \cite{S-Purcell77}. Note that in calculating the Reynolds number, the density is taken to be $\rho\sim 10^3\,$kg/m$^3$, since bacteria are mainly constituted of water and proteins. Given that the characteristic Reynolds number is almost zero, the problem is governed by the Stokes equations for the slime film \cite{S-Happel83}. 
 
\begin{figure}[htp!]
\centering
\begin{tikzpicture}
 \node (img1) {\includegraphics[scale=0.4]{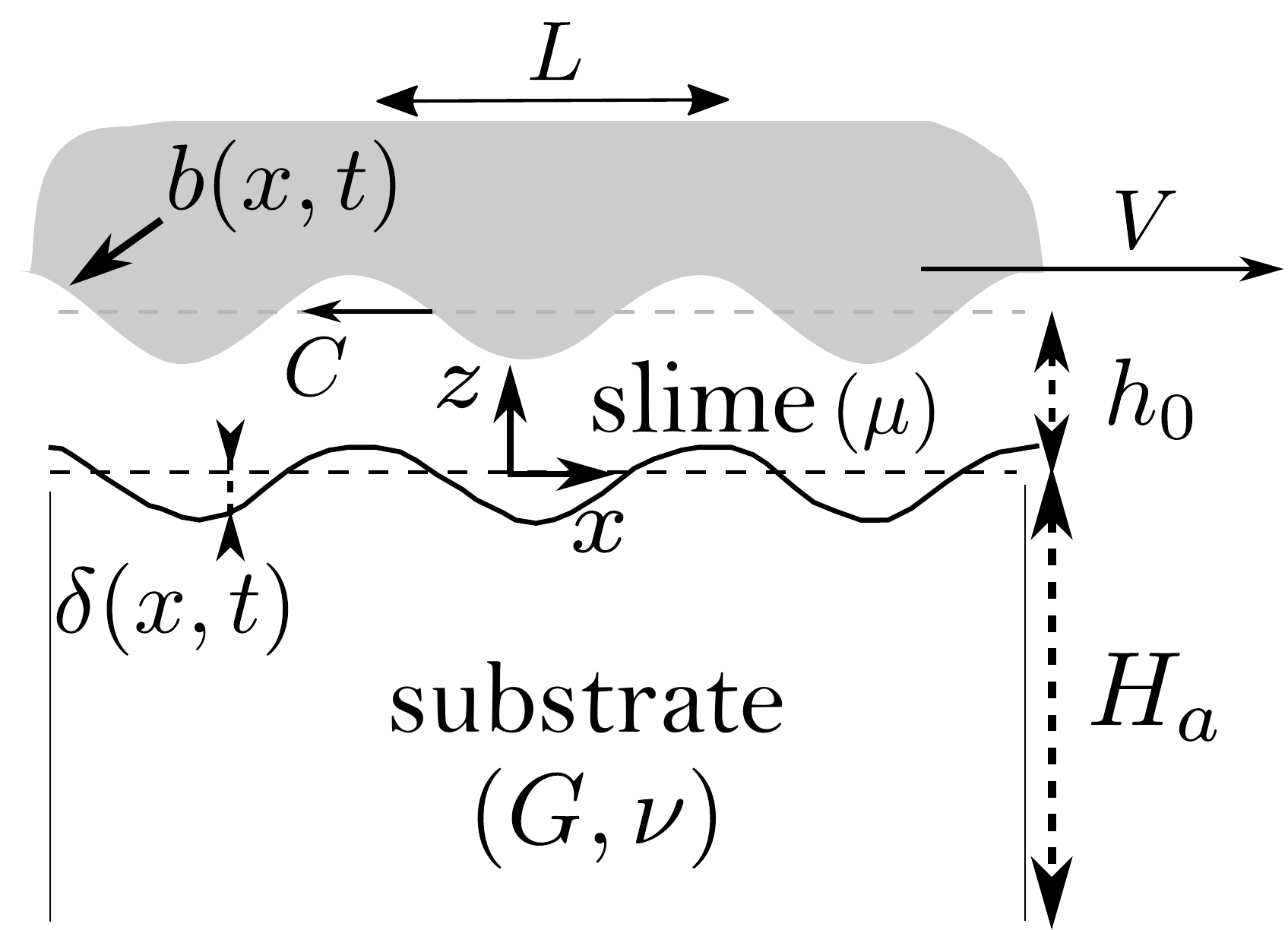}};
\end{tikzpicture}
\caption{Schematic description: a gliding bacterium (in grey) with a sinusoidal basal shape. The contact with the soft substrate is lubricated by a thin film of slime. See the text for a description of the variables.}
\label{fig:config}
\end{figure} 
 In what follows, we systematically develop the equations of motion governing the slime. 
\addtocounter{subsection}{1}  
\subsection{Stokes equations and lubrication approximation}  \label{sec:lubricationapproximation}

In this section, we derive the governing equation of motion of the thin film of slime, which we assume to behave as a viscous fluid. To begin with, let ${\boldsymbol {\sigma}}_\text{f}$ be the stress tensor in the fluid. Given low Reynolds number, the Stokes equations of motion governing the slime dynamics read
\begin{equation}
\nabla\cdot {\boldsymbol \sigma}_\text{f}=\bf{0}, 
\label{eq:zerodiv}
\end{equation}
where $\nabla$ denotes the gradient operator. 
We consider the slime as a Newtonian fluid, assuming that the strain rates exerted by the gliding cell on the film remain small during locomotion. The stress tensor is then given by the constitutive equation
\begin{equation}
{\boldsymbol \sigma}_\text{f}=-p{\bf{I}} + \mu\left(\nabla{\bf{u}}+\nabla{\bf{u}}^T\right),
\label{eq:fluidstresstensor}
\end{equation} 
where $\bf{u}$ is the velocity vector of the slime fluid, $\mu$ is the fluid dynamic viscosity, $p$ is the pressure, $\bf{I}$ is the identity tensor, $(\cdot)^T$ indicates the transpose of the matrix. Assuming the slime to be incompressible, using \eqref{eq:fluidstresstensor}, the governing equations of motion take the form
\begin{subequations}
\begin{align}
-\nabla p + \mu\nabla^2\bf{u}&=  0, \\
\nabla\cdot\bf{u}  &=  0.
\end{align}
\label{eq:Stokes}%
\end{subequations}

Given the geometric ratio of the interstitial gap $\epsilon=h_0/L\sim 10^{-2} \ll 1$, where $h_0$ is the mean thickness of the slime film and is approximately $10 \,$nm, we shall use the classical lubrication approximation \cite{S-Reynolds86, S-Leal07} for the slime film and further simplify the Stokes equations. To begin with, ignoring the rotation of the myxobacteria along its long axis, we confine the problem to 2-dimensions. To this end, rewriting Eq.~(\ref{eq:Stokes}) in component form, we obtain
\begin{subequations}
\begin{align}
-\frac{\partial p}{\partial x} + \mu\left(\frac{\partial^2 u_x}{\partial x^2}+\frac{\partial^2 u_x}{\partial z^2}\right) & = 0 , \\
- \frac{\partial p}{\partial z} +  \mu\left(\frac{\partial^2 u_z}{\partial x^2}+\frac{\partial^2 u_z}{\partial z^2}\right) & = 0 , \\
\frac{\partial u_x}{\partial x} +\frac{\partial u_z}{\partial z} & = 0,
\end{align}
\label{eq:Stokes2}%
\end{subequations}
see Fig.~\ref{fig:config} for the representation of axes. 
Next, we use the classical non-dimensional scaling involved in lubrication problems and obtain the following dimensionless variables (see Chapter~5 in \cite{S-Leal07}) 
\begin{center}
$x=L\hat{x}$, $z=\epsilon L \hat{z}$, $u_x=V \hat{u}_x$, $u_z=\epsilon V\hat{u}_z$, $p$=$\dfrac{\mu VL}{h_0^2}\hat{p}$=$\dfrac{\mu V/L}{\epsilon^2}\hat{p}$, 
\end{center}
where $\hat{( \cdot )}$ represents the corresponding variables in dimensionless form and $V$ is the speed of the bacteria that is to be determined. 
Using the aforementioned dimensionless variables, \eqref{eq:Stokes2} become
\begin{subequations}
\begin{align}
- \frac{\mu V}{h_0^2}\frac{\partial \hat{p}}{\partial \hat{x}} + \left(\epsilon^2\frac{\mu V}{h_0^2}\frac{\partial^2 \hat{u}_x}{\partial \hat{x}^2}+\frac{\mu V}{h_0^2}\frac{\partial^2 \hat{u}_x}{\partial \hat{z}^2}\right) & = 0 , \\
- \frac{1}{\epsilon}\frac{\mu V}{h_0^2}\frac{\partial \hat{p}}{\partial z} + \left(\epsilon^3\frac{\mu V}{h_0^2}\frac{\partial^2 \hat{u}_z}{\partial \hat{x}^2}+\epsilon\frac{\mu V}{h_0^2}\frac{\partial^2 \hat{u}_z}{\partial \hat{z}^2}\right) & = 0 , \\
\frac{V}{L}\frac{\partial \hat{u}_x}{\partial \hat{x}} +\frac{\epsilon V}{\epsilon L}\frac{\partial \hat{u}_z}{\partial \hat{z}} & = 0.
\end{align}
\label{eq:Stokes3}%
\end{subequations}
Neglecting all the terms of order $O(\epsilon^j)$ with $j\geq 1$, \eqref{eq:Stokes3} reduce to
\begin{subequations}
\begin{align}
-\frac{\partial p}{\partial x} +\mu\frac{\partial^2 u_x}{\partial z^2}&=0, \label{eq:momentum-x}\\
\frac{\partial p}{\partial z}&=0, \label{eq:momentum-z}\\
\frac{\partial u_x}{\partial x}+\frac{\partial u_z}{\partial z} &= 0. \label{eq:continuity}
\end{align}
\label{eq:lubrication}%
\end{subequations}
\eqref{eq:momentum-z} simply dictates that the pressure $p$ varies only in the x-direction in the lubrication approximation. \\

The boundary conditions associated with \eqref{eq:lubrication} correspond to no-slip at the bacteria-slime and slime-substrate interfaces, and are given in the reference frame of the gliding bacteria by
\begin{equation}\label{eq:bcf_1}
u_x(z=h)=0,  \ \ \ u_z(z=h)= \dfrac{\mathrm{d}h}{\mathrm{d}t},
\end{equation}
and 
\begin{equation}\label{eq:bcf_2}
u_x(z=\delta)=-V, \ \ \ u_z(z=\delta)=\dfrac{\mathrm{d} \delta}{\mathrm{d}t},
\end{equation}
where $\delta(x,t)$ is the vertical deformation of the surface of the soft substrate as indicated on Fig.~\ref{fig:config}, and $h(x,t) = h_0+b(x,t)$. In \eqref{eq:bcf_1} and \eqref{eq:bcf_2}, $\dfrac{\mathrm{d}(\cdot)}{\mathrm{d}t}$ denote the total time derivative. Note that the boundary condition at the slime-substrate interface, given by \eqref{eq:bcf_2}, assumes that in the laboratory frame, the material points of the substrate surface only move vertically at speed $\dfrac{\mathrm{d} \delta}{\mathrm{d}t}$.

\addtocounter{subsection}{1}
\subsection{Modified Reynolds equation}
In this subsection, we derive the modified form of the Reynolds equation corresponding to classical lubrication approximation \cite{Reynolds86}. To this end, integrating twice the horizontal momentum Eq.~(\ref{eq:momentum-x}) and the aforementioned boundary conditions, we obtain the horizontal component of the velocity field as
\begin{equation}
u_x(x,z,t)=\frac{1}{2\mu}\frac{\partial p}{\partial x}\left(z-h\right)\left(z-\delta\right)+V\frac{z-h}{h-\delta}.
\label{eq:uvel}
\end{equation}

We now integrate the continuity Eq.~(\ref{eq:continuity}) over the whole gap, i.e. from $z=\delta$ to $z=h$. Using Leibniz's integration rule, we obtain
\begin{subequations}
\label{eq:Leibniz}
\begin{align}
 & \int_{\delta}^h \frac{\partial u_z}{\partial z} dz=-\int_{\delta}^h \frac{\partial u_x}{\partial x} dz, \\
\implies & u_z(h)-u_z(\delta)=-\underbrace{\frac{\partial}{\partial x}\left(\int_{\delta}^h u_x dz \right)}_{\partial_xQ_h}+\frac{\partial \delta}{\partial x}, 
\end{align}
\end{subequations}
where $Q_h = \int_{\delta}^h u_x dz$ denotes the horizontal flow rate of slime across the gap between the substrate and the bacterial surface. Using the boundary conditions proposed in \eqref{eq:bcf_1} and \eqref{eq:bcf_2}, \eqref{eq:Leibniz} can be further reduced to 

\begin{subequations}\label{eq:height-flowrate}
\begin{align}
& \frac{\mathrm{d} h}{\mathrm{d} t}-\frac{\mathrm{d} \delta}{\mathrm{d} t}=-\frac{\partial Q_h}{\partial x}+\frac{\partial \delta}{\partial x}. \\
\implies & \frac{\partial h}{\partial t} -\frac{\partial \delta}{\partial t}+V\frac{\partial \delta}{\partial x}=-\frac{\partial Q_h}{\partial x}+V\frac{\partial \delta}{\partial x}. \\
\implies &  \frac{\partial }{\partial t} (h-\delta)+\frac{\partial Q_h}{\partial x}=0. \label{eq:height-flowrateC}
\end{align}
\end{subequations}
Using \eqref{eq:uvel}, the flow rate $Q_h$ is given by
\begin{equation}
Q_h=-\left[ \frac{1}{12\mu}\frac{\partial p}{\partial x}(h-\delta)^3+\frac{V}{2} (h-\delta) \right], 
\end{equation}
which upon substituting in \eqref{eq:height-flowrateC} leads to the modified Reynolds equation
\begin{equation}
-\frac{\partial }{\partial t}(h-\delta)+\frac{\partial}{\partial x}\left[ \frac{1}{12\mu}\frac{\partial p}{\partial x}(h-\delta)^3+\frac{V}{2} (h-\delta) \right] =0.
\label{eq:Reynoldsa}
\end{equation}

In what follows, we analyze the possibility of existence of traveling wave solutions of the modified Reynolds equation in \eqref{eq:Reynoldsa}. 

\addtocounter{subsection}{1} 
\subsection{Traveling wave solutions}\label{sec:travel_wave}
In the frame of reference translating with the cell, we consider the propagation of a traveling wave along the cell surface, motivated by recent experiments which reported that gliding is strongly correlated with molecular motor complexes moving with helical trajectories \cite{S-Nan11, S-Sun11, S-Nan13, S-Faure16, S-Fu18}. According to \cite{S-Nan11}, when \emph{viewed externally, the motors driving the rotation of the helical rotor generate transverse waves on the ventral surface}. In other words, the shape of the bacterial membrane, given by $h(x,t)=h_0+b(x,t)$, obeys the relation
\begin{equation}\label{eq:h_wave}
\frac{\partial h}{\partial t}=C\frac{\partial h}{\partial x}, 
\end{equation}
where $C$ is the phase speed of the wave. Hereafter, we shall consider $C$ to be a free parameter of our model. Seeking for unidirectional traveling wave solutions of the substrate deformation $\delta(x,t)$, we shall also assume
\begin{equation}
\frac{\partial \delta}{\partial t}=C_1\frac{\partial \delta}{\partial x}.
\label{eq:wave}
\end{equation}
Strictly, the speeds $C_1$ and $C$ need not be equal. However, in this work we shall only investigate the subspace of solutions such that $C_1=C$. Note that positive (resp. negative) values of $C$ correspond to left (resp. right) traveling disturbances. Using the traveling wave equations \eqref{eq:h_wave} $\&$ \eqref{eq:wave}, the modified Reynolds equation \eqref{eq:Reynoldsa} then becomes
\begin{equation}
\frac{\partial}{\partial x}\left[ \frac{\partial p}{\partial x}(h-\delta)^3+6\mu (V-2C) (h-\delta) \right] =0.
\label{eq:Reynolds}
\end{equation}
Eq.~(\ref{eq:Reynolds}) requires two boundary conditions on the pressure that we shall specify later. For now, we shall proceed to derive equations to determine the substrate deformation $\delta(x,t)$ and gliding speed $V(t)$. 

\section{Deformation of the soft elastic substrate}\label{sec:elasticsubstrate}
During the locomotion of bacteria, the substrate deforms due to the pressure loading imposed by the lubricating slime. In what follows, we derive the equations governing the deformation of the soft substrate. To obtain the deformations, we make the following assumptions:

\begin{itemize}
\item The horizontal and vertical length scales of the substrate are on the order of centimeters and millimeters respectively \cite{S-Chen90, S-Croze11}. Since both dimensions are much larger than the typical length of myxobacteria, we will represent the substrate as a semi-infinite medium. 
\item Polymer and gel substrates are generally viscoelastic. However, the relative importance of viscous to elastic effects depends on the excitation frequency of the material. Here, the characteristic frequency is that of the traveling disturbance: 
$$f= \frac{C}{L}\sim 3\, \mu \mathrm{m.s}^{-1}/1\, \mu \mathrm{m}= 3 \,\mathrm{Hz}, $$
where the estimated traveling wave speed $C \sim 3\, \mu$m.s$^{-1}$ corresponds to the experimental speed of the AlgR molecules, assuming the motors generate the traveling wave disturbance on the cell membrane \cite{S-Nan13}. In the case of substrates made of agar gels both at low and high  concentrations, dissipative effects in the material are negligible at such frequencies \cite{S-Nayar12}. Therefore, we shall consider the substrate to behave as a pure elastic half-space for bacterial locomotion.
\item Moreover, we restrict this analysis to the limit of small deformations $\delta \ll H_a$ such that linear elasticity theory is applicable. 
\item Last, we will neglect inertial effects in the substrate, and consider it to be in mechanical equilibrium at every instant. This equilibrium assumption is justified by the fact that the characteristic elastic wave speed (which exists in the presence of inertia), $\sqrt{G/\rho_s}$, is much larger than $V+C$, the gliding speed of the myxobacteria in the wave frame of reference where the shapes (of both the bacteria and the substrate) are fixed. Indeed, experimental values for \textit{M. xanthus} cells show $V\sim 4\mu$m/min, while Nan \textit{et al.} \cite{S-Nan13} reported the speed of the AglR motor complexes to be about $\sim 3\mu$m/s. As mentioned before, assuming the motors generate the traveling wave disturbance on the cell membrane, $C\sim 3\mu$m/s. Then, even for an extremely soft substrate at $0.5\%$ agar with $G\simeq30$ kPa \cite{S-Nayar12} and $\rho_s\simeq 10^3$kg/m$^3$, $(V+C) \ll \sqrt{G/\rho_s}\sim 5$m/s. 
\end{itemize}

Given the above assumptions and justifications, we shall now proceed to derive the deformation of the substrate determined by its elastic behavior using the theory of elasticity \cite{S-Landau84}. Let the three-dimensional stress tensor in the soft elastic substrate be denoted by
\begin{equation}
{\boldsymbol \sigma}_\text{s}(x,z) =	\begin{pmatrix}
       \sigma_x & \tau_{xy} & \tau_{xz}  \\
       \tau_{yx} & \sigma_y & \tau_{yz}\\
       \tau_{zx} & \tau_{zy} & \sigma_z       
     \end{pmatrix}.
\end{equation}
The equilibrium of the substrate is then governed by (see \cite{S-Landau84})
\begin{equation}
\nabla\cdot {\boldsymbol \sigma}_\text{s}=\bf{0}.
\label{eq:zerodivs}
\end{equation}

The deformation due to the pressure loading can be characterized by the (symmetric) strain tensor defined as
\begin{equation}
{\boldsymbol \epsilon}_\text{s}(x,z) =	\begin{pmatrix}
       \epsilon_x & \gamma_{xy} & \gamma_{xz}  \\
       \gamma_{yx} & \epsilon_y & \gamma_{yz}\\
       \gamma_{zx} & \gamma_{zy} & \epsilon_z       
     \end{pmatrix}
     =
     \begin{pmatrix}
       \dfrac{\partial u_\text{s}}{\partial x} & \dfrac{1}{2}\left(\dfrac{\partial u_\text{s}}{\partial y}+ \dfrac{\partial v_\text{s}}{\partial x}\right) & \dfrac{1}{2}\left(\dfrac{\partial u_\text{s}}{\partial z}+ \dfrac{\partial w_\text{s}}{\partial x}\right)  \\
       \dfrac{1}{2}\left(\dfrac{\partial u_\text{s}}{\partial y}+ \dfrac{\partial v_\text{s}}{\partial x}\right) & \dfrac{\partial v_\text{s}}{\partial y} & \dfrac{1}{2}\left(\dfrac{\partial v_\text{s}}{\partial z}+ \dfrac{\partial w_\text{s}}{\partial y}\right)\\
      \dfrac{1}{2}\left(\dfrac{\partial u_\text{s}}{\partial z}+ \dfrac{\partial w_\text{s}}{\partial x}\right)  & \dfrac{1}{2}\left(\dfrac{\partial v_\text{s}}{\partial z}+ \dfrac{\partial w_\text{s}}{\partial y}\right) & \dfrac{\partial w_\text{s}}{\partial z}      
     \end{pmatrix},
\end{equation}
where $(u_\text{s},v_\text{s},w_\text{s})$ is the displacement field of material points in the cartesian coordinate system represented in Fig.~\ref{fig:config}.

In the linear elasticity framework, the stress and the strain tensors are related through Hooke's law. This constitutive law states that for a substrate of Young's modulus $E$, Poissons' ratio $\nu$ and shear modulus $G=\dfrac{E}{2(1+\nu)}$, the components of ${\boldsymbol \epsilon}_\text{s}$ are related to those of ${\boldsymbol \sigma}_\text{s}$ according to \cite{S-Landau84}
\begin{subequations}
\begin{align}
\epsilon_x &= \frac{1}{E}[\sigma_x - \nu(\sigma_y +\sigma_z)],\\
\epsilon_y &= \frac{1}{E}[\sigma_y - \nu(\sigma_x +\sigma_z)],\\
\epsilon_z &= \frac{1}{E}[\sigma_z - \nu(\sigma_x +\sigma_y)],\\
\gamma_{xy} &=\frac{\tau_{xy}}{2G},\\
\gamma_{xz} &=\frac{\tau_{xz}}{2G},\\
\gamma_{yz} &=\frac{\tau_{yz}}{2G},
\end{align}
\label{eq:Hooke1}%
\end{subequations} 

\addtocounter{subsection}{1}
\subsection{Plane strain approximation}\label{sec:plane-strain}
Having considered a two-dimensional flow of slime, the pressure $p(x)$ exerted on the substrate is uniform in the $y$-direction and acts only on the distance spanning the bacterial cell length $\ell$ in the $x$-direction. Since the substrate thickness in the transverse $y$-dimension is of order $H_a \gg \ell$, we can use the plane strain approximation to reduce the problem from three-dimensions to two-dimensions~\cite{S-Johnson85}. 

In the plane strain approximation, the components of the strain tensor orthogonal to the plane $(x,z)$ vanish, so that
\begin{subequations}
\begin{align}
\epsilon_y=& 0,  \\
\gamma_{yx}=& \dfrac{1}{2}\left(\frac{\partial u_\text{s}}{\partial y}+\frac{\partial v_\text{s}}{\partial x}\right) =0, \\
\gamma_{yz}=& \dfrac{1}{2}\left(\frac{\partial v_\text{s}}{\partial z}+\frac{\partial w_\text{s}}{\partial y}\right) =0.
\end{align}
\end{subequations}
The only non-zero components of the strain tensor are then
\begin{eqnarray}
\epsilon_x = \frac{\partial u_\text{s}}{\partial x}, \ \ \epsilon_z=\frac{\partial w_\text{s}}{\partial z}, \ \ \text{and} \ \ \gamma_{xz} = \dfrac{1}{2}\left(\frac{\partial u_\text{s}}{\partial z} + \frac{\partial w_\text{s}}{\partial x}\right).
\label{eq:strain2d}
\end{eqnarray}
Using \eqref{eq:Hooke1}, the $y-$components of the stress tensor are thus found to be 
\begin{subequations}\label{eq:p_strain_stress}
\begin{align}
\sigma_y= &\nu(\sigma_x +\sigma_z), \\
 \tau_{yx}= & 0, \\
 \tau_{yz}= & 0. 
\end{align}
\end{subequations}

Hence, using \eqref{eq:p_strain_stress} in the plane strain approximation, the stress equilibrium equations~(\ref{eq:zerodivs}) reduce in the component form to 
\begin{subequations}\label{eq:reduced_eq}
\begin{align}
\frac{\partial \sigma_x}{\partial x} + \frac{\partial \tau_{xz}}{\partial z}=0,\\
 \frac{\partial \sigma_z}{\partial z} + \frac{\partial \tau_{xz}}{\partial x}=0.
\end{align}
\end{subequations}

Further, note that we have three strain components which are functions of only two displacements $u_\text{s}$ and $w_\text{s}$. Hence, the strain components are not all independent but are related through a \textit{compatibility condition} which can be readily obtained from (\ref{eq:strain2d}). Indeed, differentiating $\epsilon_x$ twice with respect to $z$, then $\epsilon_z$ with $x$, and $\gamma_{xz}$ once with respect to $x$ and once with respect to $z$, we obtain
\begin{equation}
\frac{\partial^2 \epsilon_x}{\partial z^2} + \frac{\partial^2 \epsilon_z}{\partial x^2} = 2\frac{\partial^2 \gamma_{xz}}{\partial x \partial z}.
\label{eq:compcond}
\end{equation}
By using Hooke's law, \eqref{eq:compcond} can also be transformed into a relation between the components of the stress tensor. Recalling that $\sigma_y=\nu(\sigma_x + \sigma_z)$ in the plane strain approximation, Hooke's law in Eq.~(\ref{eq:Hooke1}) can be rewritten as
\begin{subequations}\label{eq:reducedHooke_eq}
\begin{align}
\epsilon_x   =& \frac{1}{E}[(1-\nu^2)\sigma_x - \nu(1+\nu)\sigma_z],\\
\epsilon_z   =&  \frac{1}{E}[(1-\nu^2)\sigma_z - \nu(1+\nu)\sigma_x],\\
\gamma_{xz}  =& \frac{(1+\nu)}{E}\tau_{xz},
\end{align}
\end{subequations}
for the non-zero components of the strain tensor. Using these expressions and the reduced equilibrium equations~\eqref{eq:reduced_eq}, the strain compatibility relation in Eq.~(\ref{eq:compcond}) can be recast into a corresponding equation for the relation between stresses given by
\begin{eqnarray}
\left( \frac{\partial^2}{\partial x^2} + \frac{\partial^2}{\partial z^2} \right)(\sigma_x + \sigma_z)=0.
\end{eqnarray} 
Thus, the complete set equations to be solved to find the non-zero components of the stress tensor in the plane strain approximation are 
\begin{subequations}
\begin{align}
\frac{\partial \sigma_x}{\partial x} + \frac{\partial \tau_{xz}}{\partial z}=0,  \label{eq:stressa}\\
 \frac{\partial \sigma_z}{\partial z} + \frac{\partial \tau_{xz}}{\partial x}=0,  \label{eq:stressb}\\
\left( \frac{\partial^2}{\partial x^2} + \frac{\partial^2}{\partial z^2} \right)(\sigma_x + \sigma_z)=0. \label{eq:stressc}
\end{align}
\label{eq:stress}%
\end{subequations}

The governing equations of equilibrium~\eqref{eq:stress} are subjected to the following boundary conditions: 
\begin{itemize}
\item The stress at the surface must match the slime pressure and shear stress in the region spanned by the bacteria. Thus, for $-\ell/2 \leq x\leq \ell/2$, 
\begin{equation}\label{eq:ps_bc_1}
\sigma_z(x,z=0)=-p(x),  \text{  and   } \tau_{xz}(x,z=0)=q(x)=\mu\left(\dfrac{\partial u_x}{\partial z}+\dfrac{\partial u_z}{\partial x}\right).
\end{equation}
\item Outside the region overlapping the bacterium, the surface of the substrate is stress-free, i.e., for $x < -\ell/2$, and $x > \ell/2$ 
\begin{equation}\label{eq:ps_bc_2}
\sigma_z(x,z=0)=\tau_{xz}(x,z=0)= 0.  
\end{equation}
\item The substrate being a semi-infinite half-space, we also require the stress to vanish at distances far from the region overlapping the bacterium. Hence, ${\boldsymbol \sigma}_\text{s}(x,z)=\mathbf{0}$, for $||\mathbf{x}||\rightarrow\infty$, where $\mathbf{x}$ is a position vector corresponding to the semi-infinite substrate.
\end{itemize}

Denoting the quantities on the substrate surface at $z=0$ by $\bar{(\cdot)}$ symbol, the above described boundary conditions can be succinctly rewritten as:
\begin{itemize}
\item Within the loaded region ($-\ell/2 \leq x \leq \ell/2$), $\bar{\sigma}_z=-p(x)$, $\bar{\tau}_{xz}=q(x)=\mu\left(\dfrac{\partial u_x}{\partial z}+\dfrac{\partial u_z}{\partial x}\right)$.
\item Outside the loaded region ($x < -\ell/2$, $x > \ell/2$), $\bar{\sigma}_z = \bar{\tau}_{xz}=0$.
 \end{itemize}

\addtocounter{subsection}{1}
\subsection{Airy stress function}
In this section, we obtain an analytical solution for the elastic stresses and the surface deformation $\delta(x,t)$ of the soft elastic substrate so that they will be used in conjunction with the lubrication equations obtained in Section~\ref{sec:travel_wave}.

One method of solving the system of equations~(\ref{eq:stress}) is to introduce an auxiliary variable $\phi(x,z)$ known as the Airy stress function \cite{S-Johnson85}. It can be verified that the first two equations of the system of equations~(\ref{eq:stress}) are automatically satisfied for any function $\phi(x,z)$ such that
\begin{eqnarray}
\sigma_x = \frac{\partial^2 \phi}{\partial z^2};  \;\; \sigma_z =   \frac{\partial^2 \phi}{\partial x^2}; \;\; \tau_{xz} = -\frac{\partial^2 \phi}{\partial x \partial z}.
\end{eqnarray}
In solving for the Airy stress function $\phi(x,z)$, there exist multiple solutions that can be obtained from the equations of equilibrium~(\ref{eq:stressa},\ref{eq:stressb}). The solution of the problem is then selected to be one which satisfies the compatibility condition~(\ref{eq:stressc}). To this end, using the relations between $\sigma_x$, $\sigma_z$ and the Airy stress function in the compatibility condition \eqref{eq:stressc}, we see that $\phi(x,z)$ must satisfy the biharmonic equation
\begin{eqnarray}
\left(\frac{\partial^2}{\partial x^2}+\frac{\partial^2}{\partial z^2}\right)\left(\frac{\partial^2 \phi}{\partial x^2}+\frac{\partial^2 \phi}{\partial z^2}\right)=0,
\label{eq:biharmcartesian}
\end{eqnarray}
subject to the boundary conditions \eqref{eq:ps_bc_1} and \eqref{eq:ps_bc_2}.  

Since we treat the substrate as a half-space, it is convenient to solve for the Airy stress function $\phi$ by making use of polar coordinates $(r,\theta)$ as depicted on Fig.~\ref{fig:halfspace}. In order to obtain the equivalent form of \eqref{eq:biharmcartesian} in polar coordinates, we now proceed to give the equivalent forms of the stress equilibrium equation~\eqref{eq:reduced_eq}, the compatibility condition~\eqref{eq:compcond}, and Hooke's law~\eqref{eq:reducedHooke_eq} in $(r,\theta)$-coordinates. 

Under plane strain approximation, the stress equilibrium equations~\eqref{eq:zerodivs} of the substrate in polar coordinates read (see \cite{S-Timoshenko70})
\begin{subequations}
\begin{align}
\frac{\partial \sigma_r}{\partial r} +\frac{\sigma_r-\sigma_\theta}{r}+ \frac{1}{r}\frac{\partial \tau_{r\theta}}{\partial \theta} =0,  \label{eq:stresspola}\\
\frac{1}{r}\frac{\partial \sigma_{\theta}}{\partial \theta} +\frac{2}{r}\tau_{r\theta}+ \frac{\partial \tau_{r\theta}}{\partial r} =0.  \label{eq:stresspolb}
\end{align}
\label{eq:stresspol}%
\end{subequations}
It can be verified that \eqref{eq:stresspol} is satisfied for any stress function $\phi(r,\theta)$ such that
\begin{eqnarray}
\sigma_r =\frac{1}{r} \frac{\partial \phi}{\partial r}+\frac{1}{r^2} \frac{\partial^2 \phi}{\partial \theta^2},  \;\; \sigma_{\theta} =   \frac{\partial^2 \phi}{\partial r^2}, \;\; \tau_{r\theta} = -\frac{\partial }{\partial r} \left(\frac{1}{r}\frac{\partial \phi}{\partial \theta}\right),
\label{eq:stressfuncpolar}
\end{eqnarray}

Furthermore, corresponding to the state of stress given by $\sigma_r$, $\sigma_\theta$ and $\tau_{r\theta}$ are the strain tensor components $\epsilon_r$, $\epsilon_\theta$ and $\gamma_{r\theta}$ defined by (see \cite{S-Landau84})
\begin{subequations}
\begin{align}
\epsilon_r = & \frac{\partial u_r}{\partial r}, \\
 \epsilon_{\theta} = &  \frac{u_r}{r} +\frac{1}{r}\frac{\partial u_\theta}{\partial\theta},\\
 \gamma_{r\theta} = & \dfrac{1}{2}\left(\frac{1}{r}\frac{\partial u_r}{\partial \theta}+\frac{\partial u_\theta}{\partial r}-\frac{u_\theta}{r}\right),
\end{align}
\label{eq:strainpolar}%
\end{subequations}
where $u_r$ and $u_\theta$ are respectively the radial and azimuthal component of the displacement field in the substrate. 
As earlier, these displacements are not independent and are related through the following compatibility equation:
\begin{equation}\label{eq:compcondpol}
-\dfrac{\partial \epsilon_r}{\partial r}+\dfrac{1}{r}\dfrac{\partial^2 \epsilon_r}{\partial \theta^2}+\dfrac{\partial \epsilon_\theta}{\partial r}+\dfrac{\partial}{\partial r}\left(r\dfrac{\partial \epsilon_\theta}{\partial r} \right)=2\dfrac{\partial }{\partial \theta}\left[\dfrac{1}{r}\dfrac{\partial}{\partial r}\left(r\gamma_{r\theta}\right)\right],
\end{equation}
which can be obtained following similar procedure described for the cartesian coordinate system. 

Last, the constitutive equations (Hooke's law) between the stress and the strain components take the following form in polar coordinates
\begin{subequations}
\begin{align}
\epsilon_r  =&  \frac{1}{E}[(1-\nu^2)\sigma_r - \nu(1+\nu)\sigma_\theta],  \\
\epsilon_\theta  =&  \frac{1}{E}[(1-\nu^2)\sigma_\theta - \nu(1+\nu)\sigma_r],  \\
\gamma_{r\theta} =& \frac{(1+\nu)}{E}\tau_{r\theta}. 
\end{align}
\label{eq:stressstrainpol}%
\end{subequations}

Using the stress-strain relations~\eqref{eq:stressstrainpol}, the compatibility condition given by \eqref{eq:compcondpol} can be rewritten in terms of the stress tensor components. The newly found compatibility equation can then be rewritten as a function of the stress function $\phi(r,\theta)$ using \eqref{eq:stressfuncpolar}. After some manipulation, one again obtains the following biharmonic equation in the polar coordinate system:  
\begin{equation}
\left(\frac{\partial^2}{\partial r^2}+\frac{1}{r}\frac{\partial}{\partial r}+\frac{1}{r^2}\frac{\partial^2}{\partial \theta^2}\right)\left(\frac{\partial^2\phi}{\partial r^2}+\frac{1}{r}\frac{\partial\phi}{\partial r}+\frac{1}{r^2}\frac{\partial^2\phi}{\partial \theta^2}\right)=0.
\label{eq:biharm}
\end{equation}

Although analysis of the biharmonic equation~\eqref{eq:biharm} can be found in classical textbooks on elasticity \cite{S-Johnson85}, the solutions of $\phi$ in polar coordinates found therein are only partial solutions. Only recently \cite{S-Stampouloglou10, S-Doschoris12}, additional solutions were obtained for the biharmonic equation. Moreover, the solution for the surface deformation $\delta$ of an elastic half space, which is necessary in the context of the current problem, has been obtained in \cite{S-Johnson85} with previously obtained partial solutions. In what follows, we provide a complete derivation of the general solution of the Airy stress function as provided in \cite{S-Stampouloglou10, S-Doschoris12}, before specifying the particular solutions which satisfy the boundary conditions imposed for the gliding bacteria on the half-space substrate. In performing this exercise, we find that even the newly found general solutions for the biharmonic equation yield the same expression for the substrate deformation $\delta$ as that found in \cite{S-Johnson85} based on the partial solutions.

\begin{figure}[t!]
\centering
 \begin{tikzpicture}
  \node (img1) {\includegraphics[scale=0.6]{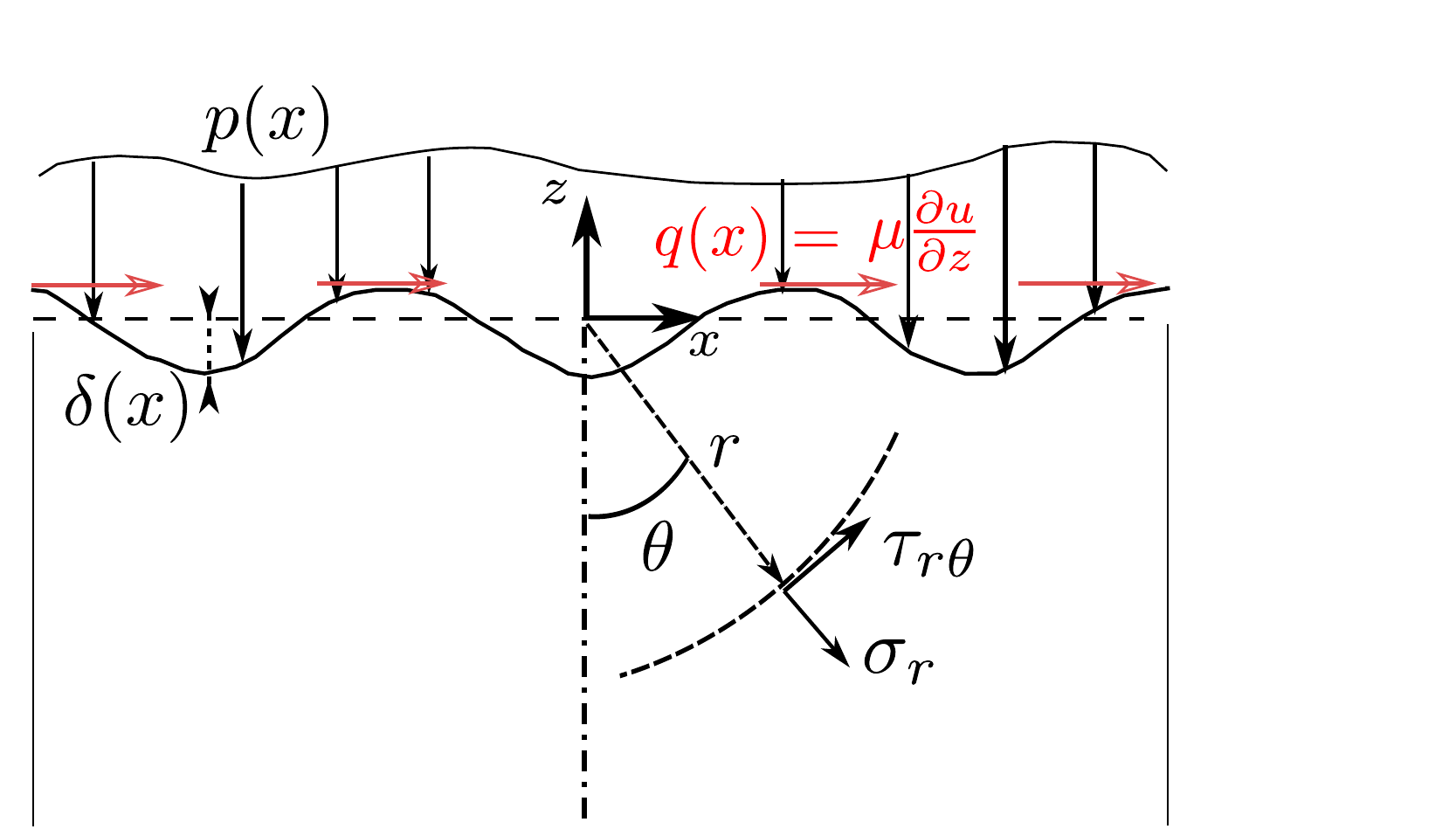}};
  \end{tikzpicture}
 \caption{Schematic representation of the substrate treated as an elastic half-space under pressure and viscous shear stress loads due to the slime. Also shown is the polar coordinate system representing the half space occupied by the substrate.} 
 \label{fig:halfspace}
 \end{figure} 

\addtocounter{subsection}{1}
\subsection{Solution to the biharmonic equation}
In this section, we derive the solution for the biharmonic equation as provided in \cite{S-Stampouloglou10, S-Doschoris12}.
To begin, using the method of separation of variables, let
\begin{equation}\label{eq:sep_var}
\phi(r,\theta)=\mathcal{R}(r)\Theta(\theta)
\end{equation}
with $\mathcal{R}(r)\neq 0$ and $\Theta(\theta)\neq 0$. Using Eq.~(\ref{eq:sep_var}), Eq.~(\ref{eq:biharm}) reduces to 
\begin{equation}
\frac{d^4\Theta}{d \theta^4}+2\mathcal{R}_1\frac{d^2\Theta}{d \theta^2}+\mathcal{R}_2\Theta=0,
\label{eq:biharm2}
\end{equation}
where 
$$\mathcal{R}_1=\frac{r^2\mathcal{R}''-r\mathcal{R}'+2\mathcal{R}}{\mathcal{R}},$$ and $$\mathcal{R}_2=\frac{r^4\mathcal{R}''''+2r^3\mathcal{R}'''-r^2\mathcal{R}''+r\mathcal{R}'}{\mathcal{R}}$$
and $(\cdot)'$ denotes derivative with respect to $r$. In order to further separate the solution into $\mathcal{R}$ and $\Theta$, we eliminate the first term on the left-hand-side of Eq.~(\ref{eq:biharm2}) by differentiating with respect to $r$. leading to
\begin{equation}
2\mathcal{R}_1' \frac{d^2\Theta}{d \theta^2}+\mathcal{R}_2'\Theta=0.
\label{eq:biharm3}
\end{equation}
Two cases can be distinguished:
\begin{enumerate}
\item For the first case, we consider $\mathcal{R}_1' \neq 0$ and find
\begin{equation}
\frac{2}{\Theta}\frac{d^2\Theta}{d \theta^2}=\frac{\mathcal{R}_2'}{\mathcal{R}_1'}=\alpha,
\label{eq:case1a}
\end{equation}
where $\alpha$ is a constant. Therefore, the function $\Theta(\theta)$ is periodic if $\alpha<0$, or a superposition of hyperbolic functions if $\alpha>0$, or a linear function if $\alpha=0$. In any of these cases, we can rewrite $\dfrac{d^2\Theta}{d \theta^2}=\dfrac{\alpha}{2}\Theta$ and $\dfrac{d^4\Theta}{d \theta^4}=\dfrac{\alpha^2}{4}\Theta$, which then reduces \eqref{eq:biharm2} to the following 4th order ODE for $\mathcal{R}$ in $r$ 
\begin{equation}
\frac{\alpha^2}{4}+\alpha\mathcal{R}_1+\mathcal{R}_2=0,
\label{eq:case1b}
\end{equation}
whose solutions are of the form $\mathcal{R}\sim r^n$ with $n\in\mathbb{Z}$, where $\mathbb{Z}$ is the space of integers \cite{S-Sadeh66, S-Doschoris12}. This leads to the characteristic equation for $n$ given by 
\begin{equation}
\frac{\alpha^2}{4}+\alpha\left(n(n-2)+2\right)+n(n-2)(n-3)=0,
\label{eq:characteq}
\end{equation}
whose determinant is $\Delta_k=4(n-1)^2$. Therefore,
$$(n-1\pm 1)^2 = -\frac{\alpha}{2},$$ 
where $-\alpha/2$ should be a positive number for real solutions to exist. We now proceed to analyze different scenarios by defining the variable $\omega=\pm\sqrt{-\alpha/2}$ and consider cases when $\omega \geq 0$ and $\omega < 0$.

\paragraph{Choice of $\omega\geq 0$:}
This leads to three solutions: one corresponding to $\omega=0$ for which $\Theta(\theta)$ is linear, and two solutions corresponding to $\omega\neq 0$ when $\Theta(\theta)$ is periodic. In the latter category, we will distinguish the cases $\omega=1$ and $\omega\geq 2$, since the former corresponds to a double root of the characteristic equation~(\ref{eq:characteq}).

\begin{itemize}
\item \underline{Solution 1}: For $\omega=0$ (i.e., $n=0$ or $n=2$),  Eq.~(\ref{eq:case1a}) shows that $\Theta=c_{01}\theta+c_{02}$. Herein, variables $c_i$, $c_{i j}$, $c_{i j k}$ where ${i,j,k}\in\mathbb{N}$, are used to denote constants. Since $\alpha=0$, Eq.~(\ref{eq:case1b}) reduces to $\mathcal{R}_2=0$.  To find additional solutions to $\mathcal{R}=c_{03}r^2$ (since $n=2$ is a double root of the characteristic equation $\mathcal{R}_2=0$), we use the method of variation of constants and search for a function $f_1(r)$ such that $\mathcal{R}=f_1(r)r^2$ is also a solution of $\mathcal{R}_2=0$. This amounts to solving 
\begin{equation}
r^4f_1''''+10r^3f_1'''+23r^2f_1''+9rf_1'=0.
\label{eq:case1c}
\end{equation}

Starting again with the ansatz $f_1'=r^m$, one finds $m\in\{-3,-1\}$ leading to $f_1=c_{04}r^{-2}+c_{05}\ln{r}+c_{06}$. However, here as well, $m=-3$ is double root. Repeating the process with the search for another function $g(r)$ such that $g(r)r^{-2}$ is a solution of Eq.~(\ref{eq:case1c}), we find $g(r)=c_{07}\ln{r}+c_{08}r^2\ln{r}+c_{09}r^2$. Given these choices, $f_1(r)$ can be recast as $f_1(r)=c_{04}r^{-2}+c_{010}\ln{r}+c_{07}r^{-2}\ln{r}+c_{011}$. Therefore, the first set of solution $\mathcal{R}(r)\Theta(\theta)$ reads
\begin{equation}\label{eq:phi_0}
\phi_0(r,\theta)=(a_0r^2+b_0r^2\ln{r}+c_0+d_0\ln{r})\theta+(\tilde{a}_0r^2+\tilde{b}_0 r^2\ln{r}+\tilde{c}_0 +\tilde{d}_0\ln{r}), 
\end{equation}
where the constants $c_{ij}$ have been combined into new ones. 
\item \underline{Solution 2}: For $\omega=1$ (i.e., $n=1$), we have a double root of the characteristic equation. To find the additional solution to $\mathcal{R}=c_{10}r$, we again use the method of variation of constants and search for $f_2(r)$ such that $\mathcal{R}=f_2(r)r$ is also a solution of Eq.~(\ref{eq:case1b}). This amounts to solving the equation
\begin{equation}
r^4f_2''''+6r^3f_2'''+3r^2f_2''-3rf_2'=0.
\end{equation}
Starting again with the ansatz $f_3'=r^m$, one finds that $m\in\{-3,-1,1\}$. This leads to $f_2(r)=c_{11}r^2+c_{12}\ln{r}+c_{13}r^{-2}+c_{14}$, which results in $\mathcal{R}=c_{11}r^3+c_{12}r\ln{r}+(c_{10}+c_{14})r+c_{13}r^{-1}$. Moreover, for $\alpha/2=-1$, Eq.~(\ref{eq:case1a}) shows that $\Theta=c_{15}\cos{\theta}+c_{16}\sin{\theta}.$ Therefore, recombining the various constants $c_{ij}$ into new ones, the second set of solutions $\mathcal{R}(r)\Theta(\theta)$ are given by
\begin{equation}\label{eq:phi_1}
\phi_1(r,\theta)=(a_1r^3+b_1r^{-1}+c_1r+d_1r\ln{r})\cos{\theta}+(\tilde{a}_1 r^3+\tilde{b}_1 r^{-1}+\tilde{c}_1 r+\tilde{d}_1 r\ln{r})\sin{\theta}.
\end{equation}

\item \underline{Solution 3}: Last, when $\omega \geq 2$, there are four types of solutions corresponding to periodic functions of $\theta$ of pulsation $\omega \in\{\pm(n-2),\pm n\}$:
\begin{equation}\label{eq:phi_2}
\begin{split}
\phi_2(r,\theta) & =\sum\limits_{n=4}^{\infty} r^n\left[c_{n1}\cos{(n-2)\theta}+\tilde{c}_{n1} \sin{(n-2)\theta}\right]+\sum\limits_{n=2}^{\infty} r^n\left[c_{n2}\cos{n\theta}+\tilde{c}_{n2} \sin{n\theta}\right] \nonumber \\
& \hspace{1.0in} + \sum\limits_{n=-\infty}^{0} r^n\left[c_{n3}\cos{(-n+2)\theta}+\tilde{c}_{n3} \sin{(-n+2)\theta}\right] \\
& \hspace{1.5in} +\sum\limits_{n=-\infty}^{-2} r^n\left[c_{n4}\cos{(-n\theta)}+\tilde{c}_{n4} \sin{(-n\theta)}\right]. 
\end{split}
\end{equation}

Applying changes of variables $(n-2)=m_1$, $-n+2=m_3$ and $-n=m_4$, to the first, third and fourth terms of the right hand side of \eqref{eq:phi_2}, we obtain 
\begin{equation}
\begin{split}
\phi_2(r,\theta) &=\sum\limits_{m_1=2}^{\infty} r^{m_1+2}\left[c_{(m_1+2)1}\cos{m_1\theta}+\tilde{c}_{(m_1+2)1} \sin{m_1\theta}\right]+\sum\limits_{n=2}^{\infty} r^n\left[c_{n2}\cos{n\theta}+\tilde{c}_{n2} \sin{n\theta}\right] \nonumber \\
& \hspace{1.0in}+\sum\limits_{m_3=2}^{\infty} r^{-m_3+2}\left[c_{(2-m_3)3}\cos{m_3\theta}+\tilde{c}_{(2-m_3)3}\sin{m_3\theta}\right] \\ 
& \hspace{1.5in} +\sum\limits_{m_4=2}^{\infty} r^{-m_4}\left[c_{(-m_4)4}\cos{(m_4\theta)}+\tilde{c}_{(-m_4)4}\sin{(m_4\theta)}\right], 
\end{split}
\end{equation}
which can be further rewritten as 
\begin{eqnarray}
\sum\limits_{n=2}^{\infty} \phi_{2n}(r,\theta) =\sum\limits_{n=2}^{\infty} \left[c_{n1}r^{n+2}+c_{n2}r^{n}+c_{n3}r^{-n+2}+c_{n4}r^{-n}\right]\cos{n\theta} \nonumber \\
+\left[\tilde{c}_{n1}r^{n+2}+\tilde{c}_{n2}r^n+\tilde{c}_{n3}r^{-n+2}+\tilde{c}_{n4}r^{-n}\right]\sin{n\theta}. \nonumber
\end{eqnarray}
\end{itemize}
This ends the analysis for possible solutions for $\omega \geq 0$. We now proceed to analyze the choice of $\omega<0$.

\paragraph{Choice of $\omega< 0$:} This leads to two solutions corresponding to $\omega=-1$ and $\omega\leq -2$. 
\begin{itemize}
\item \underline{Solution 1}: For $\omega=-1$ (i.e., $n=1$), the function $\mathcal{R}(r)$ is the same as in the case of $\omega=1$. However, $\Theta$ is not a periodic solution in contrast to the previous case, but is made of super-position of $\cosh(\theta)$ and $\sinh(\theta)$ with the same analysis as before. This leads to the following solution: 
\begin{equation}
\begin{split}
\phi_3(r,\theta) &= (a_1r^3+b_1r^{-1}+c_1r+d_1r\ln{r})\cosh({\theta}) \\
& \hspace{1in} +(\tilde{a}_1r^3+\tilde{b}_1r^{-1}+\tilde{c}_1r+\tilde{d}_1r\ln{r})\sinh({\theta}).
\end{split}
\end{equation}
\item \underline{Solution 2}: For $\omega\leq -2$, we proceed as in the case of $\omega\geq 2$ to solve $\dfrac{d^4\Theta}{d\theta^4}=\omega^4\Theta$. The solution takes a similar form, with the periodic $\cos$ and $\sin$ functions being replaced by the hyperbolic $\cosh(n\theta)$ and $\sinh(n\theta)$ functions. Again, following similar arguments as made for the case $\omega \leq -2$, we obtain the fifth and last set of solutions for the case $\mathcal{R}_1 = 0$ as
\begin{eqnarray}
\sum\limits_{n=2}^{\infty} \phi_{4n}(r,\theta) =\sum\limits_{n=2}^{\infty} \left[c_{n1}r^{n+2}+c_{n2}r^{n}+c_{n3}r^{-n+2}+c_{n4}r^{-n}\right]\cosh({n\theta}) \nonumber \\
+\left[\tilde{c}_{n1}r^{n+2}+\tilde{c}_{n2}r^n+\tilde{c}_{n3}r^{-n+2}+\tilde{c}_{n4}r^{-n}\right]\sinh({n\theta}). \nonumber
\end{eqnarray}

\end{itemize}
We now proceed onto the second case when $\mathcal{R}_1'  = 0$.

\item 
Due to Eq.~(\ref{eq:biharm3}), $\mathcal{R}_1'  = 0$ also implies $\mathcal{R}_2'=0$. Therefore, $\mathcal{R}_1=C_1$ and $\mathcal{R}_2=C_2$ where $C_1$ and $C_2$ will be found so that they correspond to the same solution $\mathcal{R}$. Injecting guess solutions of the form $\sim r^n$ into $\mathcal{R}_1-C_1=0$ and $\mathcal{R}_1-C_2=0$, we find that 
\begin{subequations}
\begin{align}
C_1 & =(n-1)^2+1, \\
C_2 & =n^2(n-2)^2.
\end{align}
\label{eq:complexC1C2}
\end{subequations}
\paragraph{First,}let us consider only real values of $C_1$ such that $C_1-1=\beta^2\geq 0$. So, we find $n\in\{-\beta+1, \beta+1\}$ and $C_2=(\beta+1)^2(\beta-1)^2$.  This leads to three sets of solutions:

\begin{itemize}
\item \underline{Solution 1}: For $\beta=0$, $C_1=1$ and $C_2=1$. So, $\mathcal{R}_1-C1=0$ becomes
\begin{equation}
r^2\mathcal{R}''-r\mathcal{R}'+\mathcal{R}=0.
\label{eq:RC1_1}
\end{equation}
Proceeding as before to solve \eqref{eq:RC1_1}, we find that $\mathcal{R}=c_{01}r+c_{02}r\ln{r}$. Since $C_2=1$ in this case, Eq.~(\ref{eq:biharm2}) for $\Theta$ becomes
\begin{equation}
\frac{d^4\Theta}{d \theta^4}+2\frac{d^2\Theta}{d \theta^2}+\Theta=0.
\label{eq:TC1_1}
\end{equation}
In order to solve \eqref{eq:TC1_1}, we start with guess solutions of the form $\sim e^{k\theta}$ and find $k=\pm i$, where $i=\sqrt{-1}$. Then applying the method of variation of constants, we obtain the solution as
$\Theta = (c_{03}+c_{04}\theta)\cos\theta+(c_{03}'+c_{04}'\theta)\sin\theta.$
So, the first set of solutions $\mathcal{R}(r)\Theta(\theta)$ takes the form 
$$\tilde{\phi_0}(r,\theta)=\left[c_{01}r+c_{02}r\ln{r}\right]\left[(c_{03}+c_{04}\theta)\cos\theta+(\tilde{c}_{03}+\tilde{c}_{04}\theta)\sin\theta\right].$$
We note that the periodic part of this solution is already accounted for in $\phi_2(r,\theta)$ and thus reduce this set of solutions to

$$\tilde{\phi_0}(r,\theta)=\left(c_{1}r+c_{2}r\ln{r}\right)\theta\cos\theta+\left(\tilde{c}_{1}r+\tilde{c}_{2}r\ln{r}\right)\theta\sin\theta.$$

\item \underline{Solution 2}: For $\beta =1$, $C_1=2$ and $C_2=0$. So, the equation $\mathcal{R}_1-C_1=0$ becomes
\begin{equation}
r^2\mathcal{R}''-r\mathcal{R}'=0.
\label{eq:RC1_2}
\end{equation}
Searching for solutions to \eqref{eq:RC1_2} in the form $\sim r^n$, we find $\mathcal{R}=c_{11}r^{2}+c_{12}$. This reduces the equation for $\Theta(\theta)$ to
\begin{equation}
\frac{d^4\Theta}{d \theta^4}+4\frac{d^2\Theta}{d \theta^2}=0.
\label{eq:TC1_2}
\end{equation}
Therefore, $\Theta=c_{1}\cos2\theta+c_{2}\sin2\theta+c_{3}\theta+c_{4}$, so that the second set of solutions read
$$\tilde{\phi_1}(r,\theta)=\left(c_{11}r^{2}+c_{12}\right)\left(c_{3}\theta+c_{4}\right)+\left(c_{11}r^{2}+c_{12}\right)\left(c_{1}\cos2\theta+c_{2}\sin2\theta\right).$$
Note that this set of solutions can be lumped with the sets $\phi_0(r,\theta)$ and $\phi_2(r,\theta)$ obtained previously. 

\item  \underline{Solution 3}: For $\beta\geq 2$, $C_1=\beta^2+1$ and $C_2=(1-\beta^2)^2$. So the equation $\mathcal{R}_1-C_1=0$ becomes
\begin{equation}
r^2\mathcal{R}''-r\mathcal{R}'+(1-\beta^2)\mathcal{R}=0.
\label{eq:RC1_3}
\end{equation}
Searching for solutions to \eqref{eq:RC1_3} in the form $\sim r^n$, we find $\mathcal{R}=c_{11}r^{1+\beta}+c_{12}r^{1-\beta}$. On the other hand, \eqref{eq:biharm2} for $\Theta$ becomes
\begin{equation}
\frac{d^4\Theta}{d \theta^4}+2(1+\beta^2)\frac{d^2\Theta}{d \theta^2}+(1-\beta^2)^2\Theta=0.
\label{eq:TC1_3}
\end{equation}
Searching for solutions to \eqref{eq:TC1_3} in the form $\sim e^{k\theta}$ leads to 
$$\Theta=c_{n1}\cos[(1+\beta)\theta]+\tilde{c}_{n1}\sin[(1+\beta)\theta]+c_{n2}\cos[(1-\beta)\theta]+\tilde{c}_{n2}\sin[(1-\beta)\theta].$$

Therefore, the third set of solutions reads
\begin{eqnarray}
\sum\limits_{\beta=2}^{\infty}\tilde{\phi}_{2\beta}(r,\theta)=\sum\limits_{\beta=2}^{\infty}\left(c_{11}r^{1+\beta}+c_{12}r^{1-\beta}\right)\left(c_{n1}\cos[(1+\beta)\theta]+\tilde{c}_{n1}\sin[(1+\beta)\theta]\right)+\nonumber \\\left(c_{11}r^{1+\beta}+c_{12}r^{1-\beta}\right)\left(c_{n2}\cos[(1-\beta)\theta]+\tilde{c}_{n2}\sin[(1-\beta)\theta]\right). \nonumber
\end{eqnarray}
Here as well, these solutions can be lumped with $\phi_1(r,\theta)$ and $\phi_2(r,\theta)$. 
\end{itemize}

\paragraph{In the second case,} we follow the approach by other authors \cite{S-Stampouloglou10} and consider complex values such that $C_1-1=-\beta^2<0$. In this case, \eqref{eq:complexC1C2} implies that
$$n\in\{-i\beta+1, i\beta+1\},$$ 
along with 
$$C_2=(i\beta+1)^2(i\beta-1)^2=(1+\beta^2)^2.$$ 
With these values of $n$, the solution to $\mathcal{R}_1-C_1=0$ is given by
$\mathcal{R}=a_{1}r^{1+i\beta}+\tilde{a}_{1}r^{1-i\beta}$. Since $r^{i\beta}=e^{i\beta\ln r}$, this can be rewritten as
$$\mathcal{R}=\left[a_2\cos(\beta\ln r)+\tilde{a}_2\sin(\beta\ln r)\right]r.$$

And \eqref{eq:biharm2} for $\Theta$ becomes
\begin{equation}
\frac{d^4\Theta}{d \theta^4}+2(1-\beta^2)\frac{d^2\Theta}{d \theta^2}+(1+\beta^2)^2\Theta=0.
\label{eq:TCC1_3}%
\end{equation}

In order to solve \eqref{eq:TCC1_3}, we start with guess solutions of the form $e^{k\theta}$ and obtain 
$$\Theta= (b_1e^{\beta\theta}+\tilde{b}_1e^{-\beta\theta})\cos\theta+(c_1e^{\beta\theta}+\tilde{c}_1e^{-\beta\theta})\sin\theta.$$

Therefore, the last set of solutions $\mathcal{R}(r)\Theta(\theta)$ in the case $\mathcal{R}_1'=0$ reads
\begin{eqnarray}
\sum\limits_{\beta=1}^{\infty}\tilde{\phi}_{3\beta}(r,\theta)=\sum\limits_{\beta=1}^{\infty}\left[ (A_1e^{\beta\theta}+\tilde{A}_1e^{-\beta\theta})r\cos(\beta\ln r)+(A_2e^{\beta\theta}+\tilde{A}_2e^{-\beta\theta})r\sin(\beta\ln r)\right]\cos\theta \nonumber \\
+\left[ (A_3e^{\beta\theta}+\tilde{A}_3e^{-\beta\theta})r\cos(\beta\ln r)+(A_4e^{\beta\theta}+\tilde{A}_4e^{-\beta\theta})r\sin(\beta\ln r)\right]\sin\theta. \nonumber
\end{eqnarray}
\end{enumerate}
 
Combining all the above derived sets of solutions, we find the general complex solution of the biharmonic equation~(\ref{eq:biharm}) in polar coordinates to be \cite{S-Sadeh66, S-Stampouloglou10, S-Doschoris12}
$$\phi(r,\theta)=\phi_0+\phi_1+\phi_3+\tilde{\phi_0}+\sum\limits_{n=2}^{\infty}\phi_{2n}+\sum\limits_{n=2}^{\infty}\phi_{4n}+\sum\limits_{n=1}^{\infty}\tilde{\phi}_{3n}.$$
In the full extended form, the solution is 
\begin{equation} \label{eq:biharmsolution}
\begin{split}
\phi(r,\theta) = & (a_0r^2+b_0r^2\ln{r}+c_0+d_0\ln{r})\theta+ (\tilde{a}_0r^2+\tilde{b}_0r^2\ln{r}+\tilde{c}_0+\tilde{d}_0\ln{r}) +  \\
& (a_1r^3+b_1r^{-1}+c_1r+d_1r\ln{r})\cos{\theta} + (\tilde{a}_1r^3+\tilde{b}_1r^{-1}+\tilde{c}_1r+\tilde{d}_1r\ln{r})\sin{\theta} +  \\
& (e_1r^3+f_1r^{-1}+g_1r+h_1r\ln{r})\cosh{\theta} + (\tilde{e}_1r^3+\tilde{f}_1r^{-1}+\tilde{g}_1r+\tilde{h}_1r\ln{r})\sinh{\theta}+ \\
& \left(C_{0}r+D_{0}r\ln{r}\right)\theta\cos\theta+\left(\tilde{C}_{0}r+\tilde{D}_{0}r\ln{r}\right)\theta\sin\theta+ \\
&  \sum\limits_{n=2}^{\infty} \left[a_{n}r^{n}+b_{n}r^{n+2}+c_{n}r^{-n}+d_{n}r^{-n+2}\right]\cos{n\theta} + \\
& \sum\limits_{n=2}^{\infty}\left[\tilde{a}_{n}r^{n}+\tilde{b}_{n}r^{n+2}+\tilde{c}_{n}r^{-n}+\tilde{d}_{n}r^{-n+2}\right]\sin{n\theta}+  \\
& \sum\limits_{n=2}^{\infty} \left[e_{n}r^{n}+f_{n}r^{n+2}+g_{n}r^{-n}+h_{n}r^{-n+2}\right]\cosh{n\theta} + \\
& \sum\limits_{n=2}^{\infty}\left[\tilde{e}_{n}r^{n}+\tilde{f}_{n}r^{n+2}+\tilde{g}_{n}r^{-n}+\tilde{h}_{n}r^{-n+2}\right]\sinh{n\theta}+ \\
&\sum\limits_{n=1}^{\infty}\left[ (A_1e^{n\theta}+B_1e^{-n\theta})r\cos(n\ln r)+\right]\cos\theta+ \\
& \sum\limits_{n=1}^{\infty}\ \left[ (A_2e^{n\theta}+B_2e^{-n\theta})r\sin(n\ln r)\right]\cos\theta + \\
& \sum\limits_{n=1}^{\infty}\left[ (\tilde{A}_1e^{n\theta}+\tilde{B}_1e^{-n\theta})r\cos(n\ln r)\right]\sin\theta + \\
& \sum\limits_{n=1}^{\infty}\left[ (\tilde{A}_2e^{n\theta}+\tilde{B}_2e^{-n\theta})r\sin(n\ln r)\right]\sin\theta. 
\end{split}
\end{equation}
 
Note that due to the definitions of the stress components in Eq.~(\ref{eq:stressfuncpolar}), we can set $\tilde{c}_0=c_1=\tilde{c}_1=0$ in the general solution without loss of generality, since the terms $\tilde{c}_0$, $c_1r\cos\theta$ and $\tilde{c}_1r\sin\theta$ do not generate any stress. Moreover, in most cases, only some of the terms of the general solution~\eqref{eq:biharmsolution} are relevant to obtain the stress, strain and displacement fields. Let us recall that for the problem at hand, our goal is to obtain the displacement of the substrate surface under the action of the slime pressure $p(x)$ and shear stress $q(x)$. To that end, we shall now proceed as follows. 

First, we shall consider an elastic half-space under the action of concentrated normal and tangential loads and identify the relevant terms of \eqref{eq:biharmsolution} for which the generated stress fields satisfy the boundary conditions described at the end of Section~\ref{sec:plane-strain}. The newly found stress and strain fields will then allow us to find the displacement of the substrate surface due to these concentrated forces. Last, by virtue of the superposition principle, we shall deduce the response of the substrate under the distributed loads as a superposition of the responses due to elementary loads $F_p=p(s)\mathrm{d}s$ and $F_q=q(s)\mathrm{d}s$. Here, $F_p$ and $F_q$ are respectively the normal and tangential elementary loads acting on an infinitesimal contact zone of length $\mathrm{d}s$.

\addtocounter{subsection}{1}
\subsection{Fundamental solution for substrate deformation under concentrated loads} 

In this section, we obtain the response of the substrate under the action of point contact forces. Let us consider an elastic half-space with the point forces $F_p$ and $F_q$ applied on the substrate surface, at the point $O$ of coordinates $(x_O,z_O)=(s,0)$. In this configuration, we shall again work in a polar reference frame whose origin is at the point $O$. In this frame, the radial coordinate becomes $r=|x-s|$, while the azimuthal angle is still defined with respect to the vertical $z-$axis. Therefore, under the action of $F_p$ and $F_q$, the stress tensor components is subject the following conditions:

\begin{itemize}
\item The boundary condition on the surface is $\bar{\sigma}_{\theta}=0$, $\bar{\tau}_{r\theta}=0$ for $\theta=\pm\pi/2$ and $r=|x-s|\neq0$
\item Far from the substrate surface, the stress components vanish: $\sigma_{r}=\sigma_{\theta}=\tau_{r\theta} \rightarrow 0$ for $r \rightarrow \infty$.
\item The forces and torques must balance at any finite distance $R$ from the origin. Along the arc of the semi-circle of radius $R$, the normal to the boundary is $\mathbf{n}=\mathbf{e}_r$. Thus, the traction on the arc in the plane $(\mathbf{e}_r,\mathbf{e}_\theta)$ reads $\boldsymbol\sigma_s\cdot\mathbf{n}=\sigma_r\mathbf{e}_r+\tau_{r\theta}\mathbf{e}_\theta$.  In the reference configuration, the equilibrium of a control volume made of the semi-circle of arbitrary radius $R$ around the origin $O$ reads:

\begin{itemize}
\item Normal force balance:
$$-F_p+\int_{-\pi/2}^{\pi/2} \left[-\sigma_r(R,\theta)\cos\theta+\tau_{r\theta}(R,\theta)\sin\theta\right]Rd\theta=0.$$
\item Tangential force balance:
$$F_q+\int_{-\pi/2}^{\pi/2} \left[\sigma_r(R,\theta)\sin\theta+\tau_{r\theta}(R,\theta)\cos\theta\right]Rd\theta=0.$$
\item Torque balance:
$$\int_{-\pi/2}^{\pi/2} \left[-R\tau_{r\theta}(R,\theta)\right] Rd\theta=0.$$
\end{itemize}
\end{itemize}
 
Since $R$ is chosen arbitrarily, the force balance equations must be independent of $R$. Thus, we deduce that $\sigma_r\sim 1/r$ and $\tau_{r\theta}\sim 1/r$. However, applying the same reasoning to the torque balance implies that $\tau_{r\theta}\sim 1/r^2$. Therefore, the tangential stress must be zero everywhere: $\tau_{r\theta}=0$. To obtain the radial stress component $\sigma_r \sim 1/r$, we then only retain parts of the Airy stress function $\phi(r,\theta)$ that lead to such a behavior. Let us recall that
\begin{equation}
\sigma_r =\frac{1}{r} \frac{\partial \phi}{\partial r}+\frac{1}{r^2} \frac{\partial^2 \phi}{\partial \theta^2},  \;\; \sigma_{\theta} =   \frac{\partial^2 \phi}{\partial r^2}, \;\; \tau_{r\theta} = -\frac{\partial }{\partial r} \left(\frac{1}{r}\frac{\partial \phi}{\partial \theta}\right).
\label{eq:stresspolar2}
\end{equation}
To satisfy $\sigma_r \sim 1/r$, the Airy stress function must then necessarily be limited to
$$\phi(r,\theta)= r\ln{r}(d_1\cos{\theta}+\tilde{d}_1\sin{\theta}) +r\ln{r}(h_1\cosh{\theta}+\tilde{h}_1\sinh{\theta})+r\theta(C_{0}\cos\theta+\tilde{C}_{0}\sin\theta).$$
Therefore, the stress components read
\begin{subequations}
\begin{align}
\sigma_r & =\frac{1}{r} \left(d_1\cos{\theta}+\tilde{d}_1\sin{\theta}\right)+\frac{1}{r} \left(h_1\cosh{\theta}+\tilde{h}_1\sin{h\theta}\right)+\frac{2}{r} \left(C_0\cos{\theta}+\tilde{C}_0\sin{\theta}\right),   \\
\sigma_{\theta} & = \frac{1}{r} \left(d_1\cos{\theta}+\tilde{d}_1\sin{\theta}\right)+\frac{1}{r} \left(h_1\cosh{\theta}+\tilde{h}_1\sinh{\theta}\right), \\ 
\tau_{r\theta} & =\frac{1}{r} \left(d_1\sin{\theta}-\tilde{d}_1\cos{\theta}\right)-\frac{1}{r} \left(h_1\sinh{\theta}+\tilde{h}_1\cosh{\theta}\right).
\end{align}
\label{eq:stresspolar2}%
 \end{subequations}  
Since we must have $\tau_{r\theta}=0$ everywhere, then $d_1=\tilde{d}_1=h_1=\tilde{h}_1=0$. Consequently, we also have $\sigma_\theta=0$ everywhere. Note that the solutions $\tau_{r\theta}=\sigma_\theta=0$ automatically satisfy the boundary conditions at the surface of the elastic half-plane. Using these simplifications, the normal and tangential force balance can be rewritten as
\begin{subequations}
\begin{align}
-F_p-2\int_{-\pi/2}^{\pi/2} (C_0\cos\theta+\tilde{C}_0\sin\theta)\cos\theta d\theta & =0,\\
F_q+2\int_{-\pi/2}^{\pi/2}  (C_0\cos\theta+\tilde{C}_0\sin\theta)\sin\theta d\theta & =0. 
\end{align}
\label{eq:forcebalance}%
 \end{subequations}   
Computing the integrals in \eqref{eq:forcebalance}, we find that 
$$C_0=-\frac{F_p}{\pi}  \;\; \mathrm{and} \;\; \tilde{C}_0=\frac{F_q}{\pi}.$$
Therefore, the polar stress components under the action of concentrated normal and tangential loads read
\begin{equation}
\sigma_r =-\frac{2}{\pi r} (F_p\cos\theta+F_q\sin\theta),  \;\; \sigma_{\theta} =  0, \;\; \tau_{r\theta} = 0
\label{eq:stressstate}
\end{equation}

In order to obtain the corresponding displacement field, we shall first obtain the strain field using the Hooke's law~\eqref{eq:stressstrainpol}. Then, the obtained strain field can be integrated to yield the displacement field in the substrate. To this end, injecting \eqref{eq:stressstate} into \eqref{eq:stressstrainpol}, we obtain:
\begin{subequations}
\begin{align}
\epsilon_r=\frac{\partial u_r}{\partial r} = - \frac{1-\nu^2}{E}\frac{2}{\pi r}(F_p\cos\theta+F_q\sin\theta), \label{eq:stressstrainpol2a}\\
\epsilon_{\theta}=\frac{u_r}{r} +\frac{1}{r}\frac{\partial u_\theta}{\partial\theta}=\frac{\nu(1+\nu)}{E}\frac{2}{\pi r}(F_p\cos\theta+F_q\sin\theta),\label{eq:stressstrainpol2b} \\
\gamma_{r\theta} = \dfrac{1}{2}\left(\frac{1}{r}\frac{\partial u_r}{\partial \theta}+\frac{\partial u_\theta}{\partial r}-\frac{u_\theta}{r}\right) =\frac{\tau_{r\theta}}{2G}=0. \label{eq:stressstrainpol2c} 
\end{align}
\label{eq:stressstrainpol2}%
 \end{subequations}   
Integrating \eqref{eq:stressstrainpol2a} gives 
$$u_r=- \dfrac{1-\nu^2}{E}\dfrac{2}{\pi}(F_p\cos\theta+F_q\sin\theta)\ln r+f(\theta).$$ 
Thus, \eqref{eq:stressstrainpol2b} and \eqref{eq:stressstrainpol2c} become
\begin{equation}\label{eq:epstheta}  
f +\frac{\partial u_\theta}{\partial\theta}  = \frac{\nu(1+\nu)}{E}\frac{2}{\pi}(F_p\cos\theta+F_q\sin\theta)+\frac{1-\nu^2}{E}\frac{2}{\pi}(F_p\cos\theta+F_q\sin\theta)\ln r, 
\end{equation}
and
\begin{equation}
\frac{df}{d\theta} +r\frac{\partial u_\theta}{\partial r}-u_\theta  =\dfrac{1-\nu^2}{E}\dfrac{2}{\pi}(-F_p\sin\theta+F_q\cos\theta)\ln r. \label{eq:gamtheta}
\end{equation}
Differentiating Eq.~(\ref{eq:epstheta}) with respect to $r$ and Eq.~(\ref{eq:gamtheta}) with respect to $\theta$ yields
$$\frac{\partial^2 u_\theta}{\partial\theta\partial r}=\frac{1-\nu^2}{E}\frac{2}{\pi r}(F_p\cos\theta+F_q\sin\theta)=-\frac{1}{r}\frac{d^2f}{d\theta^2}+\frac{1}{r}\frac{\partial u_\theta}{\partial\theta}-\dfrac{1-\nu^2}{E}\dfrac{2}{\pi}(F_p\cos\theta+F_q\sin\theta)\frac{\ln r}{r},$$
which can be rewritten as
\begin{equation}
\frac{\partial u_\theta}{\partial\theta}=\frac{d^2f}{d\theta^2}+\frac{1-\nu^2}{E}\frac{2}{\pi}(F_p\cos\theta+F_q\sin\theta)(1+\ln r).
\label{eq:epsgamtheta}
\end{equation}
Therefore, we obtain the azimuthal displacement as 
\begin{equation}
u_\theta=\dfrac{df}{d\theta}+\dfrac{1-\nu^2}{E}\dfrac{2}{\pi}(F_p\sin\theta-F_q\cos\theta)(1+\ln r)+g(r).
\label{eq:azimdisp}
\end{equation} 
On one hand, \eqref{eq:azimdisp} can be differentiated with respect to $r$ to yield
\begin{equation}
r\dfrac{\partial u_\theta}{\partial r}=\dfrac{1-\nu^2}{E}\dfrac{2}{\pi}(F_p\sin\theta-F_q\cos\theta)+rg'.
\label{eq:rdtheta}
\end{equation}
On the other hand, \eqref{eq:azimdisp} also implies that
\begin{equation}
\dfrac{df}{d\theta}-u_\theta=-\dfrac{1-\nu^2}{E}\dfrac{2}{\pi}(F_p\sin\theta-F_q\cos\theta)(1+\ln r)-g(r).
\label{eq:fprime}
\end{equation}
By combining \eqref{eq:fprime} and \eqref{eq:rdtheta} with \eqref{eq:gamtheta}, we obtain 
\begin{equation}
\dfrac{1-\nu^2}{E}\dfrac{2}{\pi}(F_p\sin\theta-F_q\cos\theta)+rg'-\dfrac{1-\nu^2}{E}\dfrac{2}{\pi}(F_p\sin\theta-F_q\cos\theta)-g(r)=0.
\end{equation} 
Therefore,
$$rg'-g=0$$
whose solution is $g(r)=C_3r$ where $C_3$ is a constant.
In order to find $f$, we consider again Eq.~(\ref{eq:epstheta}) and replace the partial derivative on the left hand side by Eq.~(\ref{eq:epsgamtheta}), to obtain
\begin{equation}
\begin{split}
f +\frac{d^2f}{d\theta^2}+\frac{1-\nu^2}{E}  \frac{2}{\pi}(F_p\cos\theta+F_q\sin\theta)(1+\ln r)  & = 
 \frac{\nu(1+\nu)}{E}\frac{2}{\pi}(F_p\cos\theta+F_q\sin\theta)+ \\ 
 &  \hspace{0.75in} \frac{1-\nu^2}{E}\frac{2}{\pi}(F_p\cos\theta+F_q\sin\theta)\ln r.
\end{split}
\end{equation}
Therefore, $f$ is the solution of the following second-order ODE
\begin{equation}
\frac{d^2f}{d\theta^2}+f= \frac{(2\nu-1)(1+\nu)}{E}\frac{2}{\pi}(F_p\cos\theta+F_q\sin\theta).
\label{eq:2ndODE}
\end{equation}
The homogeneous solution is found to be $f_h=A_1\cos\theta+A_2\sin\theta$, where $A_1$ and $A_2$ are constants to be found. The particular solution is obtained by choosing $f_p=l(\theta)(B_1\cos\theta+B_2\sin\theta)$ with $B_1$ and $B_2$ as constants. By injecting $f_p$ into the ODE~(\ref{eq:2ndODE}) and equating the $\cos$ and $\sin$ terms from both sides of the equality, we obtain the system of equations
\begin{subequations}
\begin{align}
B_1\frac{d^2l}{d\theta^2}+2B_2\frac{dl}{d\theta}=P_f, \\
B_2\frac{d^2l}{d\theta^2}-2B_1\frac{dl}{d\theta}=Q_f,
\end{align}
\label{eq:eqnforl}%
\end{subequations}
where $P_f=\dfrac{(2\nu-1)(1+\nu)}{E}\dfrac{2}{\pi}F_p$ and $Q_f=\dfrac{(2\nu-1)(1+\nu)}{E}\dfrac{2}{\pi}F_q$. 
Solving the system of Eq.~(\ref{eq:eqnforl}) yields
$$l(\theta)=\dfrac{P_f}{2B_2}\theta+D,$$
with $D$ as a constant, along with the equality $$\dfrac{P_f}{2B_2}=-\dfrac{Q_f}{2B_1}.$$

\noindent The particular solution of the ODE~(\ref{eq:2ndODE}) thus reads
$$f_p=\left(\frac{P}{2B_2}\theta+D\right)(B_1\cos\theta+B_2\sin\theta),$$
$$f_p=DB_1\cos\theta+DB_2\sin\theta+\frac{P_f}{2}\theta\sin\theta-\dfrac{Q_f}{2}\theta\cos\theta.$$
Adding $f_h$ and $f_p$, we find $f$ in the form
$$f(\theta)=C_1\cos\theta+C_2\sin\theta+\frac{(2\nu-1)(1+\nu)}{E}\frac{\theta}{\pi}(F_p\sin\theta-F_q\cos\theta).$$

\noindent Finally, one obtains the displacement field in polar coordinates as
\begin{equation}
\begin{split}
u_r(r,\theta)= & - \frac{2(1-\nu^2)}{\pi E}(F_p\cos\theta+F_q\sin\theta)\ln r+ \\
& \frac{(2\nu-1)(1+\nu)}{\pi E}\theta(F_p\sin\theta-F_q\cos\theta)+  \\
& C_1\cos\theta+C_2\sin\theta.
\end{split}
\end{equation}
\begin{equation}
\begin{split}
u_\theta(r,\theta)= & \frac{2(1-\nu^2)}{\pi E}(F_p\sin\theta-F_q\cos\theta)\ln r+ \\
& \frac{(2\nu-1)(1+\nu)}{\pi E}\theta(F_p\cos\theta+F_q\sin\theta)+ \\
& \frac{\nu(1+\nu)}{\pi E}(F_p\sin\theta -F_q\cos\theta)+ \\
& C_2\cos\theta-C_1\sin\theta+C_3r.
\end{split}
\end{equation}

\bigskip
\noindent 

Furthermore, we assume that in the absence of applied forces, there is no residual strain in the substrate. In other words, the displacement field must vanish when $F_p=F_q=0$. This implies that we must have $C_1=C_2=C_3=0$. Therefore, at the surface of the substrate (where $\theta=\pm\pi/2$), the radial and azimuthal displacements read 
\begin{eqnarray}
\bar{u}_{r^+}=u_r\left(r,\frac{\pi}{2}\right)=  -\frac{1-\nu^2}{\pi E}2F_q\ln r-\frac{(1-2\nu)(1+\nu)}{2E}F_p,  \\
\bar{u}_{\theta^+}=u_\theta\left(r,\frac{\pi}{2}\right)= \frac{1-\nu^2}{\pi E}2F_p\ln r-\frac{(1-2\nu)(1+\nu)}{2E}F_q +\frac{\nu(1+\nu)}{\pi E}F_p,
\end{eqnarray}
and
\begin{eqnarray}
\bar{u}_{r^-}=u_r\left(r,-\frac{\pi}{2}\right)= \frac{1-\nu^2}{\pi E}2F_q\ln r-\frac{(1-2\nu)(1+\nu)}{2E}F_p,  \\
\bar{u}_{\theta^-}=u_\theta\left(r,-\frac{\pi}{2}\right)=- \frac{1-\nu^2}{\pi E}2F_p\ln r-\frac{(1-2\nu)(1+\nu)}{2E}F_q-\frac{\nu(1+\nu)}{\pi E}F_p. 
\end{eqnarray}

\bigskip
Let us rewrite the surface displacements into cartesian coordinates, which is necessary for us to use it in the modified Reynolds equation. First, since the concentrated forces are applied at $x=s$, the radial coordinate reads $r=|x-s|$. Moreover, on the surface, $x\geq 0$ for $\theta\in[0,\pi/2]$, while $x\leq 0$ for $\theta\in[-\pi/2,0]$. Furthermore, given that $(\mathbf{e}_r,\mathbf{e}_\theta)$ is a rotating frame which depends on $\theta$, we have $(\mathbf{e}_r,\mathbf{e}_\theta)|_{\theta=\pi/2}=(\mathbf{e}_x,\mathbf{e}_z)$ for $\theta=\frac{\pi}{2}$. However, $(\mathbf{e}_r,\mathbf{e}_\theta)|_{\theta=-\pi/2}=(-\mathbf{e}_x,-\mathbf{e}_z)$, which implies that the cartesian displacements $(\bar{u}_{\text{s}},\bar{w}_{\text{s}})$ at the surface of the substrate verify the relationships: 
$$\bar{u}_{\text{s}^+}=\bar{u}_{r^+}; \;\; \bar{w}_{\text{s}^+}=\bar{u}_{\theta^+}; \;\; \bar{u}_{\text{s}^-}=-\bar{u}_{r^-}; \;\; \bar{w}_{\text{s}^-}=-\bar{u}_{\theta^-}.$$
Therefore, the surface displacements read in cartesian coordinates
\begin{subequations}
\begin{align}
\bar{u}_{\text{s}}(x) & =  -\frac{2(1-\nu^2)}{\pi E}F_q\ln |x-s|-\frac{(1-2\nu)(1+\nu)}{2E}\mathrm{sgn}(x-s) F_p, \\
\bar{w}_{\text{s}}(x) & = \frac{2(1-\nu^2)}{\pi E}F_p\ln |x-s|-\frac{(1-2\nu)(1+\nu)}{2E}\mathrm{sgn}(x-s)F_q+\frac{\nu(1+\nu)}{\pi E}F_p, 
\end{align}
\label{eq:surfdisp}%
\end{subequations}
where $\mathrm{sgn}(x-s)=1$ if $x-s>0$ and $-1$ if $x-s<0$. The expression of $\bar{w}_s$ contains a constant term $\dfrac{\nu(1+\nu)}{\pi E}F_p$, which doesn't depend of the coordinate $x$. As suggested by Johnson \cite{S-Johnson85}, one can choose this constant as a datum point $x_0$, i.e the position on the surface at which the substrate displacement decays to zero. Rewriting this constant term as a logarithm, the displacement fields can thus be recast into
\begin{subequations}
\begin{align}
\bar{u}_{\text{s}}(x) & =  -\frac{2(1-\nu^2)}{\pi E}F_q\ln |x-s|-\frac{(1-2\nu)(1+\nu)}{2E}\mathrm{sgn}(x-s) F_p, \\
\bar{w}_{\text{s}}(x) & = \frac{2(1-\nu^2)}{\pi E}F_p\ln \frac{|x-s|}{x_0}-\frac{(1-2\nu)(1+\nu)}{2E}\mathrm{sgn}(x-s)F_q. 
\end{align}
\label{eq:surfdispNoConst}%
\end{subequations}

\eqref{eq:surfdispNoConst} constitutes the response of the substrate surface under the action of concentrated loads acting over an infinitesimal zone of length $\mathrm{d}s$. Using the superposition principle, we next generalize this result to the case of distributed normal and tangential loads on an elastic half-space. 

\addtocounter{subsection}{1}
\subsection{Substrate deformation under distributed loads} \label{sec:elasticdistributedloads}
We now proceed to compute the displacement field in the substrate, under the action of the distributed slime pressure $p(x)$ and shear stress $q(x)$ acting over a contact zone $-\dfrac{\ell}{2}\leq x\leq \dfrac{\ell}{2}$ at the substrate surface. The response of the substrate under a distributed load can be found by superimposing the responses due to all the elementary forces composing the load \cite{S-Landau84}. Recalling that $F_p=p(s)\mathrm{d}s$ and $F_q=q(s)\mathrm{d}s$, we thus sum \eqref{eq:surfdispNoConst} in an integral sense and obtain
\begin{subequations}
\begin{align}
\bar{u}_{\text{s}}(x)= -\frac{2(1-\nu^2)}{\pi E}\int_{-\ell/2}^{\ell/2} q(s)\ln |x-s|\mathrm{d}s-\frac{(1-2\nu)(1+\nu)}{2E}\left[\int_{-\ell/2}^x p(s)\mathrm{d}s-\int_x^{\ell/2} p(s)\mathrm{d}s\right], \\
\bar{w}_{\text{s}}(x)= \frac{2(1-\nu^2)}{\pi E}\int_{-\ell/2}^{\ell/2} p(s)\ln \frac{|x-s|}{x_0}\mathrm{d}s-\frac{(1-2\nu)(1+\nu)}{2E}\left[\int_{-\ell/2}^x q(s)\mathrm{d}s-\int_x^{\ell/2} q(s)\mathrm{d}s\right].
\end{align}
\label{eq:dispfield}%
\end{subequations}

\eqref{eq:dispfield} can be further simplified owing to the lubrication approximation in the slime. In this thin film, note that the pressure scale is much larger than that of the shear stress. Indeed, on one hand, the shear stress is given by
$$q(x)=\mu\left(\dfrac{\partial u_x}{\partial z}+\dfrac{\partial u_z}{\partial x}\right)=\dfrac{\mu V}{h_0}\dfrac{\partial \hu_x}{\partial \hz}+\epsilon^2\dfrac{\mu V}{h_0}\dfrac{\partial \hu_z}{\partial \hx},$$
while on the other hand, the lubrication pressure is defined as
$$p(x)=\dfrac{1}{\epsilon}\dfrac{\mu V}{h_0}\hp.$$
Since $O\left(q/p\right)\sim \epsilon \ll 1$, we can neglect the contribution of the shear stress in the deformation of the substrate and simplify \eqref{eq:dispfield} into
\begin{subequations}
\begin{align}
\bar{u}_{\text{s}}(x) & \approx -\frac{(1-2\nu)(1+\nu)}{2E}\left[\int_{-\ell/2}^x p(s)\mathrm{d}s-\int_x^{\ell/2} p(s)\mathrm{d}s\right], \\
\bar{w}_\text{s}(x)& \approx \frac{2(1-\nu^2)}{\pi E}\int_{-\ell/2}^{\ell/2} p(s)\ln \frac{|x-s|}{x_0}\mathrm{d}s.
\end{align}
\label{eq:dispfield2}%
\end{subequations}
Furthermore, as discussed at the beginning of Section~\ref{sec:elasticsubstrate}, we consider the substrate to be in equilibrium at every instant. This hypothesis allows us to straightforwardly substitute the pressure in Eq.~(\ref{eq:dispfield2}) by the time-dependent traveling pressure $p(x,t)$ and thus obtain 
\begin{subequations}
\begin{align}
\bar{u}_{\text{s}}(x,t) & \approx -\frac{(1-2\nu)(1+\nu)}{2E}\left[\int_{-\ell/2}^x p(s,t)\mathrm{d}s-\int_x^{\ell/2} p(s,t)\mathrm{d}s\right], \label{eq:dispfieldtimea}\\
\delta(x,t)& \approx \frac{2(1-\nu^2)}{\pi E}\int_{-\ell/2}^{\ell/2} p(s,t)\ln \frac{|x-s|}{x_0}\mathrm{d}s.\label{eq:dispfieldtimeb}
\end{align}
\label{eq:dispfieldtime}%
\end{subequations}
where we also rewrote the vertical deformation to be $\delta(x,t)$. Thus, \eqref{eq:dispfieldtimea} and \eqref{eq:dispfieldtimeb} are respectively the horizontal and vertical deformations of the substrate surface due to the stresses from the thin slime secreted by the myxobacteria during gliding. 

Note that \eqref{eq:dispfieldtimea} allows for the possibility of a non-zero horizontal velocity $\dfrac{\mathrm{d}\bar{u}_{\text{s}}}{\mathrm{d}t}$ of the material points at the substrate surface, unless $\nu=1/2$. However, the horizontal velocity profile of the lubricating slime was obtained earlier by assuming that material points of the substrate surface only move vertically, at speed $\dfrac{\mathrm{d}\delta}{\mathrm{d}t}$ (see Section~\ref{sec:lubricationapproximation}). Therefore, in order to ensure a consistent formulation, we hereafter limit our analysis to incompressible substrates for which $\nu=1/2$. In this case, the substrate then deforms only vertically according to
\begin{equation}
\delta(x,t)=\frac{1-\nu}{\pi G}\int_{-\ell/2}^{\ell/2}p(x',t)\ln \frac{|x-x'|}{x_0} \mathrm{d}x',
\label{eq:thickdeformWave}
\end{equation}
where we used the definition of the shear modulus $G=E/\left[2(1+\nu)\right]$ and left the explicit dependency on the Poisson ratio. \eqref{eq:thickdeformWave} constitutes the substrate deformation which will be used in conjunction with the modified Reynolds equation. 

\section{Non-dimensionalization} \label{ND}
In this subsection, we shall recast in a dimensionless form the previously derived \eqref{eq:Reynolds} and \eqref{eq:thickdeformWave} that govern the problem at hand. To this end, we use the following scalings: 
\begin{center}
$\hat{x}=\dfrac{x}{L}$, $n=\dfrac{\ell}{L}$, $\hat{t}=\dfrac{t}{L/C}$, $\hat{z}=\dfrac{z}{h_0}=\dfrac{z}{\epsilon L}$, $ \hat{u}_x=\dfrac{u_x}{C}$, $\hat{u}_z=\dfrac{u_z}{\epsilon C}$, $\hat{V}=\dfrac{V}{C}$, $\hat{U}=\dfrac{U}{C}$.
\end{center}
\begin{center}
$\hat{p}=\dfrac{p}{P}=\dfrac{p h_0^2}{\mu CL}$, $\hat{\delta}=\dfrac{\delta}{\Delta}=\dfrac{\delta G}{(1-\nu)PL}$,
\end{center}
where $\Delta$ is, by definition, the vertical deformation scale of the substrate. 
The dimensionless height of the lubrication gap is then given by $\hat{h}(\hat{x},t)=1+b(x,t)/h_0=1+\hat{b}(x,t)$ and $\eta=\Delta/h_0$ is the so-called softness parameter. This dimensionless number $\eta$ compares the deformation of the soft substrate caused by the fluid pressure to the mean thickness of the slime film and can also be rewritten as 
\begin{equation}
\eta=\frac{(1-\nu)PL}{Gh_0}=\frac{\mu(1-\nu)L^2C}{Gh_0^3}.
\label{eq:softness}
\end{equation}

\noindent Therefore, the dimensionless equations governing the soft lubrication problem corresponding to myxobacterial gliding motility are
\begin{subequations}
\begin{align}
\frac{\partial}{\partial\hat{x}}\left[ \frac{\partial\hat{p}}{\partial\hat{x}}(\hat{h}-\eta\hat{\delta})^3+6(\hV-2) (\hat{h}-\eta\hat{\delta}) \right] & =0, \label{eq:DEa}\\
\hat{\delta}(\hat{x},t)-\frac{1}{\pi} \int_{-n/2}^{n/2} \hat{p}(\hat{x}',t) \ln\frac{|\hat{x}-\hat{x}'|}{\hat{x}_0} \mathrm{d}\hat{x}' & =0. \label{eq:DEb}
\end{align}
\label{eq:DE}%
\end{subequations}

In order to solve for the three unknowns of the problem, i.e. $\hp(\hx)$, $\hdelta(\hx)$ and $V$, \eqref{eq:DE} must be complemented with an extra equation for the gliding speed and with two boundary conditions on the pressure to solve the 2nd-order PDE in \eqref{eq:DEa} . To obtain the governing equation for the gliding speed of the myxobacteria, we use the Newton's second law of motion which requires, at zero Reynolds number, the total external force and torque on the self-propelled bacteria to vanish \cite{S-Purcell77}. In other words, the bacteria glide when the lift (or vertical) and drag (or horizontal) forces vanish. 

\section{Force and torque on the bacteria}
In this section, we give expressions for the force and torque on the myxobacteria. These loads can arise, when they exist, from surface and volume contributions. The surface force and torque on the bacteria come from the slime, while gravity exerts body force and torque on the bacteria. However, we shall show that the body force and torque can be neglected in the system at hand. 

\bigskip
In the case of surface loads, the force and the torque on the bacteria are respectively given by 
$$\bfv{F}=\int\int {\boldsymbol \sigma}_\text{f} \cdot\bfv{n}\mathrm{d}S$$
and 
$$\bfv{M}=\int\int \left(\bfv{r}-\bfv{r}_\text{c} \right)\times{\boldsymbol \sigma}_\text{f} \cdot\bfv{n}\mathrm{d}S,$$ 
where ${\boldsymbol \sigma}_\text{f} $ is the stress tensor in the slime. Here, $S$ is the surface of the bacteria membrane, $\bfv{r}=x\bfv{e}_x+z\bfv{e}_z$ is the position vector and $\bfv{r}_\text{c}$ is the position of the bacteria's center of mass. Using the reference pressure $P$ as the reference scale of stresses, the slime stress tensor reads in its dimensionless form 
\begin{equation}
{\boldsymbol \sigma}_\text{f}(\hx,\hz) =	\begin{pmatrix}
       -\hp+2\epsilon^2\dfrac{\partial \hu_x}{\partial \hX} & \epsilon\dfrac{\partial \hu_x}{\partial \hz}+\epsilon^3\dfrac{\partial \hu_z}{\partial \hX}  \\
       & \\
       \epsilon\dfrac{\partial \hu_x}{\partial \hz}+\epsilon^3\dfrac{\partial \hu_z}{\partial \hx} & -\hp+2\epsilon^2\dfrac{\partial \hu_z}{\partial \hz}         
     \end{pmatrix}.
     \label{eq:dimenionlessstress}
\end{equation}
In 2D, the dimensionless forces (resp. torque) per unit length of the $y-$direction, after being rescaled with $PL$ (resp. $PL^2$), read
\begin{subequations}
\begin{align}
\bfv{F}(\hti ) & =\int_{-n/2}^{n/2}{\boldsymbol \sigma}_\text{f}(x,z=h)\cdot\bfv{n}_b\mathrm{d}\hx, \\
\bfv{M}(\hti ) & =\int_{-n/2}^{n/2}\left[(\hx-\hx_\text{c})\bfv{e}_x+h\bfv{e}_z\right]\times\left[{\boldsymbol \sigma}_\text{f}(x,z=h)\cdot\bfv{n}_b\right]\mathrm{d}\hx,
\end{align}
\label{eq:FT}%
\end{subequations}
where $\hX_\text{c}=0$ and $\bfv{n}_\text{b}=\left(\epsilon \hat{b}'\bfv{e}_x-\bfv{e}_z\right)/\sqrt{1+\epsilon^2 \hat{b}'^2}$ is the unit normal vector to the boundary of the bacteria. Hereinafter, the prime symbol $(\cdot)'$ denotes the differentiation with respect to $\hx$. The normal vector can be approximated up to order $O(\epsilon)$ by
\begin{equation}
\bfv{n}_\text{b} \approx -\bfv{e}_z+\epsilon \hat{b}' \bfv{e}_x.
\label{eq:normalvector}
\end{equation}
Therefore, the force and torque per unit length can be rewritten as
\begin{subequations}
\begin{align}
\bfv{F}(\hti ) & =\left(\int_{-n/2}^{n/2} \hp \mathrm{d}\hX \right)\bfv{e}_z - \epsilon \left( \int_{-n/2}^{n/2} \left(\hp\hat{b}'(\hX)+\partial_z\hu_x|_{\hz=\hh}\right) \mathrm{d}\hX \right)\bfv{e}_x + O(\epsilon^2), \\
\bfv{M}(\hti ) & = \left(\int_{-n/2}^{n/2}\hX\hp \mathrm{d}\hX \right)\bfv{e}_y + O(\epsilon). 
\end{align}
\label{eq:FTapprox}%
\end{subequations}
However, Eq.~(\ref{eq:uvel}) shows that
$$\left. \frac{\partial \hu_x}{\partial \hz}\right |_{\hz=\hh}=\frac{1}{2}\frac{\partial\hp}{\partial\hX}\left(\hh-\eta\hdelta\right)+\frac{\hat{V}}{\hh-\eta\hdelta}.$$
Thus, we can approximate the force on the bacteria and rewrite it as
\begin{equation}
\bfv{F}\approx \left(\int_{-n/2}^{n/2} \hp \mathrm{d}\hX \right)\bfv{e}_z - \epsilon \left( \int_{-n/2}^{n/2} \left[\hp\hat{b}'+\frac{\hp'}{2}\left(\hh-\eta\hdelta\right)+\frac{\hat{V}}{\hh-\eta\hdelta}\right] \mathrm{d}\hX \right)\bfv{e}_x.
\label{eq:Force}
\end{equation}
From \eqref{eq:Force}, we can then obtain, up to order $O(\epsilon)$, the drag (horizontal force), the lift (vertical force) and the moment per unit length as\begin{subequations}
\begin{align}
\mcal{F}(\hti ) &=- \epsilon \int_{-n/2}^{n/2} \left(\hp\hat{b}'+\frac{\hp'}{2}\left(\hh-\eta\hdelta\right)+\frac{\hat{V}}{\hh-\eta\hdelta}\right) \mathrm{d}\hX , \\
\mcal{L}(\hti )&=\int_{-n/2}^{n/2} \hp \mathrm{d}\hX ,\\
\mcal{M}(\hti ) & =\int_{-n/2}^{n/2} \hX \hp \mathrm{d}\hX. 
\end{align}
\label{eq:LiftDragMoment}%
\end{subequations}
Having thus determined the surface forces and torque, we now turn to the volume contributions. 

The only body force on the bacteria is the buoyancy force in the vertical direction. However, this contribution can be neglected with respect to the lift from the slime. This is apparent by comparing the order of magnitude of these forces. Indeed, for a typical cylindrical bacteria of radius $R\approx 250$ nm and length $\ell\approx 5 \, \mu$m and density $\rho\approx 1300$ kg/m$^3$, the weight $\mcal{W}=\rho\pi R^2\ell g\approx 10^{-3}$ nN whereas, for $h_0\approx 10$ nm and $V\approx 4 \, \mu/$min, the lift is about $P\ell R\sim \mu VL\ell R/h_0^2\approx 8.4$ nN. Therefore, the lift force dominates in the normal direction to the gliding and the vertical force balance condition reduces to
\begin{equation}
\mcal{L}(\hti )=\int_{-n/2}^{n/2} \hp(\hX,\hti) \mathrm{d}\hX=0.
\label{eq:zeroliftcond}%
\end{equation}
In the horizontal direction, the force balance results in
\begin{equation}
\mcal{F}(\hti )=- \epsilon \int_{-n/2}^{n/2} \left(\hp\hat{b}'+\frac{\hp'}{2}\left(\hh-\eta\hdelta\right)+\frac{\hat{V}}{\hh-\eta\hdelta}\right) \mathrm{d}\hX=0.
\label{eq:zerodragcond}%
\end{equation}
\eqref{eq:zerodragcond} constitutes the governing equation for the gliding speed $\hat{V}$, for a given bacterial shape and substrate parameters while \eqref{eq:zeroliftcond} constitutes the first condition on the unknown pressure distribution underneath the bacteria. For the second condition needed to solve \eqref{eq:DEa}, we impose the pressure at the bacteria's leading edge to match the atmospheric (zero) pressure, i.e.
\begin{equation}
\hp(n/2,\hti)=0. 
\label{eq:zeropressurecond}
\end{equation}
Having already imposed two conditions on the pressure, the problem would be over-constrained if we were to require the torque to vanish as well. Therefore, we ignore here the zero-torque condition, as commonly seen in the literature of swimming sheets \cite{S-Taylor51, S-Lauga06, S-Pak16}, which comes as a consequence of having imposed the shape of the bacteria. A more complete treatment that remedies to this drawback of the model would consists in solving for the membrane shape under the requirement that its bending and tensile stresses balance those due to any non-zero torque exerted by the slime. This approach, that we will leave for future work, has been formerly applied to buoyancy-driven fluid-lubricated foils whose swimming speed and geometry shape were simultaneously computed as part of the solution \cite{S-Argentina07}. 

As a final remark before ending this section, the zero-lift condition can be used to simplify \eqref{eq:DEb} to yield
$$\hdelta(\hx)=\frac{1}{\pi }\int_{-n/2}^{n/2} \hp(\hx ') \ln\frac{|\hx-\hx'|}{\hx_0} \mathrm{d}\hx' = \frac{1}{\pi }\int_{-n/2}^{n/2} \hp(\hx ') \ln\left | \hx-\hx' \right | \mathrm{d}\hx' -\underbrace{\frac{\ln{\hx_0}}{\pi }\int_{-n/2}^{n/2} \hp(\hx ')\mathrm{d}\hx'}_{=0}. $$
Therefore, the constant $\hx_0$ will be, hereinafter, left out of the substrate deformation which simply reads
\begin{equation}
\hdelta(\hx)=\frac{1}{\pi }\int_{-n/2}^{n/2} \hp(\hx ') \ln\left | \hx-\hx'\right | \mathrm{d}\hx'. 
\label{eq:substratedeformation}
\end{equation}

This ends the derivation of the governing equations of the elasto-hydrodynamics problem of myxobacterial gliding on a thick elastic substrate. As the Reynolds equation~\eqref{eq:DEa} is nonlinear in $\hdelta$, we proceed with the numerical resolution of the elasto-hydrodynamics problem using Newton's algorithm. The linearized equations obtained at every iteration are solved numerically using the finite element method \cite{S-Zienkiewicz91}, as described below.

\section{Numerical resolution of the elasto-hydrodynamics problem}

In this section, we describe the numerical scheme used to solve the governing equations of the elasto-hydrodynamics problem, namely \eqref{eq:DEa}, \eqref{eq:substratedeformation} and \eqref{eq:zerodragcond}, under the conditions given by \eqref{eq:zeroliftcond} and \eqref{eq:zeropressurecond}. 

For the numerical resolution of the Reynolds equation under the zero-lift global constraint, it is convenient to integrate \eqref{eq:DEa} once and instead solve 
\begin{equation}
\frac{\partial\hp}{\partial\hX}(\hh-\eta \hdelta)^3+6(\hV-2)(\hh-\eta \hdelta)=\hm(\hti),
\end{equation}
with the boundary condition $\hp(n/2,\hti)=0$. The unknown $\hm$ has the dimension of a mass flux and is related to $\hat{Q}_h=Q_h/(Ch_0)$, the dimensionless slime flow rate across a section of the lubrication gap, by the following expression
\begin{equation}
\hat{Q}_h=\int_{\eta\hdelta}^{\hh} \hat{u}_x\mathrm{d}\hz=-\left[ \frac{1}{12}\frac{\partial \hp}{\partial \hx}(\hh-\eta\hdelta)^3+\frac{\hV}{2} (\hh-\eta\hdelta) \right]=-\frac{\hm}{12}-(\hh-\eta\hdelta).
\label{eq:FlowrateDimensionless}
\end{equation}
Therefore, the elasto-hydrodynamics problem is now represented by the system of equations 
\begin{subequations}
\begin{align}
Q_1(\hp,\hdelta,\hm,\hV) & = \frac{\partial\hp}{\partial\hX}(\hh-\eta \hdelta)^3+6(\hV-2)(\hh-\eta \hdelta)-\hm(\hti) & =0, \\
Q_2(\hp,\hdelta,\hm,\hV) & =\hdelta(\hX,\hti)- \frac{1}{\pi }\int_{-n/2}^{n/2} \hp(\hX',\hti) \ln\left | \hX-\hX' \right | \mathrm{d}\hX' &=0,\label{eq:coupledPB2b} \\
Q_3(\hp,\hdelta,\hm,\hV) & =\int_{-n/2}^{n/2} \hp(\hX,\hti) \mathrm{d}\hX & = 0,\\
Q_4(\hp,\hdelta,\hm,\hV) & =\int_{-n/2}^{n/2} \left(\hp\hat{b}'+\frac{\hp'}{2}\left(\hh-\eta\hdelta\right)+\frac{\hat{V}}{\hh-\eta\hdelta}\right) \mathrm{d}\hX & = 0.
\end{align}
\label{eq:coupledPB2}%
\end{subequations}
In the next subsection, we shall present how we solve \eqref{eq:coupledPB2} at each instant $\hti$ and find the solution represented by the vector $\bfv{q}  = (\hp,\hdelta,\hV,\hm)$ for a given value of the softness parameter $\eta$.

\addtocounter{subsection}{1}
\subsection{Newton's method}
The solution to the system of equations~\eqref{eq:coupledPB2}, supplemented with the boundary condition $\hp(n/2,\hti)=0$, is found iteratively : given $\bfv{q}_k=\left(\hp_{k}, \hdelta_{k}, \hm_{k}, \hV_k \right)^T$ at iteration $k$, the solution of the problem at the next iteration reads  $\bfv{q}_{k+1}=\bfv{q}_{k}+\tilde{\bfv{q}}_k$. We use Newton's method to find the correction $\tilde{\bfv{q}}_k=\left(\tilde{p}_k,\tilde{\delta}_k,\tilde{m}_k,\tilde{V}_k\right)^T$. To that end, we first linearize \eqref{eq:coupledPB2} around the solution at iteration $k$ and obtain
\begin{subequations}
\begin{align}
 \frac{\partial\tilde{p}_k}{\partial\hX}(\hh-\eta\hdelta_k)^3-3\eta\frac{\partial\hp_k}{\partial\hX}(\hh-\eta\hdelta_k)^2\tilde{\delta}_k-6(\hV_k-2)\eta\tilde{\delta}_k+6\tilde{V}_k(\hh-\eta\hdelta_k)-\tilde{m}_k &= -Q_{1k}, \\
\tilde{\delta}_k  - \frac{1}{\pi}\int_{-n/2}^{n/2}\tilde{p}_k(\hX',\hti) \ln\left | \hX-\hX' \right |  \mathrm{d}\hX' &= -Q_{2k}, \\
- \int_{-n/2}^{n/2}\tilde{p}_k \mathrm{d}\hX &= -Q_{3k},\\
\int_{-n/2}^{n/2} \left(\tilde{p}_k\hat{b}'+\frac{1}{2}\frac{\partial\tilde{p}'_k}{\partial\hX}\left(\hh-\eta\hdelta_k\right)-\frac{1}{2}\frac{\partial\hp_k}{\partial\hX}\eta\tilde{\delta}_k+\frac{\tilde{V}_k}{\hh-\eta\hdelta_k}+\frac{\eta\tilde{\delta}_k\hV_k}{\left(\hh-\eta\hdelta_k\right)^2} \right)\mathrm{d}\hX & = -Q_{4k},
\end{align}
\label{eq:LinearizedPB2}%
\end{subequations}
where $Q_{lk}=Q_l(\hp_k,\hdelta_k,\hm_k,\hV_k)$ for different values $l=1,2,3,4$, and $k$ refers to the Newton's iteration step.
By inverting \eqref{eq:LinearizedPB2}, we obtain $\tilde{\bfv{q}}_k$, and subsequently $\bfv{q}_{k+1}$. We then repeat this process of linearization and inversion to obtain the corrections at iteration $\tilde{\bfv{q}}_{k+1}$. This process is repeated until the $L^2$-norm of the correction vector $\tilde{\bfv{q}}_k$ is lower than a given tolerance, that we choose to be around $\sim 10^{-10}$. 

Given that Newton's method requires to invert the linearized system of equations~\eqref{eq:LinearizedPB2} at every iteration, we now proceed to describe how this computation is carried out numerically.

\addtocounter{subsection}{1}
\subsection{Finite element implementation}
We solve the linearized system of equations (\ref{eq:LinearizedPB2}) numerically, using the finite element method (FEM) \cite{S-Zienkiewicz91}. To that end, we choose $q(\hx)$ and $\kappa(\hx)$ as test functions for the pressure and the substrate deformation, respectively, and write the weak form of \eqref{eq:LinearizedPB2} as\begin{subequations}
\begin{align}
\left< q, \frac{\partial\tilde{p}_k}{\partial\hX}(\hh-\eta\hdelta_k)^3-3\eta\frac{\partial\hp_k}{\partial\hX}(\hh-\eta\hdelta_k)^2\tilde{\delta}_k -6(\hV_k-2)\eta\tilde{\delta}_k+6\tilde{V}_k(\hh-\eta\hdelta_k)-\tilde{m}_k\right>_{\Omega}\nonumber \\
+\mathcal{N}\left<q,\tilde{p}_k\right>_{\partial\Omega_1}+\mathcal{N}\left<q,\hp_k\right>_{\partial\Omega_1}+ \left< q, Q_1(\hp_k,\hdelta_k,\hm_k,\hV_k) \right>_{\Omega} \nonumber \\
+ \left< \kappa, \tilde{\delta}_k \right> _{\Omega}- \left< \kappa, \frac{1}{\pi}\int_{-n/2}^{n/2}\tilde{p}_k(\hX',\hti) \ln\left | \hX-\hX' \right |  \mathrm{d}\hX' \right>_{\Omega} + \left<\kappa, Q_2(\hp_k,\hdelta_k,\hm_k,\hV_k)\right>_{\Omega}  \nonumber \\
- \left<1, \tilde{p}_k\right>_{\Omega} + \left<1, \hp_k \right>_{\Omega} - \left<1, \tilde{p}_k\hat{b}'+\frac{1}{2}\frac{\partial\tilde{p}_k}{\partial\hX}\left(\hh-\eta\hdelta_k\right)-\frac{1}{2}\frac{\partial\hp_k}{\partial\hX}\eta\tilde{\delta}_k+\frac{\tilde{V}_k}{\hh-\eta\hdelta_k}+\frac{\eta\tilde{\delta}_k\hV_k}{\left(\hh-\eta\hdelta_k\right)^2}\right>_{\Omega} \nonumber \\
+ \left<1,\hp_k\hat{b}'+\frac{1}{2}\frac{\partial\hp_k}{\partial\hX}\left(\hh-\eta\hdelta_k\right)+\frac{\hat{V}_k}{\hh-\eta\hdelta} \right>_{\Omega}=0, \nonumber
\end{align}
\label{eq:weakform}%
\end{subequations}
where $<a,b>_{\Omega}$ is the canonical hermitian inner product between two fields $a$ and $b$ over $\Omega$, i.e. $\int_\Omega a^*b \mathrm{d}x$ where the star symbol denotes the complex transpose. 
Also, here $<a,b>_{\partial\Omega_1}$ is the inner product $a^*b|_{\hx=n/2}$ at the leading edge. We enforce the pressure value at the leading edge by a penalty technique using a very large weigth $\mathcal{N}=10^{30}$ on the nodes of that boundary. \\

Next, we expand the pressure and deformation fields in the basis of shape functions (chosen to be the same as the test functions): 
$$ \tilde{p}_k(\hX,\hti)=\sum_{j=1}^N \tilde{p}_{k_j}(\hti) q_j(\hX)  \hspace{0.25cm}, \hspace{0.25cm} \tilde{\delta}_k(\hX,\hti)=\sum_{j=1}^N \tilde{\delta}_{k_j}(\hti) \kappa_j(\hX),$$
where $N$ is the number of nodes. Inserting this decomposition into the weak form, the problem to solve can be recast as

\begin{equation}
\begin{pmatrix}
A_k & B_k & L_k & D_{1k} \\
C_k & E_k & 0 &  0\\
-L_k^T & 0 & 0&  0 \\
D_{k1} & D_{k2} & 0 &  D_{k}\\
\end{pmatrix} 
\begin{pmatrix}
\tilde{p}_{k_j} \\
 \tilde{\delta}_{k_j} \\
  \tilde{m}_k \\
  \tilde{V}_k
\end{pmatrix}
=
-\begin{pmatrix}
\left< q, Q_1(\hp_k,\hdelta_k,\hm_k,\hV_k) \right>_{\Omega}-\mathcal{N}\left<q,(\hp_k)\right>_{\partial\Omega_1} \\
\left<\kappa, Q_2(\hp_k,\hdelta_k,\hm_k,\hV_k) \right>_{\Omega} \\
\left<1, \hp_k\right>_{\Omega}  \\
\left<1, Q_4(\hp_k,\hdelta_k,\hm_k,\hV_k)\right>_{\Omega}
\end{pmatrix},
\label{eq:matrixform}
\end{equation}
where the terms of Eq.~(\ref{eq:matrixform}) are given by
$$A_k= \left< q, \frac{dq_j}{\partial\hX}(\hh-\eta\hdelta_k)^3\right>_{\Omega}+\mathcal{N}\left<q,q_j\right>_{\partial\Omega_1}$$
$$B_k= -\left< q, \left(3\eta\frac{\partial\hp_k}{dx}(\hh-\eta\hdelta_k)^2+6(\hV_k-2)\eta\right)\kappa_j\right>_{\Omega},$$
$$C_k=- \left< \kappa, \underbrace{\frac{1}{\pi}\int_{-n/2}^{n/2}q_j (\hX') \ln\left | \hX-\hX' \right |  \mathrm{d}\hX' }_{\phi_j}\right>_{\Omega},$$
$$E_k=\left< \kappa, \kappa_j \right>_{\Omega},$$
$$L_k=-\left< q, 1 \right>_{\Omega},$$
$$D_{1k}=\left< q, 6\tilde{V}_k\left(\hh-\eta\hdelta_k\right)\right>_{\Omega},$$
$$D_{k1}=- \left<1,q_j\hat{b}'+\frac{1}{2}\frac{\partial q_j}{\partial\hX}\left(\hh-\eta\hdelta_k\right)\right>_{\Omega},$$
$$D_{k2}=- \left<1,-\frac{1}{2}\frac{\partial\hp_k}{\partial\hX}\eta\kappa_j+\frac{\eta\hV_k\kappa_j}{\left(\hh-\eta\hdelta_k\right)^2}\right>_{\Omega},$$
$$D_{k}=- \left<1,\frac{1}{\hh-\eta\hdelta_k}\right>_{\Omega},$$
and 
$$q_4=\hp_k\hat{b}'+\frac{1}{2}\frac{\partial\hp_k}{\partial\hX}\left(\hh-\eta\hdelta_k\right)+\frac{\hat{V}_k}{\hh-\eta\hdelta}.$$

Note that for every node, the Green's function integral in $C_k$ is computed using a quadrature formula. If $N_e$ is the number of element segments on the $x$-boundary of the domain and $N_g$ the number of Gauss quadrature points, then the shape function $\phi_j(x)$ is computed as
\begin{equation}
\phi_j(x)=\frac{1}{\pi}\sum_{e=1}^{N_e}\sum_{g=1}^{N_g} q_j (x_{e_g}) \ln\left | x-x_{e_g} \right | \omega_g,
\label{eq:gauss}
\end{equation}
where $\omega_g$ are the weights of the quadrature formula. $\phi_j(x)$ is computed at all nodes $x_j$ before being interpolated on all segments $[x_j,x_{j+1}]$. However, the logarithm in \eqref{eq:gauss} implies that $\phi_j(x_j)$ is well defined only if $x_j\neq x_{e_g}$ for any element on the boundary. Therefore, a careful choice of the type of finite elements and of the order of Gauss quadrature must be made, lest the constraint $x_j\neq x_{e_g}$ be violated.
For instance, if the shape functions are chosen as quadratic ($P2$), an element $e$ defined by the interval $x\in [e^i,e^j]$ has three nodes: $x_1=e^i$, $x_2=(e^i+e^j)/2$, $x_3=e^j$. Then if one chooses $N_g=3$, i.e. a three-point quadrature, the $x_{e_j}$ points are $x_{e_1}=(e^i+e^j-\sqrt{3/5})/2$,  $x_{e_2}=(e^i+e^j)/2$,  $x_{e_3}=(e^i+e^j+\sqrt{3/5})/2$. Therefore, the integral cannot be computed continuously since $x_2=x_{e_2}$. However, these numerical constraints are satisfied for P1-elements which we thus choose for our subsequent computations. For such linear elements, exact quadrature formula exist even for $N_g=1$. Hereafter, we set $N_g=2$ once for all. 

\bigskip
Last, regarding the computational domain (or mesh), we use a 2D rectangular domain given by $\Omega=\left[-n/2, n/2\right ]\times\left[-\varepsilon_y,\varepsilon_y\right]$ where $\epsilon_y\ll n$. The domain is discretized with triangles as shown in Fig.~\ref{fig:mesh}. The use of a 2D domain to solve the 1D system of equations~\eqref{eq:LinearizedPB2} is simply a constraint of the software package FreeFem++ \cite{S-Hecht12}, the finite element solver we use. However, in order to ensure that the discretized problem remains 1D, we impose periodic conditions in the $y$-direction, thereby allowing variations in the $x$-direction only. As a proof of the method, let us recall that we use P1-elements on each triangle. Thus, the 3 shape functions are given by $N_1=x$, $N_2=y$, and $N_3=1-x-y$. Any field $X(x,y)$ then reads
$$X=\sum_{i=1}^3 X_iN_i(x,y)=(X_1-X_3)x+(X_2-X_3)y+X_3.$$
Therefore, imposing the periodicity in the $y$-direction enforces the condition $X_2=X_3$. Thence, 
$$X(x)=(X_1-X_2)x+X_2=\sum_{i=1}^2 X_i\tilde{N}_i(x),$$
where $\tilde{N}_1=x$ and $\tilde{N}_2=1-x$, which are precisely the P1-elements in 1D. 
\begin{figure}[ht!]
\centering
 \begin{tikzpicture}
  \node (img1)  {\includegraphics[scale=0.45]{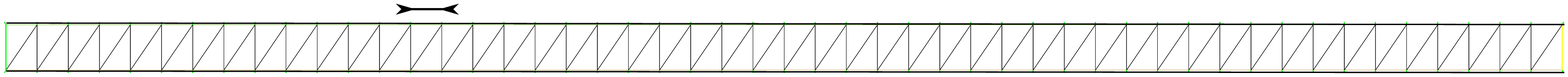}};
  \node[above=of img1, node distance=0cm, xshift=-3.6cm, yshift=-1.2cm] {$\Delta x$};
  \node[left=of img1, node distance=0cm, xshift=1cm,yshift=-0.1cm] {$y$};
  \node[below=of img1, node distance=0cm, yshift=1cm] {$x$};
\end{tikzpicture}
   \caption{{Typical mesh for the computation of the elasto-hydrodynamics problem.}}      
   \label{fig:mesh}
 \end{figure} 

Next, we proceed to show that our FEM implementation converges well as we increase the number of elements and the order of quadrature used for numerical integrations. We will also show that our FEM code can reproduce results from previous elasto-hydrodynamics related works \cite{S-Maha04, S-Skotheim05}. 

\addtocounter{subsection}{1}
\subsection{Numerical validation}
\subsubsection{Convergence study}
\begin{figure}[b!]
\centering
 \begin{tikzpicture}
 \node[xshift=-4cm, yshift=0.0cm] (img1)  {\includegraphics[scale=0.7]{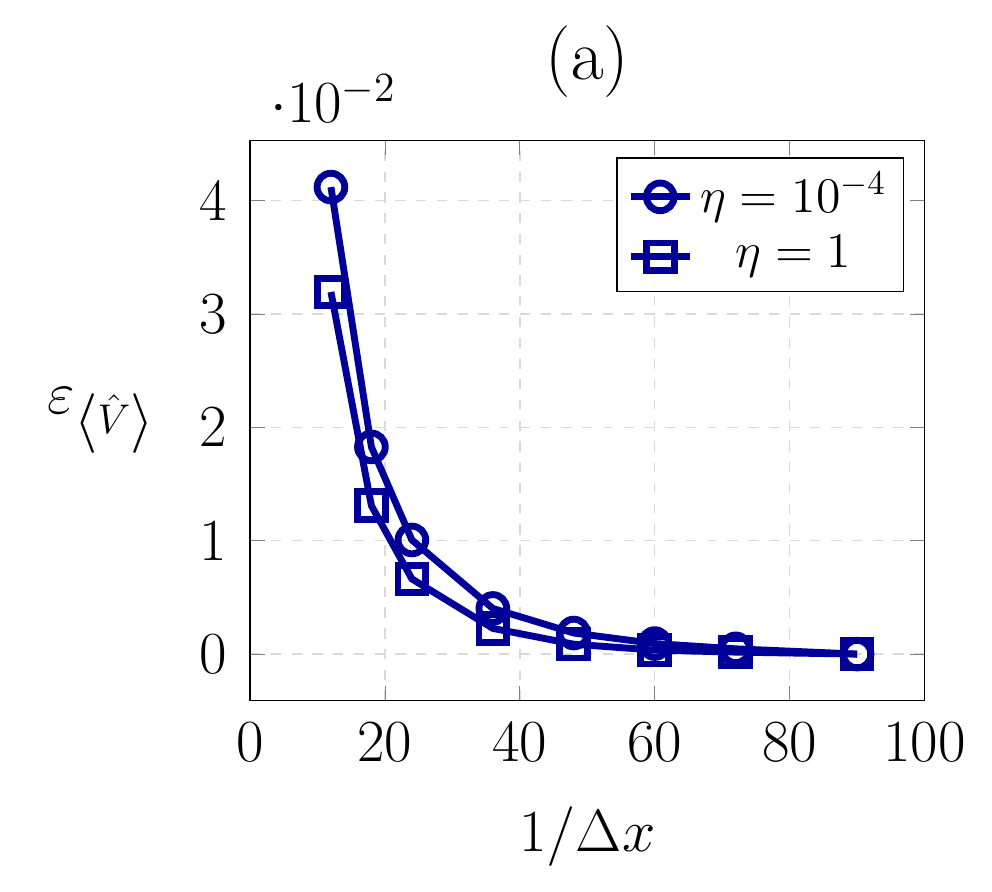}};
 \node[right=of img1,xshift=-1cm, yshift=0.0cm] (img2)  {\includegraphics[scale=0.7]{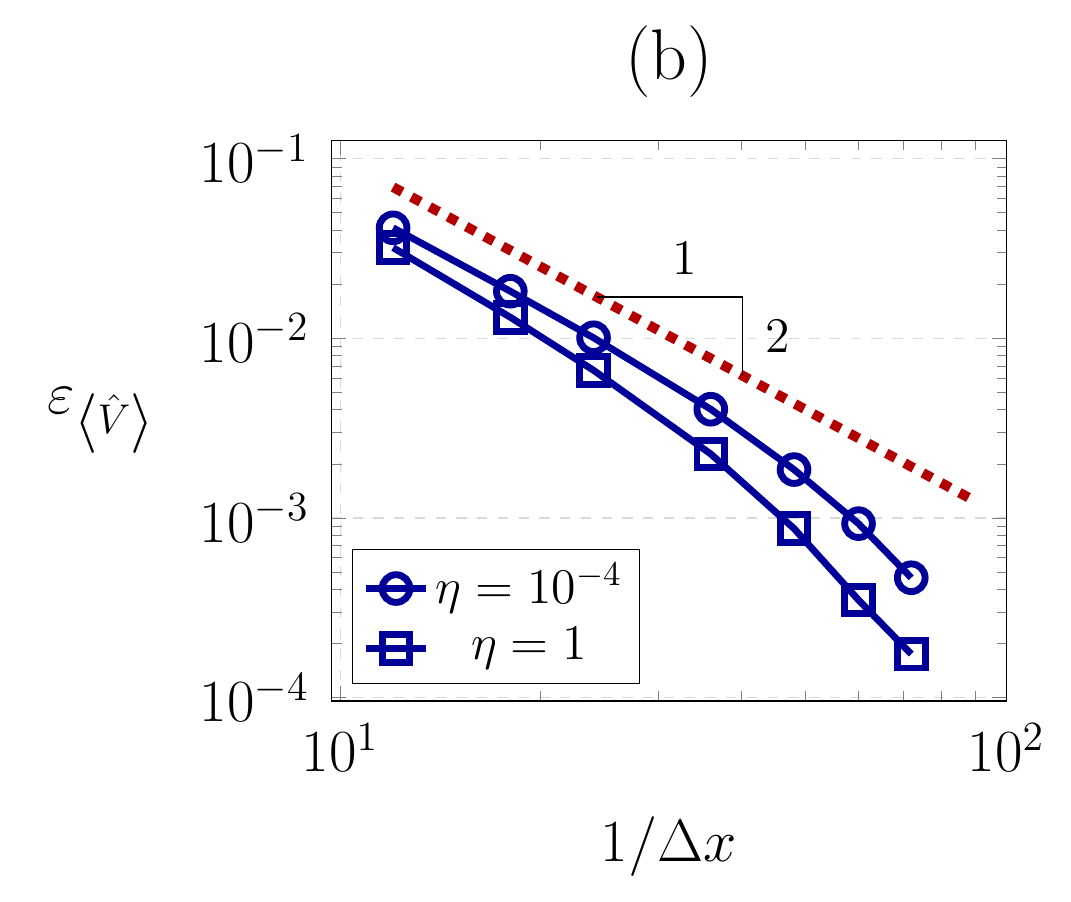}};
\end{tikzpicture}
 \caption{{Relative error on the gliding speed as a function of the inverse mesh size, in (a) linear scale and in (b) logarithmic scale. The right subfigure shows that for both stiff and soft substrate, the error decreases with the mesh size almost quadratically.}}
 \label{fig:convergence}
 \end{figure} 

Before going into the details of results, we evaluate the convergence performance of our FEM code. For this purpose, the basal geometry of the bacteria is chosen as
\begin{equation}
\hh(\hX,\hti)=1+\hA\sin\left[2\pi (\hX+\hti) \right],
\end{equation}
where $\hA=A/h_0=0.15$ is the amplitude. For two values of the softness parameter, $\eta_1=10^{-4}$ and $\eta_2=1$, we compute the gliding velocity for different mesh sizes $\Delta x\in\left[0.01,0.1\right]$. In order to show the convergence of our FEM code, let us define the relative error on the time-averaged gliding speed as 
\begin{equation}
\varepsilon_{\hV}=\dfrac{\|\bl\hV\br-\bl\hV\br_{ref}\|}{\bl\hV\br_{ref}},
\label{eq:error}
\end{equation}
where $\bl\cdot\br=\int_0^1(\cdot)\mathrm{d}\hti$ is the time-average operator and $\big\langle\hV\big\rangle_{ref}$ is the reference velocity obtained at our most refined mesh, for each of the softness parameters. The convergence of the numerical scheme is demonstrated on Fig.~\ref{fig:convergence} where the relative error on the speed is plotted as a function of the inverse mesh size. We find that the gliding speed converges slightly faster in the case of soft substrates. However, since $\eta_1$ is the lowest value of the softness parameter we will investigate, we choose for the subsequent computations the mesh size $\Delta x\sim0.025$, thus bounding the relative error to remain less than $0.2\%$.
 
Given that the reference solution of our convergence study was obtained with our own code, we shall also ensure that our FEM code accurately reproduces previously obtained results, from the elasto-hydrodynamics literature.
 
\subsubsection{Comparison with previous work}
In this subsection, we verify that our code reproduces accurately the results from Skotheim and Mahadevan \cite{S-Skotheim05} who computed the lift force on a large and soft cylinder sliding on a rigid surface. As noticed in \cite{S-Skotheim05}, this problem is equivalent to that of a rigid slider moving a soft semi-infinite substrate, which we implemented and compared with results from \cite{S-Skotheim05}. In both cases, the shape of the contact surface of the cylinder is given by 
$$\hh(\hX)=1+\hX^2,$$
and the deformation of the substrate by
$$\hdelta(\hX)= -\int_{-\infty}^{\infty}  \hp(\hX') \ln\frac{|Y|}{\left (\hX-\hX'\right )^2} \mathrm{d}\hX' $$
for a 3D cylinder of finite length $Y$. We computed the lift force as a function of $\eta$ and compared with the result with that from Skotheim and Mahadevan. Simulating the Reynolds equation coupled to that of the substrate deformation in the particular setting of this problem, we obtain results which are in excellent agreement with those of the aforementioned authors, as shown in Fig.~\ref{fig:validation}. 

\begin{figure}[t!]
\centering
 \begin{tikzpicture}
  \node (img1)  {\includegraphics[scale=0.75]{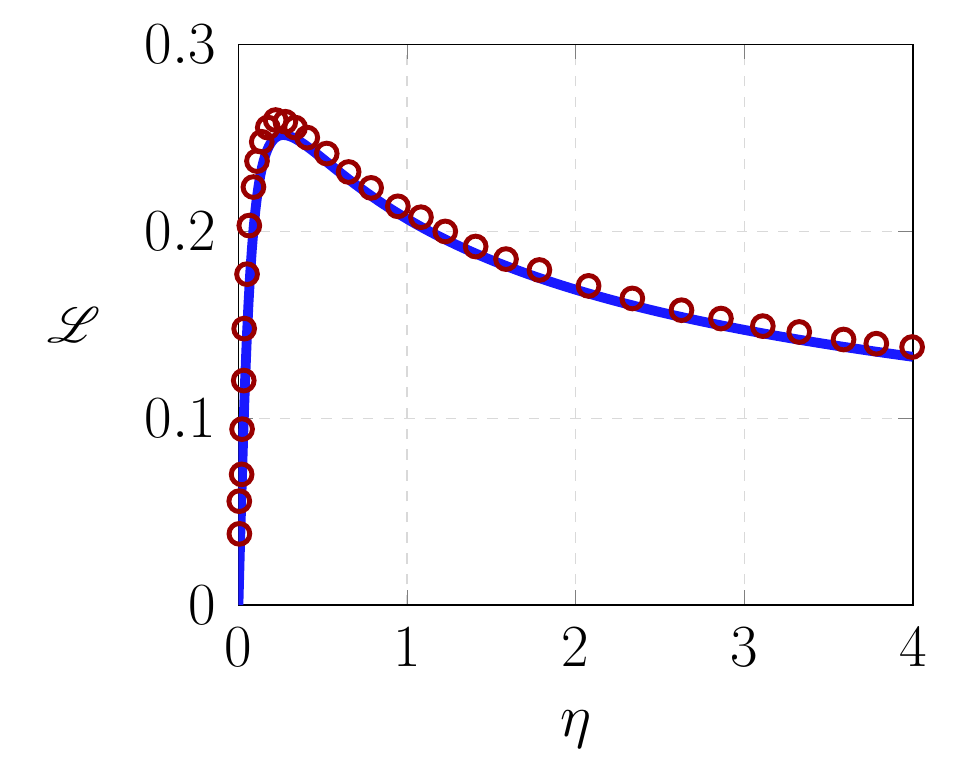}};
  \end{tikzpicture}
 \caption{{Lift force as a function of $\eta$ in the case $Y=1000$. The blue line is the result of our numerical code while the red symbols are extracted data from a previous work \cite{S-Skotheim05}.}}
 \label{fig:validation}
 \end{figure}
 
Having validated our FEM code, we show in the next sections a selection of numerical results that are supplementary to those of the main paper. 

\addtocounter{subsection}{1}
\subsection{Trajectories for different values of $\eta$}
We show on Fig.~\ref{fig:traj} the trajectories $\hx(\hti)$ of the center of mass of the myxobacteria in three configurations: a very stiff ($\eta=10^{-3}$), soft ($\eta=1$) and very soft ($\eta=10^3$) substrate. For all these cases, $\hA=0.15$ and $n=5$. Each trajectory is found to be the superposition of a non-zero mean component and an oscillatory perturbation that causes the cell to glide intermittently. On the very stiff substrate, the trajectory resembles that of a stick-slip motion. Surprisingly, this type of trajectory was also recently observed in a colony of myxobacteria gliding through the use of a different motility apparatus, namely type IV pili \cite{S-Gibiansky13}. However, in that work, Gibiansky \textit{et al.} showed that the stick-slip movements were aperiodic, unlike the periodic trajectory shown in Fig.~\ref{fig:traj}. Nevertheless, as suggested by these authors and now confirmed by our work, the stick-slip character of myxobacteria gliding may neither be due to chemotaxis nor due to the exact nature of the motility apparatus. Instead, it may depend on the cell-substrate interaction mediated by a lubricating slime \cite{S-Gibiansky13}.

\begin{figure}[t!]
\centering
 \begin{tikzpicture}
 \node[xshift=0cm, yshift=0.0cm] (img1)  {\includegraphics[scale=0.7]{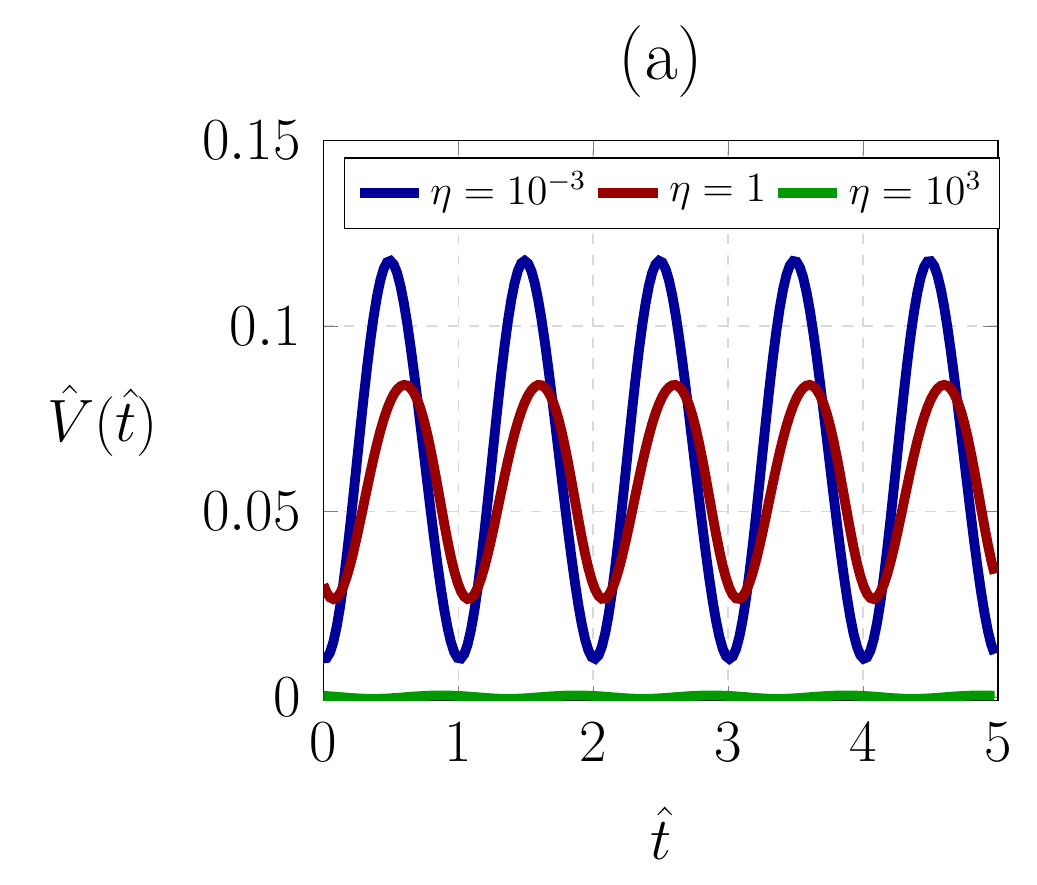}};
   \node[right=of img1,xshift=-1.cm, yshift=0.0cm] (img1)  {\includegraphics[scale=0.7]{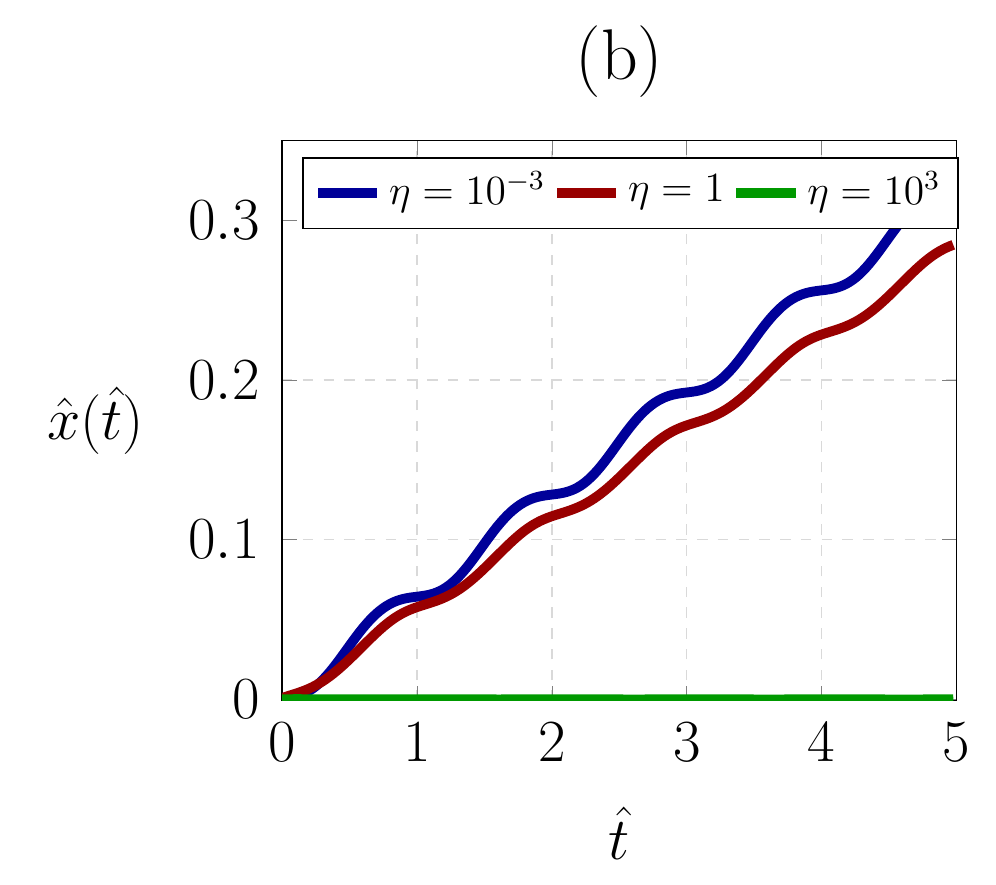}};
\end{tikzpicture}
 \caption{{Instantaneous (a) speed and (b) trajectories of the center of mass of the bacteria on substrates with different softness numbers. The simulation parameters are $\hA=0.15, n=5$.}}
 \label{fig:traj}
 \end{figure}

\addtocounter{subsection}{1}
\subsection{Influence of the dimensionless bacterial length $n$}

In the main text, we have shown in Fig.~2 that, for very stiff substrates (i.e small $\eta$), we recover the theoretical swimming speed for a Taylor's sheet moving near a rigid wall \cite{S-Taylor51, S-Katz74}, as given by
\begin{equation}
\bl \hV_{\eta=0}\br= \frac{3}{2+1/\hA^2}.
\label{eq:TaylorLub}
\end{equation}
However, Fig.~\ref{fig:speedwavenumb} shows that the precision of the agreement between $\bl \hV\br$ and $\bl \hV_{\eta=0}\br$ depends on the bacterial length. Given that Taylor's swimmer is an infinitely long sheet, we find accordingly that longer bacteria (i.e large $n$) have a gliding speed closer to the theoretical prediction. 
\begin{figure}[h!]
\centering
 \begin{tikzpicture}
  \node (img1)  {\includegraphics[scale=0.75]{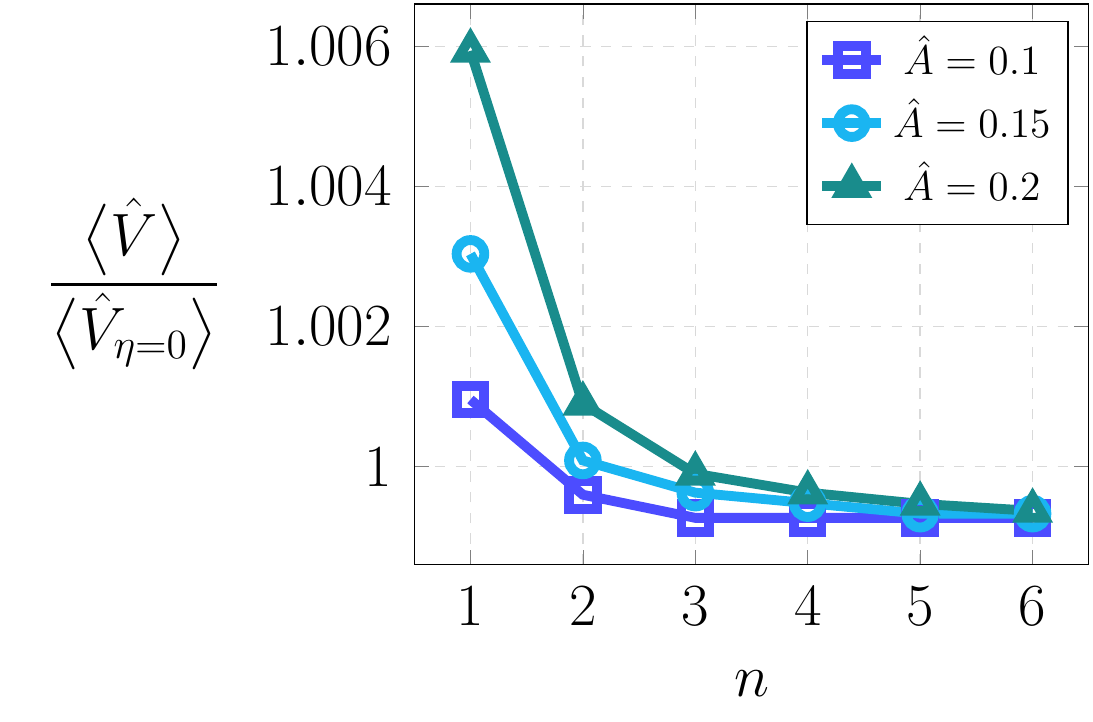}};
 \end{tikzpicture}
 \caption{{Comparison between the gliding speed from the elasto-hydrodynamic problem and the theoretical speed of Taylor's swimmer near a rigid wall, for different bacterial lengths. Here the softness parameter was set to $\eta=10^{-4}$.}}
 \label{fig:speedwavenumb}
 \end{figure}

\addtocounter{subsection}{1}
\subsection{Instantaneous pressure and substrate deformation}

Fig.~\ref{fig:pres} shows the distribution of the slime pressure at different instants of the periodic cycle. 
\begin{figure}[b!]
\centering
 \begin{tikzpicture}
 \node[xshift=0cm, yshift=0.0cm] (img1)  {\includegraphics[scale=0.82]{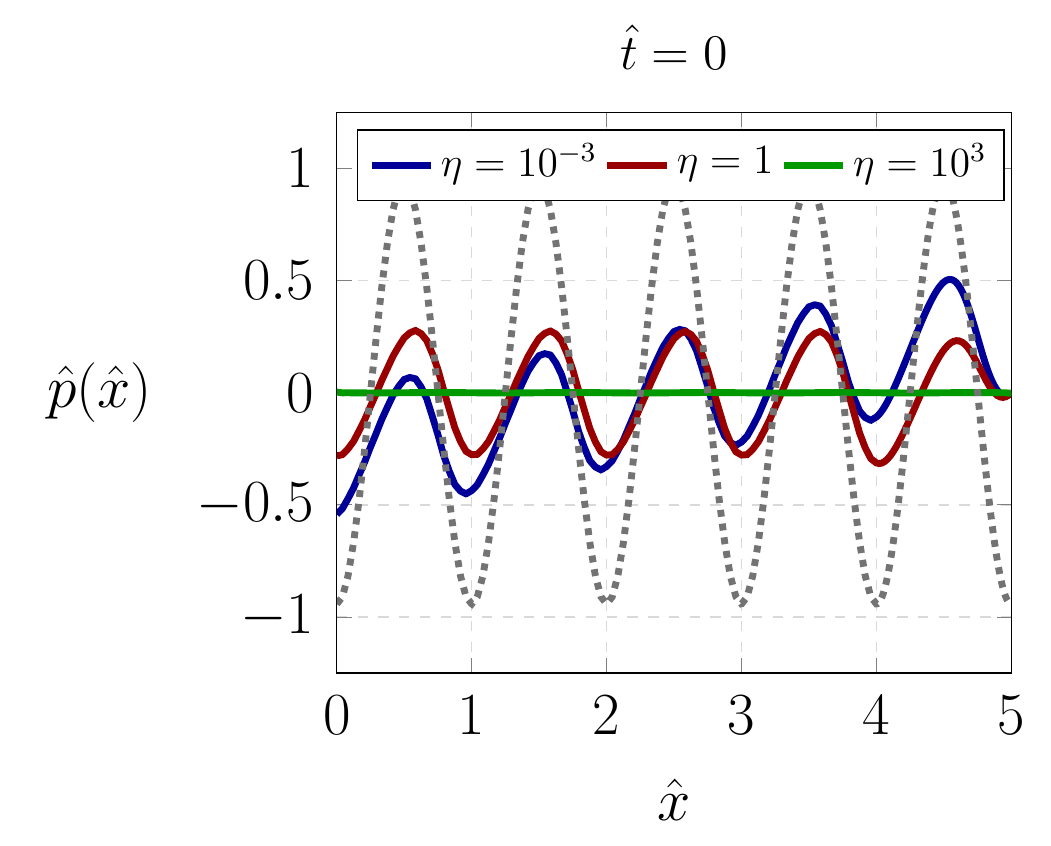}};
   \node[right=of img1,xshift=-1.6cm, yshift=0.0cm] (img2)  {\includegraphics[scale=0.82]{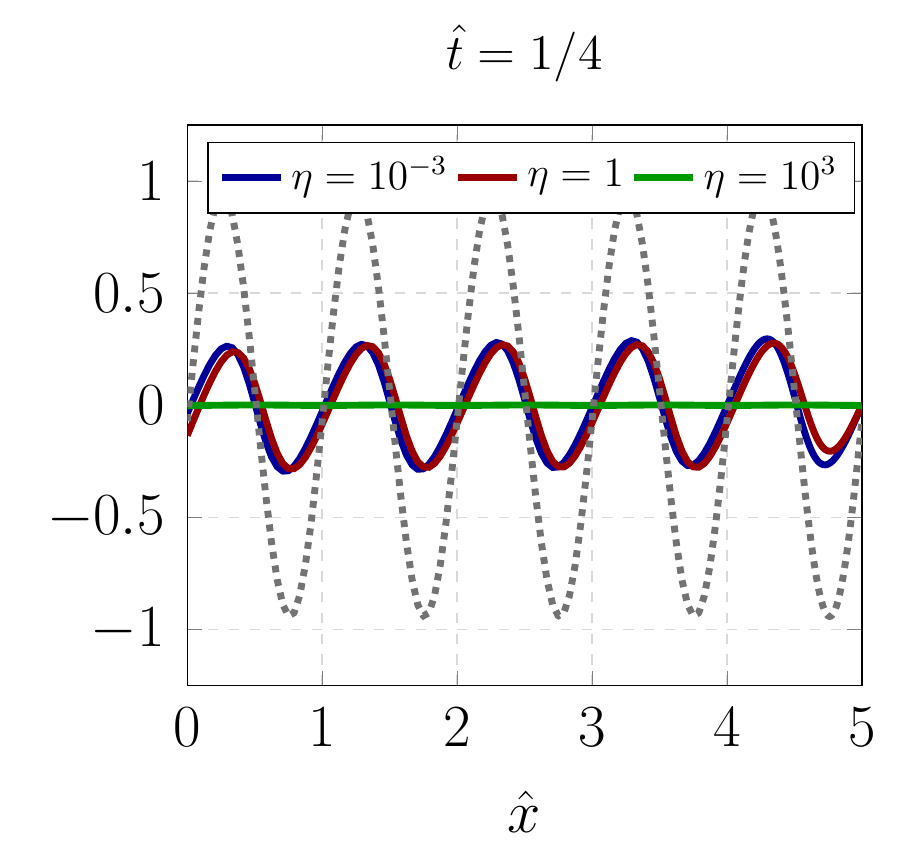}};
    \node[below=of img1,xshift=0.cm, yshift=1.0cm] (img3)  {\includegraphics[scale=0.82]{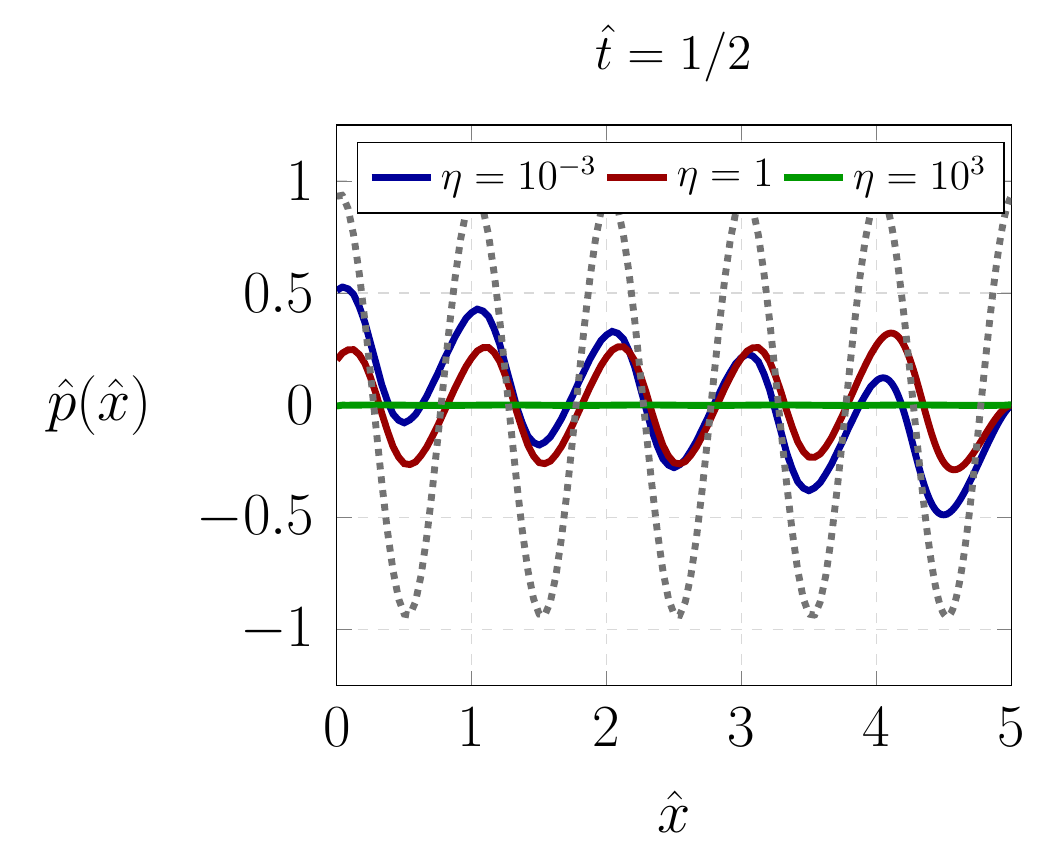}};
   \node[right=of img3,xshift=-1.6cm, yshift=0.0cm] (img4)  {\includegraphics[scale=0.82]{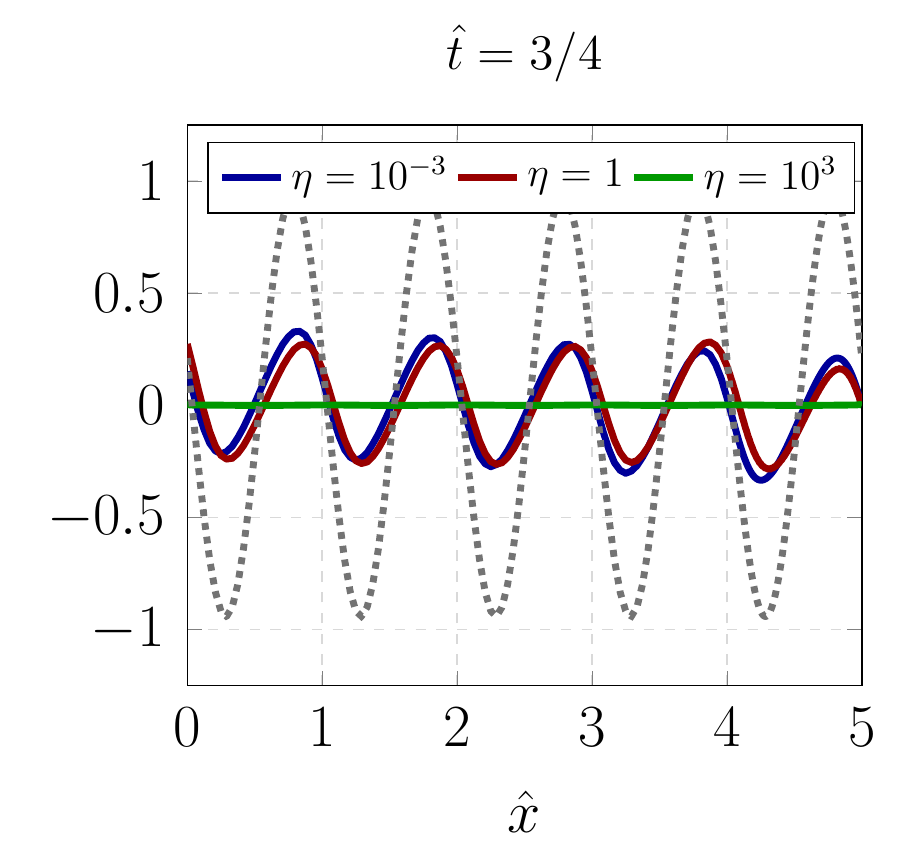}};
\end{tikzpicture}
 \caption{{Snapshots of the pressure field in the lubricating film at different instants of the periodic cycle $\hti=0, 1/4, 1/2, 3/4$ and for different values of the softness parameter $\eta$. The pressure being periodic as the oscillations of the bacterial membrane, we checked that the distribution at $\hti=1$ is exactly the same at $\hti=0$. Here, the dotted line shows the deformation of the bacterial shape $-\dfrac{\partial\hat{b}}{\partial\hx}$. The parameters of the bacterial geometry are $\hA=0.15$ and $n=5$.}}
 \label{fig:pres}
 \end{figure}
At every instant, increasing the softness parameter $\eta$ has three major consequences. First, the amplitude of the pressure decays with $\eta$. In agreement with the decrease of the gliding speed with $\eta$ (see Fig.~2 in the main paper), this diminution is moderate for $\eta\leq 1$ and very abrupt when $\eta>1$. Second, there is a net (oscillating) front-back pressure gradient that decreases to zero as the softness parameter increases. Last, as $\eta$ increases, the pressure field $\hp(\hx)$ and the deformation of the bacterial shape $\dfrac{\partial\hat{b}}{\partial\hx}$ gradually shift from being in phase to being out of phase. To illustrate this shift, note that for $\eta=10^{-3}$, the peaks of pressure and those of the membrane deformations occur at the same locations. However, for $\eta=1$, these peaks no longer appear at the same locations on the $\hx-$axis (see plots at $\hti=,1/2$), meaning that $\hp(\hx)$ and $\dfrac{\partial\hat{b}}{\partial\hx}$ are less in phase. 
These observations imply that, the pressure-dependent terms $I_1=-\hp\dfrac{\partial\hat{b}}{\partial\hx}$ and $I_2=-\dfrac{1}{2}\dfrac{\partial\hat{p}}{\partial\hx}(\hh-\eta\hdelta)$ decrease to zero with the softness parameter.

\begin{figure}[b!]
\centering
 \begin{tikzpicture}
 \node[xshift=0cm, yshift=0.0cm] (img1)  {\includegraphics[scale=0.82]{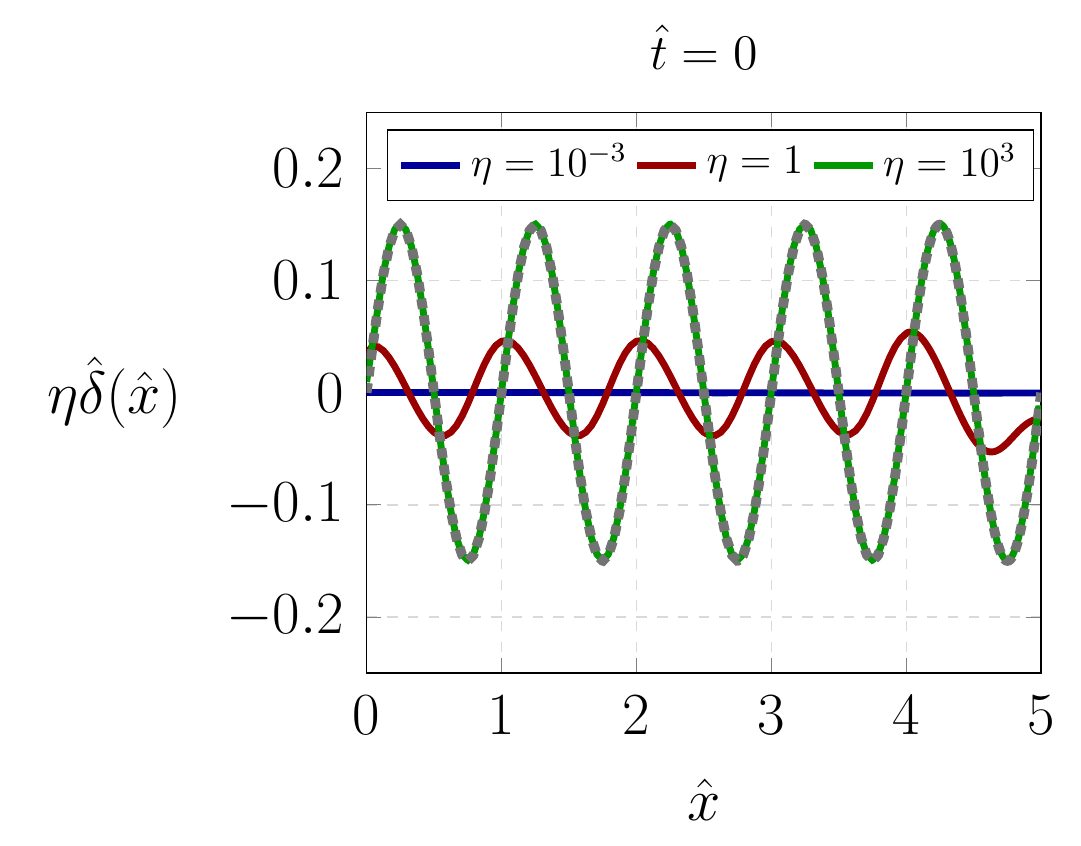}};
   \node[right=of img1,xshift=-1.6cm, yshift=0.0cm] (img2)  {\includegraphics[scale=0.82]{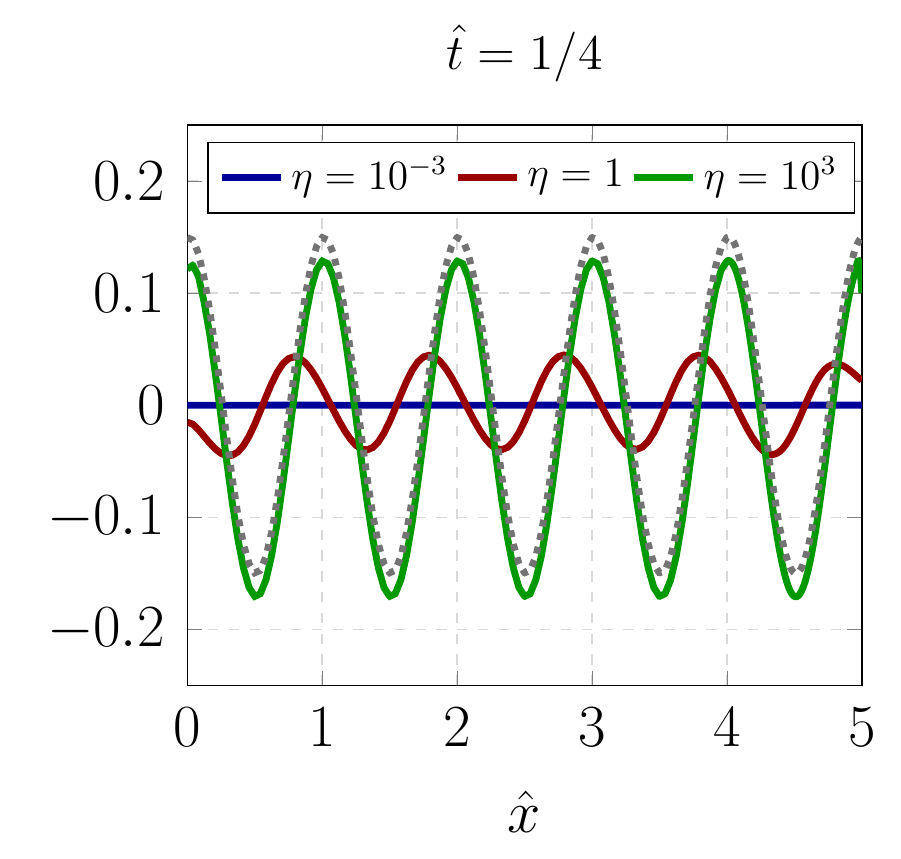}};
   \node[below=of img1,xshift=0.cm, yshift=1.0cm] (img3)  {\includegraphics[scale=0.82]{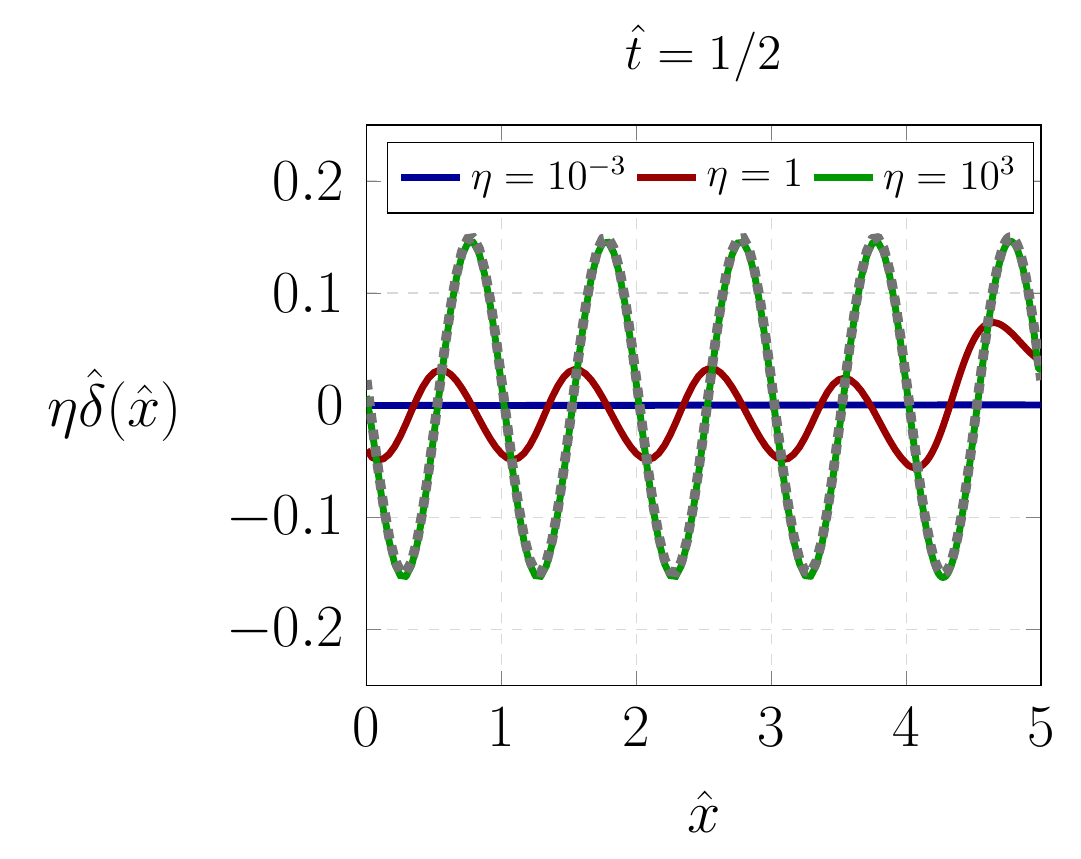}};
   \node[right=of img3,xshift=-1.6cm, yshift=0.0cm] (img4)  {\includegraphics[scale=0.82]{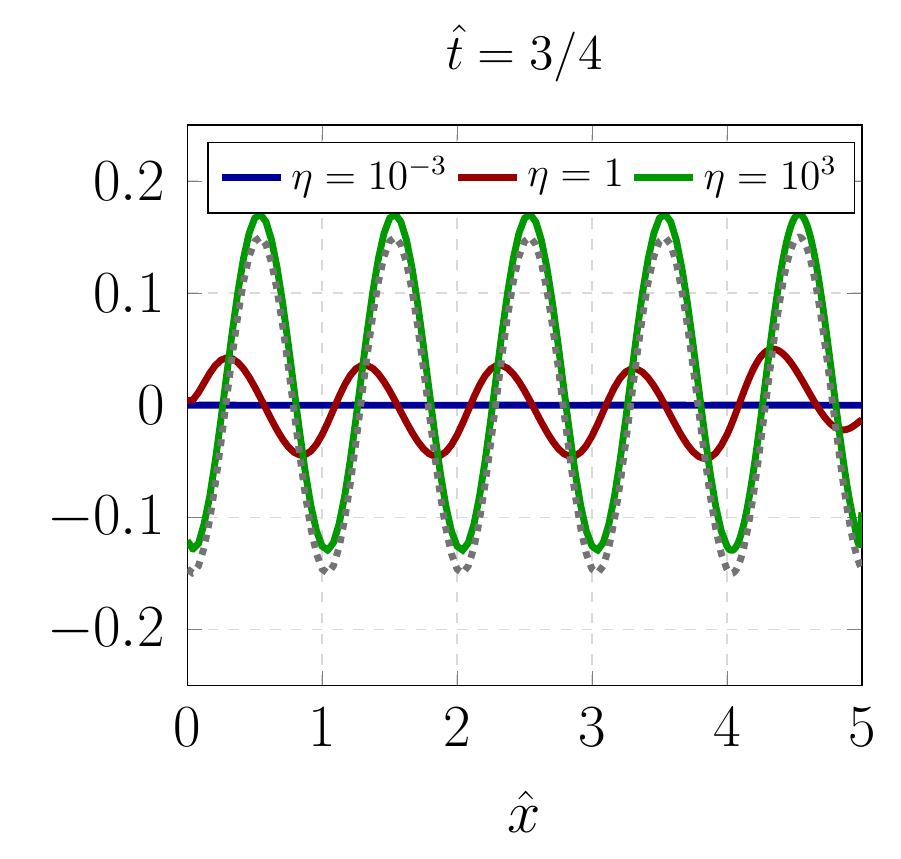}};
\end{tikzpicture}
 \caption{{Snapshots of the substrate deformation at differents instants of the periodic cycle $\hti=0, 1/4, 1/2, 3/4$ and for different values of the softness parameter $\eta$. Here, the dotted line shows the bacterial shape $\hat{b}(\hx)$. The parameters of the bacterial geometry are $\hA=0.15$ and $n=5$.}}
 \label{fig:substratedef}
 \end{figure}
  
Fig.~\ref{fig:substratedef} shows snapshots of the substrate at different instants of the periodic cycle. At every instant, we find, as expected, that the amplitude of the substrate deformation increases with the softness parameter. However, Fig.~\ref{fig:substratedef} shows that the substrate deformation does not increase indefinitely (since the pressure meanwhile decreases), but instead converges to the shape of the myxobacteria. For large values $\eta\rightarrow\infty$, the substrate is such that $\eta\hdelta(\hx,\hti) \approx\hat{b}(\hx,\hti)$. \\

As for the phase behavior, we see here as well the phase shift previously described between the pressure and the membrane deformation. Clearly, on very soft substrates ($\eta=10^3$), the deformation $\hdelta(\hx)$ and the bacterial shape $\hat{b}(\hx)$ are in phase. However, they are almost exactly out-of-phase for a harder substrate, as shown by the case $\eta=1$. Besides the phase behavior, the amplitude of the substrate deformation also shows a striking contrast between the large and the low values of the softness parameter $\eta$. Fig.~\ref{fig:substratedef} shows that the substrate amplitude is almost zero for very low values of $\eta$ (as expected, since we approach the rigid wall limit). But for very soft substrates, the substrate amplitude almost equals that of the bacterial shape. These different limits for the substrate amplitude can be derived from the depth averaged mass balance, given by \eqref{eq:height-flowrateC}. To derive the limiting behaviors,  \eqref{eq:height-flowrateC} in its dimensionless form reads

$$\dfrac{\partial \hat{b}}{\partial \hti}=\eta\dfrac{\partial \hdelta}{\partial \hti}-\dfrac{\partial \hat{Q}_h(\eta)}{\partial \hx},$$
where $\hat{Q}_h(\eta)$ is the dimensionless flow rate given by Eq.~(\ref{eq:FlowrateDimensionless}).  

At very low values of $\eta$, using the ansatz $\bfv{s}=\bfv{s}_0+\eta \bfv{s}_1+O(\eta^2)$, where $\bfv{s}=[\hp,\hdelta,\hV,\hat{Q}_h]^T$, the mass balance at zeroth order in $\eta$ reads
\begin{equation}
\dfrac{\partial \hat{b}}{\partial \hti}= - \dfrac{\partial \hat{Q}_{h_0}}{\partial \hx}.
\label{eq:membtoflow}
\end{equation}
Furthermore, it is straightforward to see from the low-$\eta$ expansion above that the substrate deformation is given by $\dfrac{\delta}{h_0}=\eta\hdelta=\eta\hdelta_0+\eta^2\hdelta_1+O(\eta^3)$. Therefore, as shown on Fig.~\ref{fig:substratedef}, the substrate deformation vanishes in the limit $\eta\rightarrow 0$, corresponding to very stiff substrates. We conclude that at the lowest order in $\eta$, the substrate remains undeformed while, as stipulated by \eqref{eq:membtoflow}, the membrane oscillations are converted into a slime flow in the gap. 

In the limit of large $\eta$, using the ansatz $\bfv{s}=\eta^{-1} \bfv{\tilde{s}}_1+O(\eta^{-2})$, we obtain at lowest order
\begin{equation}
\dfrac{\partial \hat{b}}{\partial \hti}= \dfrac{\partial \tilde{\hdelta}_1}{\partial \hti}.
\label{eq:membtosubs}
\end{equation}
Therefore, in the limit $\eta\rightarrow \infty$, the substrate deformation reads $\hdelta= \eta^{-1}\hat{b}+O(\eta^{-2})$, or equivalently $\eta\hdelta\approx \hat{b}$. As a result, the slime pressure compatible with such deformation, in the linear elasticity framework, decreases to zero as $\hp\sim \hdelta \sim 1/\eta$.   

\section{Elasto-capillary deformation of a soft substrate}

Contrary to the prediction of the elasto-hydrodynamics model, experiments show that \textit{M. xanthus} cells can glide with a non-zero speed even on very soft substrates. Hereafter, we show that this mismatch is due to capillary effects which can no longer be neglected in the limit of very soft substrates. There are two motivations to account for capillary effects. First, in the vicinity of the triple line of the slime-substrate-air interfaces, the surface tension of the slime-air can cause a ridge to form, as well documented in the literature of soft solids \cite{S-Shanahan86, S-Shanahan88, S-Marchand12, S-Lubbers14, S-Style17}. Such a ridge can create a curvature of the slime-air interface (see Fig.~\ref{fig:setupec}), which can then induce a pressure difference at the leading edge of the cell. Therefore, the zero pressure condition at the leading edge, $\hp(n/2,\hti)=0$, does not hold for very soft substrates. Second, the singularity of the displacement at $\hx=s$ existing in the case of the pure elastic half-space (see \eqref{eq:surfdisp}) is removed by accounting for the capillary effects. Indeed, this singularity is due to the fact that very close to the application point of the load, the deformation is so large that the force balance cannot be written on the reference (undeformed) configuration. In the deformed state, one must also account, in the vertical force balance, for the surface tension of the slime-substrate interface. \\

In the next subsections, we shall derive the expression of the elasto-capillary deformation of the substrate surface in a dimensional form before carrying out the non-dimensionalization of the full elasto-capillary-hydrodynamics problem. Proceeding as before, we shall first obtain the substrate deformation in the case of concentrated loads $F_p=p(s)\mathrm{d}s$ and $F_q=q(s)\mathrm{d}s$ that include the effects of the surface tension between the slime and the substrate, and then we shall generalize the results using the superposition principle for distributed loads. 

\begin{figure}[b!]
\centering
 \begin{tikzpicture}
  \node (img1)  {\includegraphics[scale=0.55]{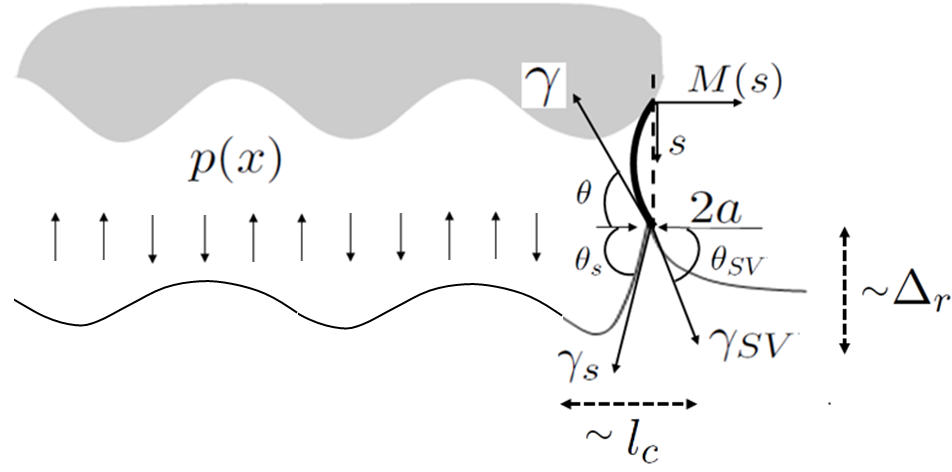}};
 \end{tikzpicture}
 \caption{{Schematic description of the elasto-capillary problem: a gliding bacterium (in grey) with a sinusoidal ventral shape. A ridge is formed and balanced by the surface tensions at the slime-air, substrate-slime and air-substrate interfaces. The latter two tensions are assumed equal in the rest of the problem. See the text for a description of the variables.}}
 \label{fig:setupec}
 \end{figure}
 
\addtocounter{subsection}{1}
\subsection{Solution for a concentrated line load}
In the limit of small deformations, the angle of action of $\gamma_\text{s}$, the slime-substrate interfacial tension, corresponds to the slope of the substrate deformation $\dfrac{\partial \bar{w}_s}{\partial x}$, where $\bar{w}_s$ is the vertical deformation of the substrate surface, given by Eq.~(\ref{eq:surfdisp}). By differentiating Eq.~(\ref{eq:surfdisp}) and adding the contribution of the surface tension to the vertical force $F_p$, we obtain \cite{S-Limat12}:
\begin{subequations}
\begin{align}
\frac{\partial \bar{u}_\text{s}}{\partial x} & \approx -\frac{2(1-\nu^2)}{\pi E} \frac{F_q}{|x-s|}-\frac{(1-2\nu)(1+\nu)}{E}\delta(x-s)\left(F_p-2\gamma_\text{s}\frac{\partial \bar{w}_\text{s}}{\partial x}\right), \label{eq:surfdisp2a} \\
\frac{\partial \bar{w}_\text{s}}{\partial x} & \approx \frac{2(1-\nu^2)}{\pi E}\frac{1}{|x-s|}\left(F_p-2\gamma_\text{s}\frac{\partial \bar{w}_\text{s}}{\partial x}\right)-\frac{(1-2\nu)(1+\nu)}{E}\delta(x-s)F_q. 
\end{align}
\label{eq:surfdisp2}%
\end{subequations} 
In \eqref{eq:surfdisp2}, we only added the vertical projection of the slime-substrate surface tension, since we consider only small angles of action. Therefore, the horizontal load remains $F_q$, while the vertical one becomes $\left(F_p-2\gamma_s\dfrac{\partial \bar{w}_s}{\partial x}\right)$. Note that the factor $2$ in front of $\gamma_s$ originates from the fact that the surface tension acts twice, on each side of the (symmetric) ridge located at $x=s$ where the concentrated load is applied. Let us stress that this is an approximate approach to remove the singularity at $x=s$.
A more rigorous derivation would require including the surface tension effects from the start rather than first obtaining the slope in absence of the surface tension, and then adding the correction as we do here. Nevertheless, Limat \textit{et al.} showed that this approximation is very satisfactory when compared to the full solution \cite{S-Limat12, S-Dervaux15}. The derivation of the full solution is more cumbersome, yet it adds little precision to the elegant approximation first derived by Limat \cite{S-Limat12}, which we now set to reproduce in the context of our problem.

To obtain the elasto-capillary deformation of the substrate surface, we make the following assumptions. First, as demonstrated in the case of the pure elastic substrate (see Section~\ref{sec:elasticdistributedloads}), we can neglect the contribution of the shear stress from the slime to the deformation of the substrate (i.e $F_q\approx 0)$. Moreover, we assume the substrate to be incompressible ($\nu=1/2$). As a result, $\bar{u}_\text{s}\approx 0$ and we can restrict our analysis to the vertical deformation of the surface. In this case, the slope of the substrate surface reads
$$\left(|x-s|+ \frac{2(1-\nu^2)}{\pi E}2\gamma_\text{s}\right)\frac{\partial \bar{w}_\text{s}}{\partial x}  \approx \frac{2(1-\nu^2)}{\pi E} F_p-\frac{(1-2\nu)(1+\nu)}{E}|x-s|\delta(x-s)F_q. $$
Let $\ell_\text{s}=\dfrac{2(1-\nu^2)}{E}2\gamma_\text{s}=\dfrac{(1-\nu)}{G}2\gamma_\text{s},$ be the elasto-capillary length scale, then
$$\left(|x-s|+\ell_\text{s}/\pi\right)\frac{\partial \bar{w}_\text{s}}{\partial x}  \approx \frac{2(1-\nu^2)}{\pi E} F_p-\frac{(1-2\nu)(1+\nu)}{E}|x-s|\delta(x-s)F_q, $$
$$\implies \frac{\partial \bar{w}_\text{s}}{\partial x}  \approx \frac{2(1-\nu^2)}{\pi E} \frac{F_p}{\left(|x-s|+\ell_\text{s}/\pi\right)}+\frac{(1-2\nu)(1+\nu)}{E}\frac{|x-s|}{\left(|x-s|+\ell_s/\pi\right)}\delta(x-s)F_q. $$
Therefore,
\begin{equation}
\bar{w}_\text{s}(x)=\frac{2(1-\nu^2)}{\pi E}F_p\ln \frac{|x-s|+\ell_\text{s}/\pi}{x_1},
\label{eq:surfdisp3}%
\end{equation}
where $x_1$ is a constant. Having obtained the response due the point load $F_p$, we now proceed to compute the response due to the slime pressure $p(x)$, acting over a contact length $-\ell/2\leq x\leq \ell/2$. 

\addtocounter{subsection}{1}
\subsection{Solution for distributed loads}
The elasto-capillary response of the substrate under a distributed load can be found by superimposing the responses due to all the elementary force composing the load \cite{S-Landau84}. Recalling that $F_p=p(s)\mathrm{d}s$, we thus sum \eqref{eq:surfdisp3} in an integral sense and obtain

\begin{equation}
\bar{w}_\text{s}(x)=\frac{2(1-\nu^2)}{\pi E}\int_{-\ell/2}^{\ell/2} p(s)\ln \frac{|x-s|+\ell_\text{s}/\pi}{x_1}\mathrm{d}s.
\end{equation}
Furthermore, we consider the substrate at equilibrium at every instant, thereby substituting with no further complication the pressure in Eq.~(\ref{eq:dispfield2}) by the time-dependent traveling pressure $p(x,t)$. Last, we use the shear modulus definition $G=E/2(1+\nu)$ and find the substrate deformation as
\begin{equation}
\delta(x,t)=\frac{1-\nu}{\pi G}\int_{-\ell/2}^{\ell/2}p(x',t)\ln \left( \frac{|x-x'|+\ell_\text{s}/\pi}{x_1}\right) \mathrm{d}x'.
\label{eq:thickdeformWave2}
\end{equation}

Equation \eqref{eq:thickdeformWave2} is to be solved with the Reynolds equation~\eqref{eq:Reynolds} governing the pressure distribution in the slime. However, as stated earlier, the deformation of a very soft substrate at the leading edge of the bacteria can cause a non-zero curvature of the slime-air interface, as shown in Fig.~\ref{fig:setupec}. Due to the pressure difference across a curved interface (Laplace's law), the slime pressure at the leading edge to the left of the curved interface will be different from the atmospheric reference pressure, i.e. $p(\ell/2,t)\neq 0$. We now proceed to obtain the appropriate boundary condition at the leading edge of the bacteria.

\addtocounter{subsection}{1}
\subsection{Boundary condition at the leading edge}
In order to obtain an equation for $p(\ell/2,t)$, let us consider the pressure jump across the slime-air interface, $M(s)$, as indicated in Fig.~\ref{fig:setupec}. In the limit of small curvatures, the pressure jump is given by the Laplace equation
\begin{equation}\label{eq:Laplace}
-p(\ell/2)\approx \gamma\frac{d^2M}{ds^2},
\end{equation}
where $\gamma$ is the slime-air surface tension. We find the shape $M(s)$ by integrating this equation with the boundary conditions $M(s=0)=M(s=s_t)=0$, where $s_t=h(\ell/2)-\delta(\ell/2)$ is the position of the tip of the ridge, in the $(s,M)$-coordinate system (see Fig.~\ref{fig:setupec}). Therefore, 
\begin{equation}\label{eq:M_int}
M(s)=-\frac{p(\ell/2)}{2\gamma}s(s-s_t).
\end{equation}
Moroever, the contact angle $\theta$ at the slime-substrate-air ridge is related to the slope of the slime-air interface through
$$\theta=\frac{\pi}{2}-\underbrace{\tan^{-1}\left[\left. \frac{dM}{ds}\right|_{s_t}\right]}_{\theta_M},$$ 
where $\theta_M=\tan^{-1}\left[-\dfrac{p(\ell/2)}{2\gamma}s_t\right]$ using \eqref{eq:M_int}. Therefore, $\cos\theta=\sin\theta_M$ and $\sin\theta=\cos\theta_M.$ 

Next, we assume that the slime-air interface spreads at the top of the ridge over a small extent $2a$ \cite{S-deGennes85, S-Dervaux15, S-Lubbers14}. In so doing, we define $\Gamma$, the distributed surface tension over the length $2a$ such that
\begin{equation}
\gamma\bfv{e}_p=\Gamma\int_{\ell/2-a}^{\ell/2+a} \bfv{n}\mathrm{d}x.
\label{eq:slimeairgamma}%
\end{equation}
In Eq.~(\ref{eq:slimeairgamma}), $\bfv{e}_p=(-\cos\theta,\sin\theta)^T=(-\sin\theta_M,\cos\theta_M)^T$ is the orientation vector defining the contact angle of the slime on the substrate, while $\bfv{n}=\left(-\dfrac{\partial\delta}{\partial x},1\right)^T\left[\left(\dfrac{\partial\delta}{\partial x}\right)^2+1\right]^{-1/2}$ is the normal vector to the substrate. Therefore, the horizontal and vertical projections of Eq.~(\ref{eq:slimeairgamma}) read respectively
\begin{subequations}
\begin{align}
-\gamma\sin\theta_M \simeq \Gamma\int_{\ell/2-a}^{\ell/2+a}-\frac{\partial\delta}{\partial x}\mathrm{d}x &=-\Gamma(\delta(\ell/2+a)-\delta(\ell/2-a))\approx -\Gamma\left. \frac{\partial\delta}{\partial x}\right|_{\ell/2-a}2a,\\
\gamma\cos\theta_M & \simeq \Gamma\int_{\ell/2-a}^{\ell/2+a}\mathrm{d}x = 2a\Gamma.
\end{align}
\end{subequations}
Combining these two equations yields the relation
\begin{equation}
\tan\theta_M=\left.\frac{\partial\delta}{\partial x}\right|_{\ell/2-a}.
\label{eq:thetaM}%
\end{equation}
Following the work of Dervaux $\&$ Limat \cite{S-Dervaux15}, we can obtain the expression of the substrate deformation $\left. \dfrac{\partial\delta}{\partial x}\right|_{\ell/2-a}$ due to the distributed pressure $\Gamma$ to be
\begin{subequations}
\begin{align}
\left. \frac{\partial\delta}{\partial x}\right|_{\ell/2-a} & = \frac{2(1-\nu^2)}{\pi E}\int_{\ell/2-a}^{\ell/2+a} \frac{\Gamma}{|\ell/2-a-x'|+\ell_\text{s}/\pi} \mathrm{d}x'+ \frac{2(1-\nu^2)}{\pi E}\int_{-\ell/2}^{\ell/2} \frac{p(x')}{|\ell/2-a-x'|+\ell_\text{s}/\pi} \mathrm{d}x'  \\
 & \simeq  \frac{2(1-\nu^2)}{\pi E}\Gamma \ln\frac{2a+\ell_\text{s}/\pi}{\ell_\text{s}/\pi}+\frac{\ell_\text{s}}{\pi}\int_{-\ell/2}^{\ell/2} \frac{p(x')}{|\ell/2-a-x'|+\ell_\text{s}/\pi} \mathrm{d}x'  \\
 & \simeq \frac{2(1-\nu^2)\gamma\cos\theta_M}{\pi E 2a} \ln\left(1+\frac{2a}{\ell_\text{s}/\pi}\right)+\frac{\ell_\text{s}}{\pi}\int_{-\ell/2}^{\ell/2} \frac{p(x')}{|\ell/2-a-x'|+\ell_\text{s}/\pi} \mathrm{d}x'  \\
 & \simeq\frac{\gamma\cos\theta_M}{2\gamma_\text{s}}\frac{\ell_\text{s}/\pi}{2a}\ln\left(1+\frac{2a}{\ell_\text{s}/\pi}\right)+\frac{\ell_\text{s}}{\pi}\int_{-\ell/2}^{\ell/2} \frac{p(x')}{|\ell/2-a-x'|+\ell_\text{s}/\pi} \mathrm{d}x'. 
\end{align}
\end{subequations}
Therefore, Eq.~(\ref{eq:thetaM}) becomes
\begin{equation}
\dfrac{\tan\theta_M}{\cos\theta_M}=\frac{\gamma}{2\gamma_\text{s}}\frac{\ell_\text{s}/\pi}{2a}\ln\left(1+\frac{2a}{\ell_\text{s}/\pi}\right)+\frac{\ell_\text{s}/\pi}{\cos\theta_M}\int_{-\ell/2}^{\ell/2} \frac{p(x')}{|\ell/2-a-x'|+\ell_\text{s}/\pi} \mathrm{d}x'. 
\end{equation}
Using the definition $\theta_M=\tan^{-1}\left(-\dfrac{p(\ell/2)}{2\gamma}\left[h(\ell/2)-\delta(\ell/2)\right]\right)$, we obtain the following equation
$$-\dfrac{p(\ell/2)\left[h(\ell/2)-\delta(\ell/2)\right]}{2\gamma}\sqrt{1+\left(\dfrac{p(\ell/2)\left[h(\ell/2)-\delta(\ell/2)\right]}{2\gamma}\right)^2}= $$
$$ \frac{\gamma}{2\gamma_\text{s}}\frac{\ell_\text{s}/\pi}{2a}\ln\left(1+\frac{2a}{\ell_\text{s}/\pi}\right) +\sqrt{1+\left(\dfrac{p(\ell/2)\left[h(\ell/2)-\delta(\ell/2)\right]}{2\gamma}\right)^2}\underbrace{\frac{\ell_\text{s}}{\pi}\int_{-\ell/2}^{\ell/2} \frac{p(x')}{|\ell/2-a-x'|+\ell_\text{s}/\pi} \mathrm{d}x'}_{Z},$$
where we used the identity $1/\cos(\tan^{-1}(x))=\sqrt{1+x^2}$. The previous equation can also be recast as
\begin{center}
$$\left(\dfrac{p(\ell/2)}{2\gamma}\left[h(\ell/2)-\delta(\ell/2)\right]\right)^4+2Z\left(\dfrac{p(\ell/2)}{2\gamma}\left[h(\ell/2)-\delta(\ell/2)\right]\right)^3+(1+Z^2)\left(\dfrac{p(\ell/2)}{2\gamma}\left[h(\ell/2)-\delta(\ell/2)\right]\right)^2$$
$$ + 2Z\left(\dfrac{p(\ell/2)}{2\gamma}\left[h(\ell/2)-\delta(\ell/2)\right]\right)+Z^2-\left[\frac{\gamma}{2\gamma_\text{s}}\dfrac{\ell_\text{s}/\pi}{2a}\ln\left(1+\frac{2a}{\ell_\text{s}/\pi}\right)\right]^2=0.$$
\end{center}
Note that when $Z=0$, which is true in the limit of extremely soft substrates ($\ell_s\gg \ell$) and under the zero-lift constraint $\left(\int_{-\ell/2}^{\ell/2} p(x)\mathrm{d}x=0\right)$, the previous equation becomes
\begin{equation}
p(\ell/2)^4+p(\ell/2)^2\left(\dfrac{2\gamma}{h(\ell/2)-\delta(\ell/2)}\right)^2-\left[\frac{\gamma}{2\gamma_\text{s}}\dfrac{\ell_\text{s}/\pi}{2a}\ln\left(1+\frac{2a}{\ell_\text{s}/\pi}\right)\right]^2\left(\dfrac{2\gamma}{h(\ell/2)-\delta(\ell/2)}\right)^4=0.
\label{eq:poly4}
\end{equation}

Equation~(\ref{eq:poly4}) admits 4 solutions among which the only relevant one (i.e. real and negative) is
\begin{equation}
p(\ell/2)=-\dfrac{\gamma}{h(\ell/2)-\delta(\ell/2)}\left(2\sqrt{1+\left[\frac{\gamma}{\gamma_\text{s}}\dfrac{\ell_\text{s}/\pi}{2a}\ln\left(1+\frac{2a}{\ell_\text{s}/\pi}\right)\right]^2}-2\right)^{1/2}.
\label{eq:pres_capillary}
\end{equation}
Later in this paper, it will appear clear why we impose a negative pressure at the leading edge as a physical condition. We shall see that the gliding thrust on very soft substrates is given by a term $\propto -\dfrac{\partial p}{\partial x}h$. Since the pressure built up beneath the bacteria is zero almost everywhere on very soft substrates, it must be negative at the leading edge in order for the thrust to be in the positive direction (i.e, from left to right in our setup). Otherwise, the bacteria would glide in the opposite direction, turning the leading edge into the trailing one. This would imply that slime is not deposited in the wake of the cell but instead remain pinned, during gliding, between the bacteria and the substrate. Given that such scenario is in contradiction with experimental observations, we thus require the pressure at the leading edge to be negative.

This ends the derivation of the boundary condition at the leading of the elasto-capillary-hydrodynamics problem. We now proceed to recast the full problem in its dimensionless form. 

\addtocounter{subsection}{1}
\subsection{Non-dimensionalization}
There are 10 input variables for the problem 
$$A,C,\ell,L,h_0,a,\gamma,\gamma_\text{s},(1-\nu)/G,\mu,$$
which correspond to 3 dimensions: $[L]$, $[T]$, $[M]$. Therefore, we can define the following 7 dimensionless quantities
\begin{center}
$\epsilon=\dfrac{h_0}{L}$, $n=\dfrac{\ell}{L}$, $\hA=\dfrac{A}{h_0}$, $\ha=\dfrac{a\pi}{L}$,\\
$\xi=\dfrac{2\gamma_\text{s}(1-\nu)}{GL}$, $\eta=\dfrac{\mu(1-\nu) CL^2}{Gh_0^3}$, $Ca=\dfrac{\mu C}{\gamma}$.
\end{center}
As before, we used $L$ and $C$ as the reference length and velocity. Note that \textit{this does not mean that the gliding velocity scales as $C$}, but only that it is made dimensionless by normalizing with the wave speed. In addition to the dimensionless variables from Section~\ref{ND}, there are new dimensionless parameters such as $\ha$, the interface half-thickness and $\xi=\ell_\text{s}/L$, the elasto-capillary number.  

\bigskip
While the dimensionless Reynolds equation is still given by Eq.~(\ref{eq:DEa}), the substrate deformation with the capillarity correction is non-dimensionalized as follows:
$$\frac{\delta(x/L,t/(L/C))}{\Delta} =\frac{1}{\Delta}\frac{(1-\nu)PL}{\pi G}\int_{-\ell/(2L)}^{\ell/(2L)}\hp(\hx',\hti)\ln \left[\frac{L}{x_1}\left(\frac{|x-x'|}{L}+\frac{\ell_\text{s}}{L \pi}\right)\right] \mathrm{d}\hx'.$$
By applying the definition of $\Delta=(1-\nu)PL/G=(1-\nu)\mu CL^2/(Gh_0^2)$ and the zero-lift condition $\int_{-n/2}^{n/2}\hp(\hx) \mathrm{d}\hx=0$, we obtain the dimensionless substrate deformation as
$$\hdelta(\hX,\hti) =\frac{1}{\pi}\int_{-n/2}^{n/2}\hp(\hx',\hti)\ln \left(|\hx-\hx'|+\frac{\xi}{\pi}\right) \mathrm{d}\hx'.$$
Next, the non-dimensionalization of the boundary condition for the pressure at the leading edge, given by \eqref{eq:pres_capillary}, is obtained as follows:
\begin{subequations}
\begin{align}
\frac{p(\ell/2)}{\mu CL/h_0^2} &=-\frac{1}{L/h_0^2}\frac{(\gamma/\mu C)}{h(\ell/2)-\delta(\ell/2)}\left(2\sqrt{1+\left[\frac{\gamma}{\gamma_\text{s}}\dfrac{\ell_\text{s}/\pi}{2a}\ln\left(1+\frac{2a}{\ell_\text{s}/\pi}\right)\right]^2}-2\right)^{1/2}\nonumber \\
\implies \hp(n/2) & = -\frac{\epsilon/Ca}{\hat{h}(n/2)-\eta\hdelta(n/2)}\left(2\sqrt{1+\left[\mcal{R}\dfrac{\ell_\text{s}/L}{2a\pi/L}\ln\left(1+\frac{2a\pi/L}{\ell_\text{s}/L}\right)\right]^2}-2\right)^{1/2} \nonumber \\
\implies \hp(n/2) & = -\frac{\epsilon/Ca}{\hat{h}(n/2)-\eta\hdelta(n/2)}\left(2\sqrt{1+\left[\mcal{R}\dfrac{\xi}{2\ha}\ln\left(1+\frac{2\ha}{\xi}\right)\right]^2}-2\right)^{1/2},
 \end{align}
\end{subequations}
where $\mcal{R}=\gamma/\gamma_\text{s}$ is the ratio of slime-air to slime-substrate surface tensions. Therefore, the leading edge pressure reads
\begin{equation}
\hp(n/2) = -\frac{\epsilon/Ca}{\hat{h}(n/2)-\eta\hdelta(n/2)}\left(2\sqrt{1+\left[\mcal{R}\dfrac{\xi}{2\ha}\ln\left(1+\frac{2\ha}{\xi}\right)\right]^2}-2\right)^{1/2}.
\label{eq:pres_capillaryND}
\end{equation}

With the aforementioned simplifications and non-dimensionalization, the dimensionless system of equations for the full elasto-capillary-hydrodynamic problem now reads
\begin{subequations}
\begin{align}
Q_1 & = \frac{\partial\hp}{\partial\hX}(\hh-\eta \hdelta)^3+6(\hV-2)(\hh-\eta \hdelta)-\hm(\hti) & =0, \label{eq:coupledPB3a} \\
Q_2 & =\hdelta(\hX,\hti)- \frac{1}{\pi}\int_{-n/2}^{n/2}\hp(\hx',\hti)\ln \left(|\hx-\hx'|+\frac{\xi}{\pi}\right) \mathrm{d}\hx' &=0, \label{eq:substratePB3} \\
Q_3 & =\int_{-n/2}^{n/2} \hp(\hX,\hti) \mathrm{d}\hX & = 0, \label{eq:coupledPB3c} \\
Q_4 & =\int_{-n/2}^{n/2} \left(\hp\hat{b}'+\frac{\hp'}{2}\left(\hh-\eta\hdelta\right)+\frac{\hat{V}}{\hh-\eta\hdelta}\right) \mathrm{d}\hX & = 0, \label{eq:coupledPB3d} \\
Q_5 & =\hp(n/2) +\frac{\epsilon/Ca}{\hat{h}(n/2)-\eta\hdelta(n/2)}\left(2\sqrt{1+\left[\mcal{R}\dfrac{\xi}{2\ha}\ln\left(1+\frac{2\ha}{\xi}\right)\right]^2}-2\right)^{1/2}& = 0.\label{eq:coupledPB3e}%
\end{align}
\label{eq:coupledPB3}%
\end{subequations}

The elasto-capillary number $\xi$ compares compares capillary stresses at the slime-substrate interface to the elastic stresses in the bulk of the substrate. Depending on its value, the problem may be in regimes dominated by elasticity or by capillary effects. For very soft substrates in particular, we will show in the next subsection that it is possible to obtain a semi-analytical solution for the gliding speed. 

\newpage
\addtocounter{subsection}{1}
\subsection{Asymptotic regimes}
\subsubsection{Elasticity-dominated regime}
In the limit where elastic forces dominate over the capillary ones, 
$$\xi=\dfrac{2\gamma_\text{s}(1-\nu)}{GL}\rightarrow 0.$$
In this case, substituting $\xi=0$ into \eqref{eq:coupledPB3}, we obtain the system of \eqref{eq:recoveredEHD} given below. 
\begin{subequations}
\begin{align}
 \frac{\partial\hp}{\partial\hX}(\hh-\eta \hdelta)^3+6(\hV-2)(\hh-\eta \hdelta)-\hm(\hti) & =0, \\
\hdelta(\hX,\hti)- \frac{1}{\pi}\int_{-n/2}^{n/2}\hp(\hx',\hti)\ln \left(|\hx-\hx'|\right) \mathrm{d}\hx' &=0, \\
\int_{-n/2}^{n/2} \hp(\hX,\hti) \mathrm{d}\hX & = 0,\\
\int_{-n/2}^{n/2} \left(\hp\hat{b}'+\frac{\hp'}{2}\left(\hh-\eta\hdelta\right)+\frac{\hat{V}}{\hh-\eta\hdelta}\right) \mathrm{d}\hX & = 0, \\
\hp(n/2) & = 0.
\end{align}
\label{eq:recoveredEHD}
\end{subequations}
These equations show that the leading edge pressure, given by \eqref{eq:coupledPB3e}, as well as the capillary correction in the substrate deformation, given by \eqref{eq:substratePB3}, both converge to zero. Therefore, in the elasticity-dominated regime, given by $\xi\rightarrow 0$, the previously derived elasto-hydrodynamic equations (see \eqref{eq:coupledPB2}) are fully recovered.

\subsubsection{Capillarity-dominated regime}
In the opposite limit of very soft substrates, capillary effects at the slime-substrate interface dominate over the weak elastic ones in the substrate bulk. In this limit, 
$$\xi=\dfrac{2\gamma_\text{s}(1-\nu)}{GL}\rightarrow\infty.$$ 
Therefore, we can expand Eq.~(\ref{eq:substratePB3}) using Taylor series, and make use of the zero-lift condition $\int_{-n/2}^{n/2}\hp(\hx,\hti)\mathrm{d}\hx=0$ to obtain
\begin{subequations}
\begin{align}
\hdelta(\hX,\hti) & = \frac{1}{\pi}\int_{-n/2}^{n/2}\hp(\hx',\hti)\ln \left(|\hx-\hx'|+\frac{\xi}{\pi}\right) \mathrm{d}\hx', \\
			& = \frac{1}{\pi}\int_{-n/2}^{n/2}\hp(\hx',\hti)\ln \left[\frac{\xi}{\pi}\left(\frac{\pi}{\xi}|\hx-\hx'|+1\right)\right] \mathrm{d}\hx', \\
			& =  \frac{1}{\xi}\int_{-n/2}^{n/2}\hp(\hx',\hti)|\hx-\hx'|\mathrm{d}\hx' +O\left(\frac{1}{\xi^2}\right).
\end{align}
\end{subequations}

Moreover, note that the elasto-capillary and softness numbers are related through
$$\xi=\frac{(1-\nu)2\gamma_\text{s}}{GL}=\frac{(1-\nu)\mu CL^2}{Gh_0^3}\frac{2\gamma_\text{s}}{\gamma}\frac{\gamma}{\mu C}\frac{h_0^3}{L^3}=\frac{2\epsilon^3}{\mcal{R}Ca}\eta, $$
As a result, although a very soft substrate is characterized by $\eta\rightarrow\infty$, its normalized deformation ($\eta\hdelta=\delta/h_0$) remains finite and reads
$$\eta\hdelta(\hX,\hti)\simeq \frac{\mcal{R}Ca}{2\epsilon^3}\int_{-n/2}^{n/2}\hp(\hx',\hti)|\hx-\hx'|\mathrm{d}\hx'=\hdelta_{\infty}(\hX,\hti).$$
In this capillary-dominated regime, we can also obtain a simplified expression for the pressure through a Taylor expansion of Eq.~(\ref{eq:pres_capillaryND}) as 
$$\hp(n/2)\simeq  -\frac{\epsilon}{Ca}\frac{\chi}{\hat{h}(n/2)-\eta\hdelta(n/2)},$$
where $\chi=\left(2\sqrt{1+\mcal{R}^2}-2\right)^{1/2}$. Therefore, the pressure at the leading edge scales as $Ca^{-1}$. In other words, on very soft substrates where the capillary number is small, the bacteria motion will be mainly driven by the slime-air surface tension. 

Furthermore, by assuming the pressure field to vary over the length scale of the capillary ridge $l_c\ll L$ (see Fig.~\ref{fig:setupec} for the definition of $l_c$), the drag-free condition~\eqref{eq:coupledPB3d} can be approximated as
\begin{equation}
\int_{-n/2}^{n/2} \left(\frac{1}{2}\frac{\partial\hp}{\partial\hx}\left(\hh-\eta\hdelta\right)+\frac{\hat{V}}{\hh-\eta\hdelta}\right) \mathrm{d}\hX \simeq 0.
\label{eq:dragfreesimp}
\end{equation}
This relation can be used to rewrite Reynolds equation~\eqref{eq:coupledPB3a} in the following integral form 
\begin{subequations}
\begin{align}
 \frac{\partial\hp}{\partial\hX}(\hh-\eta \hdelta)^3 & +6(\hV-2)\underbrace{(\hh-\eta \hdelta)}_{(\hh-\hdelta_{\infty})}-\hm  =0, \\
\implies \frac{\partial\hp}{\partial\hX}(\hh-\hdelta_{\infty}) & +6\frac{(\hV-2)}{(\hh-\hdelta_{\infty})}-\frac{\hm}{(\hh-\hdelta_{\infty})^2}  =0, \\
\implies  \frac{\partial\hp}{\partial\hX}(\hh-\hdelta_{\infty}) & = -6\frac{(\hV-2)}{(\hh-\hdelta_{\infty})}+\frac{\hm}{(\hh-\hdelta_{\infty})^2} \label{eq:phmd}.
\end{align}
\end{subequations}
Substituting this expression of \eqref{eq:phmd} into \eqref{eq:dragfreesimp} and multiplying both sides of the resulting equation by -2, we obtain:
\begin{equation}
\int_{-n/2}^{n/2}\left(\frac{4\hV}{(\hh-\hdelta_{\infty})}-\frac{12}{(\hh-\hdelta_{\infty})}-\frac{\hm}{(\hh-\hdelta_{\infty})^2}\right)\mathrm{d}\hX \simeq 0.
\label{eq:dragfreesimp2}
\end{equation}
Defining $\zeta_j$ as the integral,
$$\zeta_j=\int_{-n/2}^{n/2}\frac{1}{(\hh-\hdelta_{\infty})^j}\mathrm{d}\hX,$$
the gliding speed can be obtained from \eqref{eq:dragfreesimp2} as
\begin{equation}
\hV=3+\frac{\hm \zeta_2}{4\zeta_1}.
\label{eq:gspeed1}
\end{equation}
In order to eliminate $\hm$, we can also use \eqref{eq:phmd} to write
$$\underbrace{\int_{-n/2}^{n/2}\frac{\partial\hp}{\partial\hx}\mathrm{d}\hx}_{\left[\hp(n/2)-\hp(-n/2)\right]=\Delta \hp}=\int_{-n/2}^{n/2}-\frac{6(\hV-2)}{(\hh-\hdelta_{\infty})^2}+\frac{\hm}{(\hh-\hdelta_{\infty})^3}\mathrm{d}\hX,$$
which leads to
\begin{equation}
\hV=2+\frac{\hm \zeta_3-\Delta \hp}{6\zeta_2}.
\label{eq:gspeed2}
\end{equation}
Using Eq.~(\ref{eq:gspeed1}) and Eq.~(\ref{eq:gspeed2}), we obtain
\begin{equation}
\hm=\frac{6\zeta_2+\Delta \hp}{\zeta_3-\frac{3\zeta_2^2}{2\zeta_1}}.
\end{equation}
Therefore, in the limit of extremely soft substrates, the gliding speed $\hV$ takes the asymptotic expression
\begin{equation}
\hV_{\infty}\approx 2-\Delta\hp\frac{(1-\alpha)}{\beta}+\alpha,
\label{eq:gspeed}
\end{equation}
where 
\begin{center}
$\alpha=\left(1-\dfrac{3\zeta_2^2}{2\zeta_1\zeta_3}\right)^{-1}$, and  $\beta=6\zeta_2$.
\end{center}

In order to verify the prediction of \eqref{eq:gspeed} and to solve the problem in non-asymptotic regimes, where elastic and capillary effects are comparable, we numerically solve the governing equations~\eqref{eq:coupledPB3}.
However, as we will show in the next section, the governing equations of the elasto-capillary-hydrodynamics problem give rise to various scaling laws. These scaling relationships are important to the numerical resolution of \eqref{eq:coupledPB3} because they guide the discretization of the computational domain. Such inference from scaling laws is similar to that of boundary layer (BL) problems in fluid mechanics, where the scaling of the BL thickness provides insight into the level of mesh refinement near boundaries, if one is to capture the BL physics properly. 

\section{Scaling analysis}
In this section, we derive some scaling laws existing in the full elasto-capillary-hydrodynamics problem. Starting from the equations in their dimensional form, we will show how, in the capillarity-dominated regime, one can obtain scaling relationships between the dimensions of the capillary ridge and the bacteria speed as a function of the dimensionless parameters of the problem.

\bigskip
Let $v_c$, $\Delta_r$ and $l_c$ be the scales of gliding speed, the ridge height and its extent, respectively. For each of these unknowns, we can find a relation in the limit of very soft substrates, where $G\rightarrow0$ and the elasto-capillary length is so large that $\ell_\text{s}\gg l_c$. 

The first relation comes from balancing the capillary-induced depression (thrust) at the leading edge with the lubrication viscous stress (friction) at the scale of the entire body, \emph{i.e.},
$${p(\ell/2)\sim \frac{\mu v_c \ell}{h_0^2}.}$$
However, recall that in the limit of large $\ell_s\gg a$, the leading edge pressure, given by Eq.~(\ref{eq:pres_capillary}), simplifies into
$$p(\ell/2)\simeq -\frac{\gamma \chi}{h(\ell/2)-\delta(\ell/2)},$$
where $\chi=\left(2\sqrt{1+(\gamma/\gamma_s)^2}-2\right)^{1/2}$. Therefore, the aforementioned balance reads
\begin{equation}
\frac{\gamma\chi}{h_0} \sim \frac{\mu v_c \ell}{h_0^2}.
\label{eq:R1}
\end{equation}
This balance is confirmed by Fig.~7 of the main text, which shows that as the bacterial length increases, the gliding speed $v_c$ decreases due to an increased friction. Therefore,
$$v_c\sim \frac{\gamma}{\mu C}\frac{Ch_0\chi}{\ell}.$$

The second relation comes from requiring that mass balance be verified. As implied by the Reynolds equation~\eqref{eq:Reynolds}, the slime flow induced by the vertical motion of the substrate at the speed C (since we look for solutions $\frac{\partial\delta}{\partial t}=C\frac{\partial\delta}{\partial x}$) must be converted in the horizontal gliding motion of the cell. That is, the gliding velocity and the traveling wave verify
$$O(Vh) \sim O(C\delta),$$
$$v_ch_0 \sim C\Delta_r.$$
Therefore,
\begin{equation}
\Delta_r\sim \frac{\gamma}{\mu C}\frac{h_0^2\chi}{\ell}.
\label{eq:R2}
\end{equation}

Last, the scale of the ridge extent $l_c$ is obtained by searching for the length over which the slime-substrate tension balances the lubrication pressure in order to maintain an equilibrium shape to the ridge. We thus have
\begin{equation}
\gamma_s\frac{\Delta_r}{l_c^2} \sim \frac{\mu v_c l_c}{h_0^2}.
\label{eq:R3}
\end{equation}
Combining Eq.~(\ref{eq:R1}), (\ref{eq:R2}) and (\ref{eq:R3}), we thus obtain
$$l_c^3\sim \frac{\Delta_r h_0^2\gamma_s}{\mu v_c}\sim \frac{\gamma_s}{\mu C}\underbrace{\frac{\Delta_r C}{h_0 v_c}}_{1}h_0^3.$$
Hence,
$$l_c \sim h_0 \left( \dfrac{\gamma_s}{\mu C} \right)^{1/3} \sim h_0\mcal{R}^{-1/3}\left( \dfrac{\gamma}{\mu C} \right)^{1/3},$$
where $\mcal{R}=\gamma/\gamma_s$ is the ratio of the interfacial tensions. 

In summary, we obtain the following scaling laws
\begin{subequations}
\begin{align}
v_c  &\sim  C\dfrac{\epsilon \chi}{n}Ca^{-1}, \label{eq:SpeedScaling} \\
\Delta_r & \sim  h_0\dfrac{\epsilon\chi}{n}Ca^{-1},\\
l_c & \sim  h_0\mcal{R}^{-1/3}Ca^{-1/3}.\label{eq:ridge_extent}
\end{align}
\label{eq:scaling}%
\end{subequations}
In Eq.~(\ref{eq:scaling}), $\epsilon=h_0/L$, $n=\ell/L$ and $Ca$ is the previously defined capillary number based on the wave speed.
Since $\chi=\left(2\sqrt{1+\mcal{R}^2}-2\right)^{1/2}$, the ridge height $\Delta_r$ thus increases with $\mcal{R}$. Therefore, among the physically admissible values of the surface tension ratio, there exists a maximum $\mcal{R}_{max}$ at which the substrate ridge is so high that its closes the lubrication gap at the leading edge.

The scaling laws given by \eqref{eq:scaling} have been verified numerically and shown to be valid for large values of the softness parameters (see Fig.~6 of the main paper). In the next section, we describe the numerical method used to solve the full elasto-capillary-hydrodynamics problem, not only in the capillary regime where the scaling laws hold, but also for mildly soft substrates where elastic and capillary effects are comparable.

\section{Numerical resolution of the elasto-hydro-capillary problem}
The system of equations (\ref{eq:coupledPB3}) is solved iteratively by Newton's method similar to the previous elasto-hydrodynamics problem. We use the fact that
\begin{equation}
\xi=\eta\frac{2\epsilon^3}{\mcal{R}Ca},
\label{eq:xieta}
\end{equation}
and rewrite the system as a function of $\eta$ only. At each iteration, we solve the system of linearized equations~(\ref{eq:LinearizedPB3}) using the finite element method. 

\addtocounter{subsection}{1}
\subsection{Newton's method} 
Let the unknown field $\bfv{w}$ be
$$\bfv{w}(\hx,\hti)=\left[\hp(\hx,\hti),\hdelta(\hx,\hti),\hm(\hti),\hV(\hti),\hp(n/2,\hti)\right]^T.$$
For different values of $\eta$, we solve the system (\ref{eq:coupledPB3}) by starting with a guess solutions $\bfv{w}_k$ and at the iteration $k$ and searching for the solution of the problem at the next iteration $\bfv{w}_{k+1}=\bfv{w}_k+\tilde{\bfv{w}}_k$. In order to obtain the corrections $\tilde{\bfv{w}}_k$, the system of equations~(\ref{eq:coupledPB3}) is linearized and inverted. In its linearized form, it reads\\

\begin{subequations}
\begin{align}
 \frac{\partial\tilde{p}_k}{\partial\hX}(\hh-\eta\hdelta_k)^3-3\eta\frac{\partial\hp_k}{\partial\hX}(\hh-\eta\hdelta_k)^2\tilde{\delta}_k-6(\hV_k-2)\eta\tilde{\delta}_k+6\tilde{V}_k(\hh-\eta\hdelta_k)-\tilde{m}_k & = -Q_1(\bfv{w}_k), \\
\tilde{\delta}_k  - \frac{1}{\pi}\int_{-n/2}^{n/2}\tilde{p}_k(\hX',\hti) \ln\left(|\hx-\hx'|+\frac{2\eta\epsilon^3}{\pi\mcal{R}Ca}\right)  \mathrm{d}\hX' &= -Q_2(\bfv{w}_k), \\
\int_{-n/2}^{n/2}\tilde{p}_k \mathrm{d}\hX &= -Q_3(\bfv{w}_k),\\
\int_{-n/2}^{n/2} \left(\tilde{p}_k\hat{b}'+\frac{1}{2}\frac{\partial\tilde{p}'_k}{\partial\hX}\left(\hh-\eta\hdelta_k\right)-\frac{1}{2}\frac{\partial\hp_k}{\partial\hX}\eta\tilde{\delta}_k+\frac{\tilde{V}_k}{\hh-\eta\hdelta_k}+\frac{\eta\tilde{\delta}_k\hV_k}{\left(\hh-\eta\hdelta_k\right)^2} \right)\mathrm{d}\hX & = -Q_4(\bfv{w}_k),\\
\tilde{p}(n/2)_k +\frac{\eta\tilde{\delta}_k(n/2)}{\left[\hat{h}(n/2)-\eta\hdelta_k(n/2)\right]^2} \\ \nonumber
 \times \frac{\epsilon}{Ca}\left(2\sqrt{1+\mcal{R}^2\left[\dfrac{\eta\epsilon^3}{\ha\mcal{R} Ca}\ln\left(1+\frac{\ha\mcal{R} Ca}{\eta\epsilon^3}\right)\right]^2}-2\right)^{1/2} & = - Q_5(\bfv{w}_k)
\end{align}
\label{eq:LinearizedPB3}%
\end{subequations}
Upon finding $\tilde{\bfv{w}}_k$ from the inversion of \eqref{eq:LinearizedPB3}, we obtain $\bfv{w}_{k+1}$ and proceed to solve for $\tilde{\bfv{w}}_{k+1}$. These iterations are repeated until the correction verifies $\| \tilde{\bfv{w}}_k\|_{L_2} <=tol.$ Here, we set the tolerance to $tol=1\times10^{-12}$, which is satisfied after 5 or 6 iterations, depending on the value of the parameter $\eta$ and the initial solution.

\addtocounter{subsection}{1}
\subsection{Finite element Implementation}
Proceeding as before, we choose $q(\hx)$ and $\kappa(\hx)$ as test functions for the pressure and deformation fields, respectively. The weak form of the linearized system (\ref{eq:LinearizedPB3}) then reads

\newpage
\begin{subequations}
\begin{align}
\left< q, \frac{\partial\tilde{p}_k}{\partial\hX}(\hh-\eta\hdelta_k)^3-3\eta\frac{\partial\hp_k}{\partial\hX}(\hh-\eta\hdelta_k)^2\tilde{\delta}_k -6(\hV_k-2)\eta\tilde{\delta}_k+6\tilde{V}_k(\hh-\eta\hdelta_k)-\tilde{m}_k\right>_{\Omega} \\
+\mathcal{N}\left<q,(\tilde{p}_k-\tilde{p}(n/2)_k)\right>_{\partial\Omega_1}+\mathcal{N}\left<q,(\hp_k-\hp(n/2)_{k})\right>_{\partial\Omega_1}+ \left< q, Q_1(\bfv{w}_k) \right>_{\Omega} \\
+ \left< \kappa, \tilde{\delta}_k \right> _{\Omega}- \left< \kappa, \frac{1}{\pi}\int_{-n/2}^{n/2}\tilde{p}_k(\hX',\hti) \ln\left(|\hx-\hx'|+\frac{2\eta\epsilon^3}{\pi\mcal{R}Ca}\right)  \mathrm{d}\hX' \right>_{\Omega} + \left<\kappa, Q_2(\bfv{w}_k)\right>_{\Omega} \\
- \left<1, \tilde{p}_k\right>_{\Omega} + \left<1, \hp_k \right>_{\Omega} \nonumber \\
 - \left<1, \tilde{p}_k\hat{b}'+\frac{1}{2}\frac{\partial\tilde{p}_k}{\partial\hX}\left(\hh-\eta\hdelta_k\right)-\frac{1}{2}\frac{\partial\hp_k}{\partial\hX}\eta\tilde{\delta}_k+\frac{\tilde{V}_k}{\hh-\eta\hdelta_k}+\frac{\eta\tilde{\delta}_k\hV_k}{\left(\hh-\eta\hdelta_k\right)^2}\right>_{\Omega} \\
+ \left<1,\hp_k\hat{b}'+\frac{1}{2}\frac{\partial\hp_k}{\partial\hX}\left(\hh-\eta\hdelta_k\right)+\frac{\hat{V}_k}{\hh-\eta\hdelta} \right>_{\Omega} \\
+\left<1,\tilde{p}(n/2)_k\right>_{\Omega} \\
 +\left<1,\frac{\eta\tilde{\hdelta}_k(n/2)}{\left[\hat{h}(n/2)-\eta\hdelta_k(n/2)\right]^2}\frac{\epsilon}{Ca}\left(2\sqrt{1+\mcal{R}^2\left[\dfrac{\eta\epsilon^3}{\ha\mcal{R} Ca}\ln\left(1+\frac{\ha\mcal{R} Ca}{\eta\epsilon^3}\right)\right]^2}-2\right)^{1/2}\right>_{\Omega} \\
 +\left<1,\hp(n/2)_{k}\right>_{\Omega} \\
 +\left<1, +\frac{1}{\hat{h}(n/2)-\eta\hdelta_k(n/2)}\frac{\epsilon}{Ca}\left(2\sqrt{1+\mcal{R}^2\left[\dfrac{\eta\epsilon^3}{\ha\mcal{R} Ca}\ln\left(1+\frac{\ha\mcal{R} Ca}{\eta\epsilon^3}\right)\right]^2}-2\right)^{1/2}\right>_{\Omega}=0.
\end{align}
\label{eq:weakform}%
\end{subequations}
Again, we expand the pressure and deformation fields in the basis of shape functions (chosen to be the same as the test functions) and write
$$ \tilde{p}_k(\hX,\hti)=\sum_{j=1}^N \tilde{p}_{k_j}(\hti) q_j(\hX)  \hspace{0.25cm}, \hspace{0.25cm} \tilde{\delta}_k(\hX,\hti)=\sum_{j=1}^N \tilde{\delta}_{k_j}(\hti) \kappa_j(\hX).$$
Inserting this decomposition into the weak form, the problem to solve can be recast as:
\begin{equation}
\begin{pmatrix}
A_k & B_k & L_k & D_{1k} & J_{1k} \\
C_k & E_k & 0 &  0 & 0 \\
-L_k^T & 0 & 0&  0 & 0 \\
D_{k1} & D_{k2} & 0 &  D_{k} & 0\\
J_{k1} & 0 & 0 &  0 & J_{k}
\end{pmatrix} 
\begin{pmatrix}
\tilde{p}_{n_j} \\
 \tilde{\delta}_{n_j} \\
  \tilde{m}_k \\
  \tilde{V}_k \\
\tilde{p}(n/2)_k
\end{pmatrix}
=
-\begin{pmatrix}
\left< q, Q_1(\bfv{w}_k) \right>_{\Omega} -\mathcal{N}\left<q,(\hp_k-\hp(n/2))\right>_{\partial\Omega_1}\\
\left<\kappa, Q_2(\bfv{w}_k) \right>_{\Omega} \\
\left<1, \hp_k\right>_{\Omega}  \\
\left<1, q_4(\bfv{w}_k)\right>_{\Omega} \\
\left<1, q_5(\bfv{w}_k)\right>_{\Omega} \\
\end{pmatrix},
\label{eq:matrixform2}
\end{equation}
where the terms in Eq.~(\ref{eq:matrixform2}) are given by
$$A_k= \left< q, \frac{dq_j}{\partial\hX}(\hh-\eta\hdelta_k)^3\right>_{\Omega}+\mathcal{N}\left<q,q_j\right>_{\partial\Omega_1},$$
$$B_k= -\left< q, \left(3\eta\frac{\partial\hp_k}{dx}(\hh-\eta\hdelta_k)^2+6(\hV_k-2)\eta\right)\kappa_j\right>_{\Omega},$$
$$C_k=- \left< \kappa, \underbrace{\frac{1}{\pi}\int_{-n/2}^{n/2}q_j (\hX') \ln\left(|\hx-\hx'|+\frac{\eta\epsilon^3}{\pi\mcal{R}Ca}\right)  \mathrm{d}\hX' }_{\phi_j}\right>_{\Omega},$$
$$E_k=\left< \kappa, \kappa_j \right>_{\Omega},$$
$$L_k=-\left< q, 1 \right>_{\Omega},$$
$$D_{1k}=\left< q, 6\tilde{V}_k\left(\hh-\eta\hdelta_k\right)\right>_{\Omega},$$
$$D_{k1}=- \left<1,q_j\hat{b}'+\frac{1}{2}\frac{\partial q_j}{\partial\hX}\left(\hh-\eta\hdelta_k\right)\right>_{\Omega},$$
$$D_{k2}=- \left<1,-\frac{1}{2}\frac{\partial\hp_k}{\partial\hX}\eta\kappa_j+\frac{\eta\hV_k\kappa_j}{\left(\hh-\eta\hdelta_k\right)^2}\right>_{\Omega},$$
$$D_{k}=- \left<1,\frac{1}{\hh-\eta\hdelta_k}\right>_{\Omega},$$
$$J_{1k}=-\mathcal{N}\left<q,1\right>_{\partial\Omega_1},$$
$$J_{k1}=-\mathcal{N}\left<1,+\frac{\eta\phi_j(n/2)}{\left[\hat{h}(n/2)-\eta\hdelta_k(n/2)\right]^2}\frac{\epsilon}{Ca}\left(2\sqrt{1+\mcal{R}^2\left[\dfrac{\eta\epsilon^3}{\ha\mcal{R} Ca}\ln\left(1+\frac{\ha\mcal{R} Ca}{\eta\epsilon^3}\right)\right]^2}-2\right)^{1/2}\right>_{\partial\Omega_1},$$
$$J_k=1,$$
$$q_4=\hp_k\hat{b}'+\frac{1}{2}\frac{\partial\hp_k}{\partial\hX}\left(\hh-\eta\hdelta_k\right)+\frac{\hat{V}_k}{\hh-\eta\hdelta},$$
$$q_5=\hp(n/2)_{k} +\frac{1}{\hat{b}(n/2)-\eta\hdelta_k(n/2)}\frac{\epsilon}{Ca}\left(2\sqrt{1+\mcal{R}^2\left[\dfrac{\eta\epsilon^3}{\ha\mcal{R} Ca}\ln\left(1+\frac{\ha\mcal{R} Ca}{\eta\epsilon^3}\right)\right]^2}-2\right)^{1/2}.$$
These equations are solved on a computational domain using the finite element open solver FreeFem++ \cite{S-Hecht12}. 

\addtocounter{subsection}{1}
\subsection{Mesh convergence}
The scaling laws given by \eqref{eq:scaling} show that an accurate computation of the substrate ridge and the gliding speed requires to refine the mesh in the vicinity of the leading edge. Given that the extent of the ridge scales as $l_c\sim \mcal{R}^{-1/3}Ca^{-1/3}$, then for large values of $Ca$ and $\mcal{R}$, we can expect the ridge to be confined to a ``boundary layer'' (BL) near the leading edge of the bacteria. In order to accurately capture the sharp contrast between the deformation from the substrate ridge (BL) and the rest of the quasi-undeformed part of the substrate (bulk), we need to refine the mesh near the leading edge, in the computation domain. A typical case of this mesh refinement is shown in Fig.~\ref{fig:meshCa}. 
\begin{figure}[hb]
\centering
 \begin{tikzpicture}
  \node (img1)  {\includegraphics[scale=0.65]{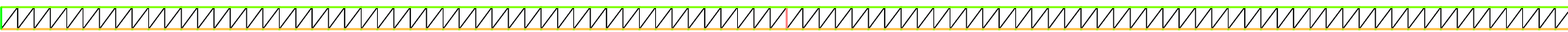}};
  \node[below=of img1, node distance=0cm, yshift=1cm] {$x$};
\end{tikzpicture}
   \caption{{Typical mesh for the computation of the elasto-capillary-hydrodynamics problem. The mesh is finer in the boundary layer prior to the leading edge (right of the rectangular domain). }}   
   \label{fig:meshCa}
 \end{figure} 
 
Strictly speaking, the scaling laws~(\ref{eq:scaling}) are verified in the limit of very soft substrates, where capillary effects dominate. Therefore, in order to choose a spatial discretization that can be used for all the range of softness numbers, say $[\eta_{min}, \eta_{max}]$, we must carry out the mesh convergence test at the largest value $\eta_{max}$. In the following, we set $[\eta_{min}, \eta_{max}]=[10^{-3},5\times10^3]$ and compute the error on the gliding speed $\varepsilon_{\hV}$ and on the lubrication gap at the leading edge $\varepsilon_{gap}$. These quantities are defined by
$$\varepsilon_{\hV}=\dfrac{\bl \hV\br-\bl \hV\br_{ref}}{\bl \hV\br_{ref}} \hspace{0.3cm} \mathrm{,} \hspace{0.3cm} \varepsilon_{gap}=\dfrac{\bl\hat{s}_t\br-\bl\hat{s}_t\br_{ref}}{\bl\hat{s}_t\br_{ref}},$$
where $\bl\cdot\br=\int_0^1(\cdot)\mathrm{d}\hti$ is the time-average operator and $\bl\hat{s}_t\br=1-\eta\bl\hdelta(n/2)\br$. Here, $\bl \hV\br_{ref}$ and $\bl \hat{s}_t\br_{ref}$ are respectively the reference time-averaged gliding speed and gap, and correspond to our most refined mesh.

Fig.~\ref{fig:meshconvCa} shows the convergence of the results with $N_{BL}$, the number of nodes in the refined part near the right end of the computational domain. This region covers a $1/5$-th of the domain length. Furthermore, in the legend of Fig.~\ref{fig:meshconvCa}, $N_B$ is the number of nodes of the remaining $4/5$-th of the computational domain. It can easily be deduced from Fig.~\ref{fig:meshconvCa} that although the precision of the results depends on both $N_B$ and $N_{BL}$, they converge fast with the former and more slowly with the latter. Note that here, the aforementioned reference gliding speed and lubrication gap are given by their values in the case of the most refined mesh, i.e. with $N_B\approx 39$ and $N_{BL}=270$.
\begin{figure}[t!]
\centering
 \begin{tikzpicture}
 \node[xshift=0cm, yshift=0.0cm] (img1)  {\includegraphics[scale=0.7]{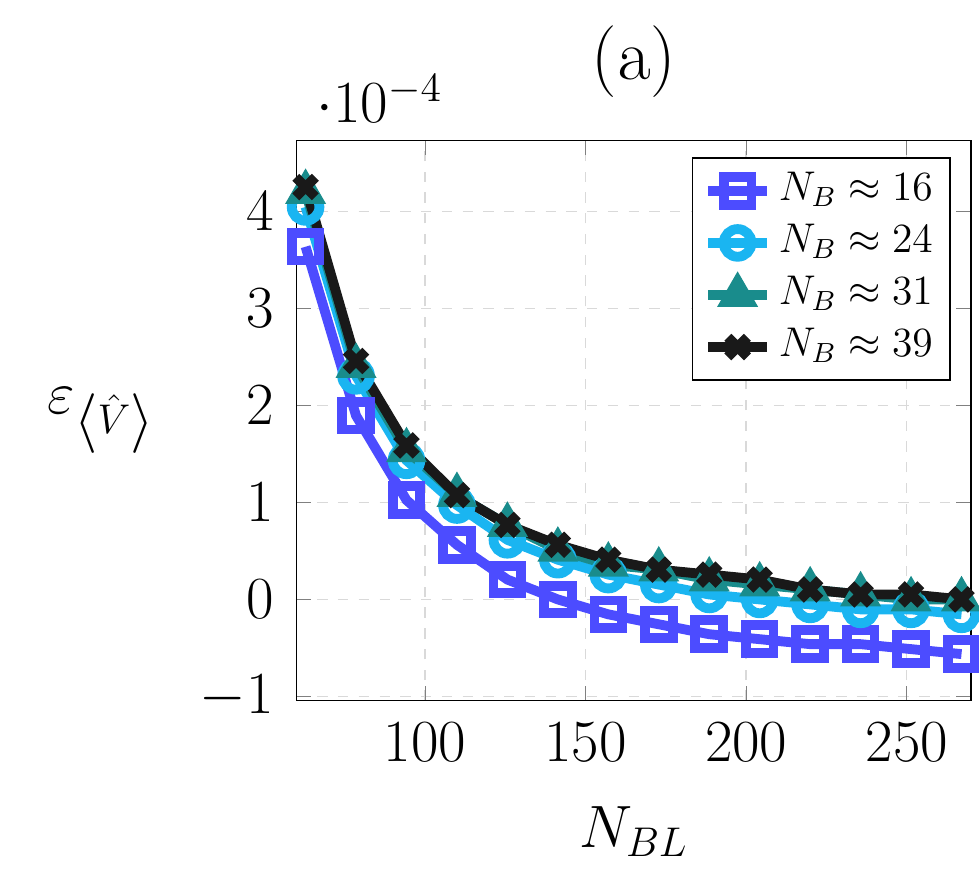}};
   \node[right=of img1,xshift=-1.cm, yshift=0.0cm] (img1)  {\includegraphics[scale=0.7]{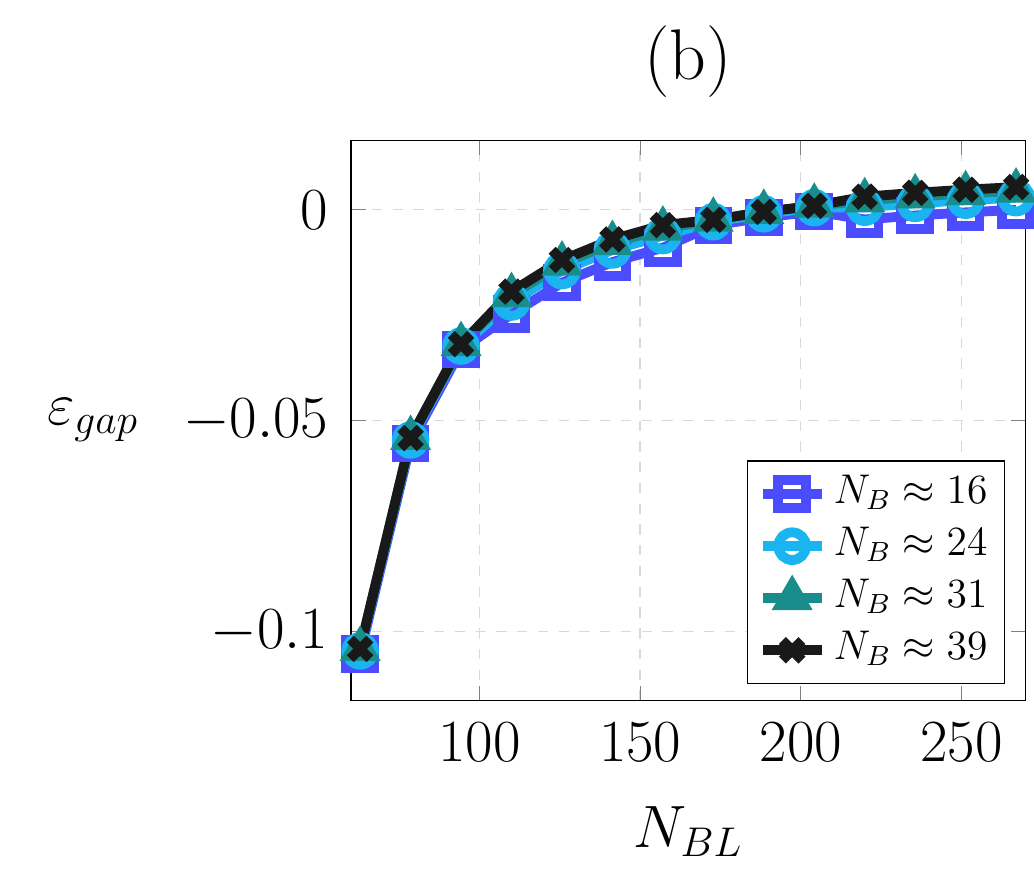}};
\end{tikzpicture}
 \caption{{Time-averaged gliding speed (a) and gap (b) as a function of the number of nodes in the ``boundary layer'' as described in Fig.~\ref{fig:meshCa}. Here, $\hA=0.25, n=5, \ha\approx 3\times10^{-3}, Ca\approx 5\times 10^{-3}, \mcal{R}=0.16$, $\eta=5\times 10^3$.  }}
 \label{fig:meshconvCa}
 \end{figure}

For the subsequent computations (shown in the present document and in the main paper), the gliding speed (resp. lubrication gap) was obtained with a relative error $\leq 0.01\%$ (resp. $\leq 1\%$).

\bigskip
Having verified the convergence of our FEM code, we show in the next sections a selection of numerical results that are supplementary to those of the main paper. 
 
\addtocounter{subsection}{1}
\subsection{Trajectories for different values of $\eta$}
We show on Fig.~\ref{fig:traj2} the trajectories $\hx(\hti)$ of the center of mass of the myxobacteria for three values of the softness parameter, corresponding to very stiff ($\eta=10^{-3}$), soft ($\eta=1$) and very soft ($\eta=10^3$) substrates. For all three cases, $\hA=0.25$ and $n=5$. As in the case of purely elastic substrate, the computed trajectories are, in general, superpositions of a non-zero mean component plus an oscillatory perturbation that causes the bacteria to glide intermittently. However, while this oscillatory component never vanishes (as evidenced by the instantaneous velocity), its influence decreases with the substrate softness. Therefore, as $\eta$ increases, the quasi stick-slip behavior progressively becomes a more persistent motion, as clearly seen in Fig.~\ref{fig:traj2}.
\begin{figure}[b!]
\centering
 \begin{tikzpicture}
 \node[xshift=0cm, yshift=0.0cm] (img1)  {\includegraphics[scale=0.7]{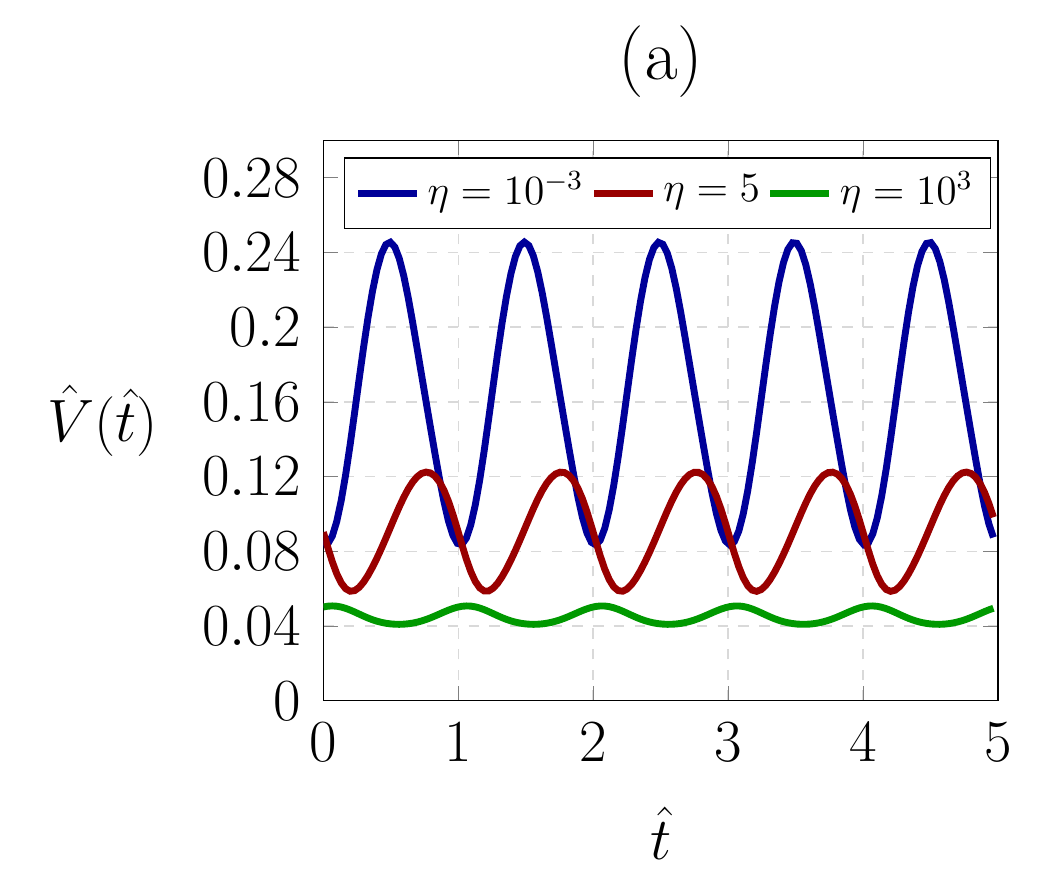}};
   \node[right=of img1,xshift=-1.cm, yshift=0.0cm] (img1)  {\includegraphics[scale=0.7]{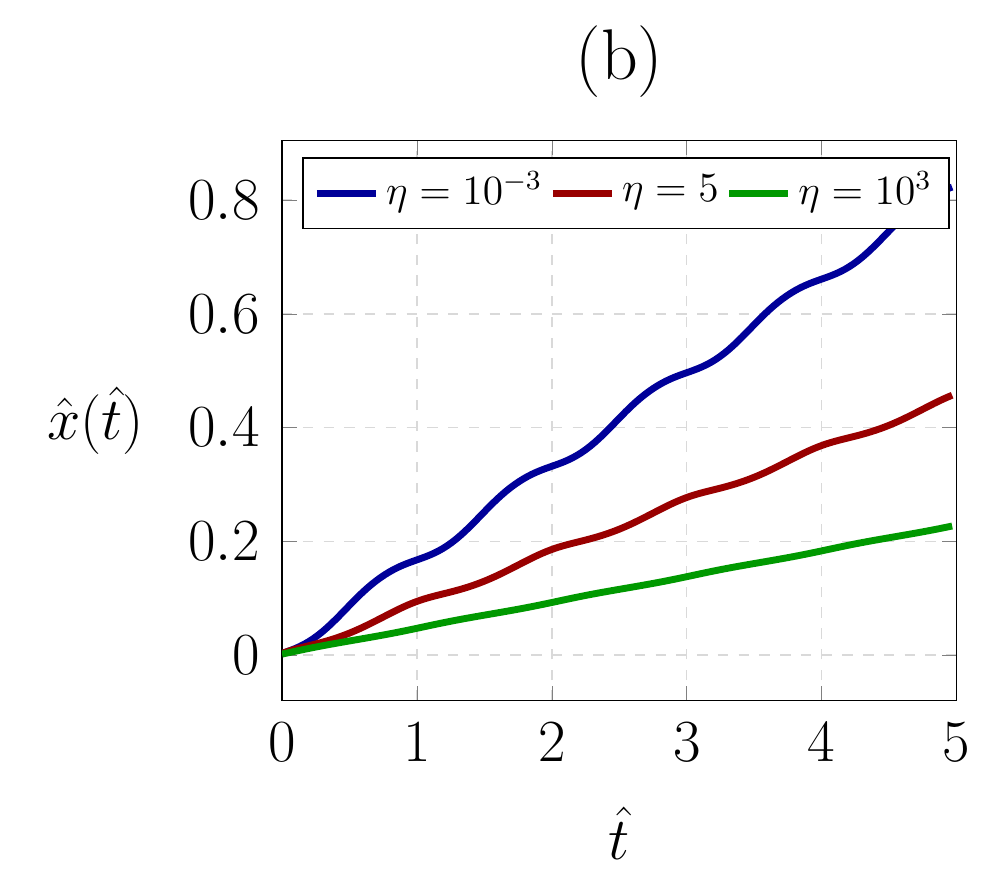}};
\end{tikzpicture}
 \caption{{Instantaneous (a) speed and (b) trajectories of the center of mass of the bacteria on substrates with different softness numbers. Here, $\hA=0.25, n=5, \ha\approx3\times10^{-3},Ca\approx1.67\times 10^{-3}, \mcal{R}=0.1$.}}
 \label{fig:traj2}
 \end{figure}
 
\begin{figure}[t!]
\centering
 \begin{tikzpicture}
 \node[xshift=0cm, yshift=0.0cm] (img1)  {\includegraphics[scale=0.7]{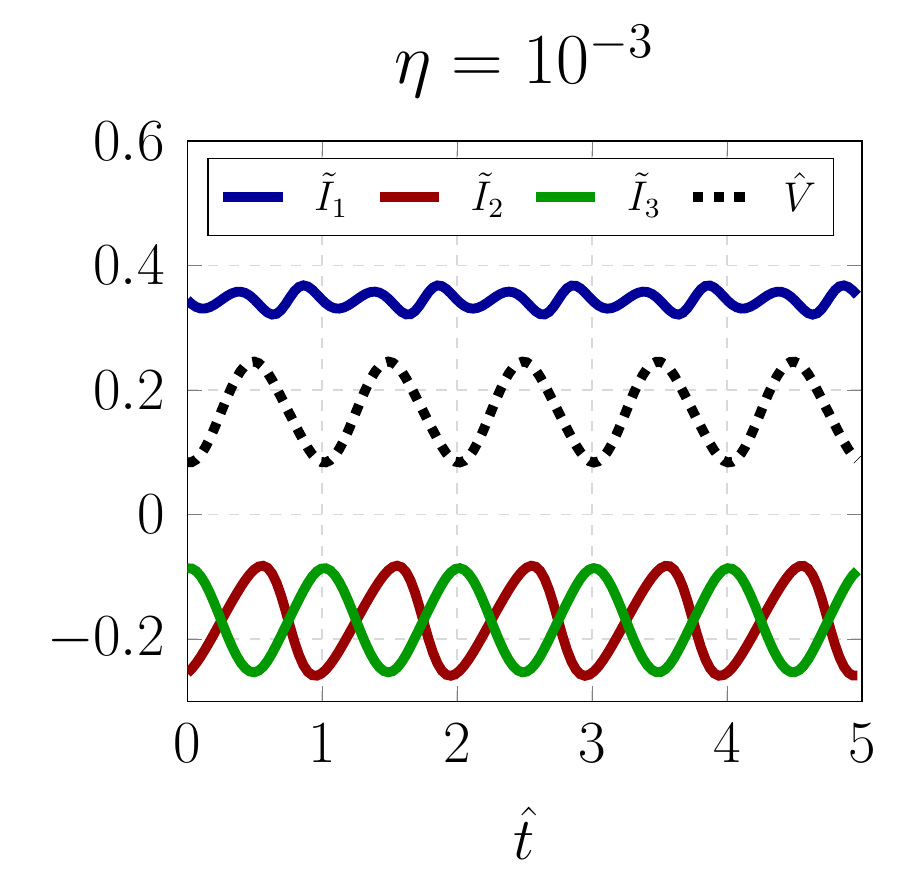}};
   \node[right=of img1,xshift=-1.cm, yshift=0.0cm] (img2)  {\includegraphics[scale=0.7]{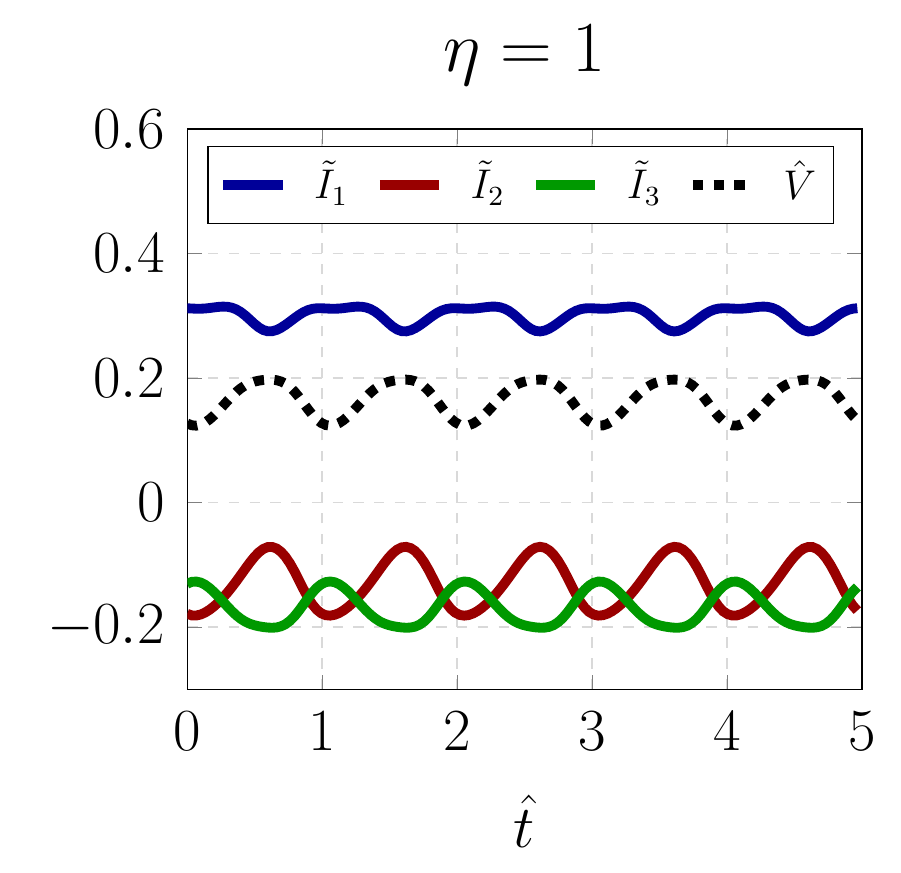}};
   \node[below=of img1,xshift=0.cm, yshift=1.0cm] (img3)  {\includegraphics[scale=0.7]{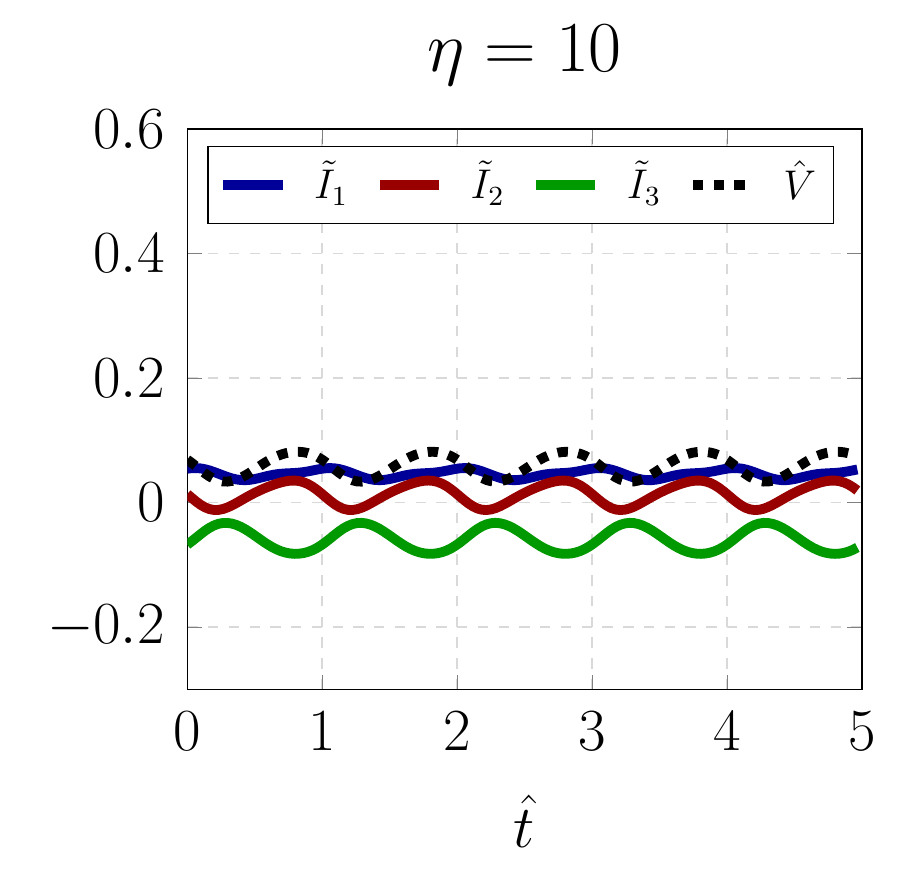}};
   \node[right=of img3,xshift=-1.cm, yshift=0.0cm] (img4)  {\includegraphics[scale=0.7]{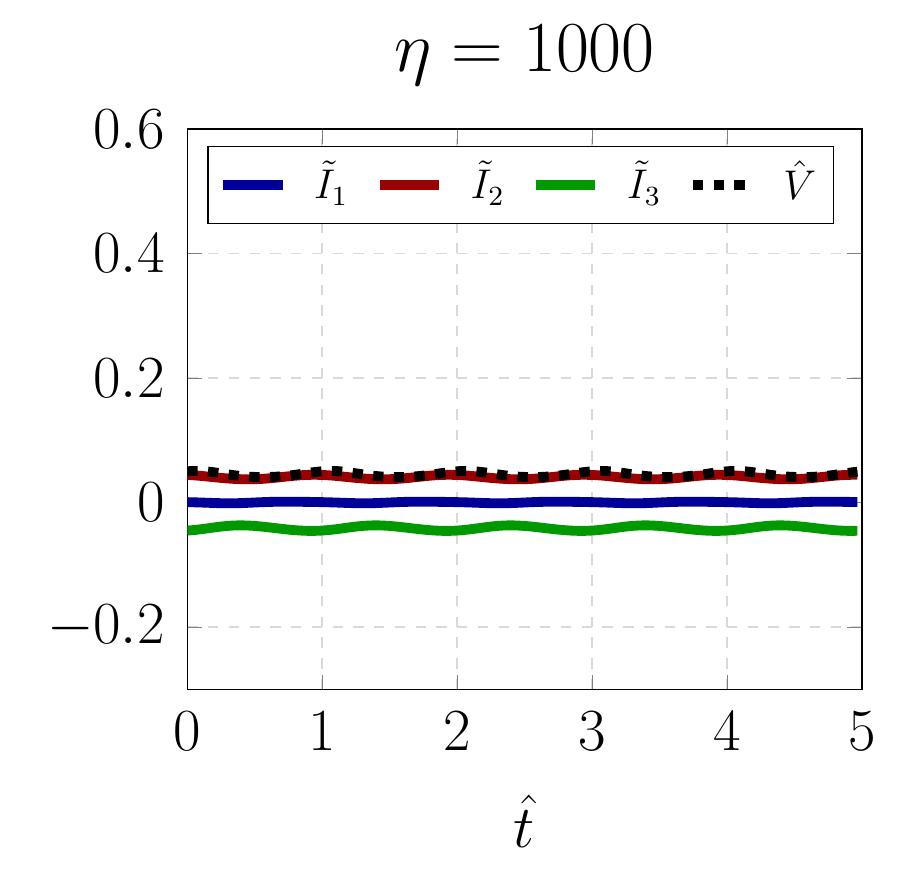}};
\end{tikzpicture}
 \caption{{Time evolution of the gliding velocity and the spatial averages $\tilde{I}_j$ (defined in the text), for different values of the softness parameters. The parameters of the bacterial geometry $\hA=0.25, n=5, \ha\approx3\times10^{-3},Ca\approx 2\times 10^{-3}, \mcal{R}=0.1$.}}
 \label{fig:Ij}
 \end{figure}
 
\bigskip
This transition from intermittent to persistent gliding results from a strong correlation between the motion and the pressure gradient that drives the lubricating flow of slime. This is evidenced in Fig.~\ref{fig:Ij} which shows the time evolution of the gliding speed and that of the spatial averages $\tilde{I}_j(\hti)$ of the contributions to the horizontal force balance. These averages are defined by:
$$\tilde{I}_j(\hti)=\dfrac{1}{n}\int_{-n/2}^{n/2}I_j(\hx,\hti)\mathrm{d}\hx,$$
where the terms $I_j$ read
\begin{center}
$I_1(\hx,\hti)= -\hp\dfrac{\partial\hat{b}}{\partial \hx}$, $I_2(\hx,\hti)=-\dfrac{1}{2}\dfrac{\partial\hp}{\partial \hx}\left(\hh-\eta\hdelta\right)$, $I_3(\hx,\hti)=\dfrac{-\hV}{\hh-\eta\hdelta}$.
\end{center} 
As expected, the friction term $\tilde{I}_3\propto \hV$ simply reflects the evolution of the gliding speed. However, although the source of the thrust changes from $I_1$ to $I_2$ as the softness number increases, we find that it is always the pressure gradient term $I_2$ which dictates the time evolution of the velocity. On very stiff substrates (e.g. $\eta=10^{-3}$) where $\tilde{I}_2$ is always negative (friction), the maximum velocity is reached when this friction term is minimum, and vice versa. Conversely, on very soft substrates (e.g. $\eta=10^{3}$) where $\tilde{I}_2$ is always positive (thrust), the maximum velocity is attained when this thrust term is maximum. Therefore, the time evolution of the speed $\hV(\hti)$ directly depends on that of $\tilde{I}_2$. Given that the trajectories $\hx(\hti)$ is obtained as $\hx(0)+\int_0^{\hti} \hV(\hti')\mathrm{d}\hti'$, the nature of $\tilde{I}_2$ explains why $\hx(\hti)$ is more intermittent on stiffer substrates. Indeed, for a stiff substrate, the term $\tilde{I}_2$ directly arises from the bacterial shape oscillations and thus experiences strong time oscillations. These oscillations cause the gliding speed $\hV$ to increase and decrease almost by a factor of 3 (see Fig.~\ref{fig:traj2}(a) for $\eta=10^{-3}$). As a result, the bacteria will intermittently glide and then slows down considerably, in almost a stick-slip manner. However, in the case of a soft substrate, this term $\tilde{I}_2$ mainly arises from the Laplace pressure jump across the curved slime-air interface. Given that this is a time-independent quantity, the term $\tilde{I}_2$ will also be quasi-constant and thence the speed $\hV(\hti)$. Consequently, the resulting trajectory $\hx(\hti)$ on very soft substrates is more persistent (linear) as shown in Fig.~\ref{fig:traj2}(b).

\addtocounter{subsection}{1}
\subsection{Influence of the capillary number $Ca$}
\begin{figure}[t!]
\centering
\includegraphics[scale=0.7]{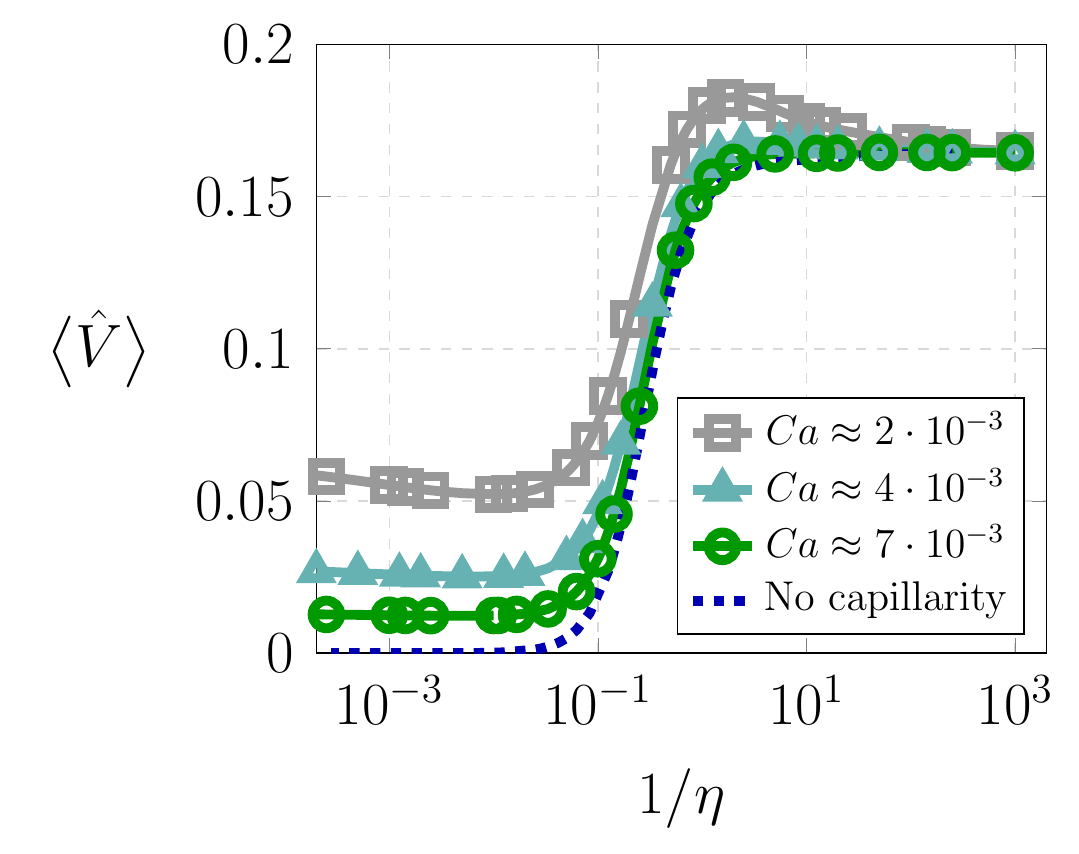}
 \caption{{Time-averaged gliding speed as a function of the softness parameter. Here $\hA=0.25$, $n=5$, $\epsilon=0.008$, $\mcal{R}=0.1$, $\ha\approx 3\times 10^{-3}$.}}
 \label{fig:speedCas}
 \end{figure}
Fig.~\ref{fig:speedCas} shows in log-log axis how the average gliding speed depends on the capillary number. For large $Ca \propto \gamma^{-1}$, the surface tension is small compared to the viscous forces and capillary corrections to the gliding speed remain small but do not vanish. However, as the substrate gets softer, the capillarity effects are more strongly pronounced and ultimately dominate at larger $\eta$, regardless of the value of $Ca$. 

\begin{figure}[h!]
\centering
\includegraphics[scale=0.7]{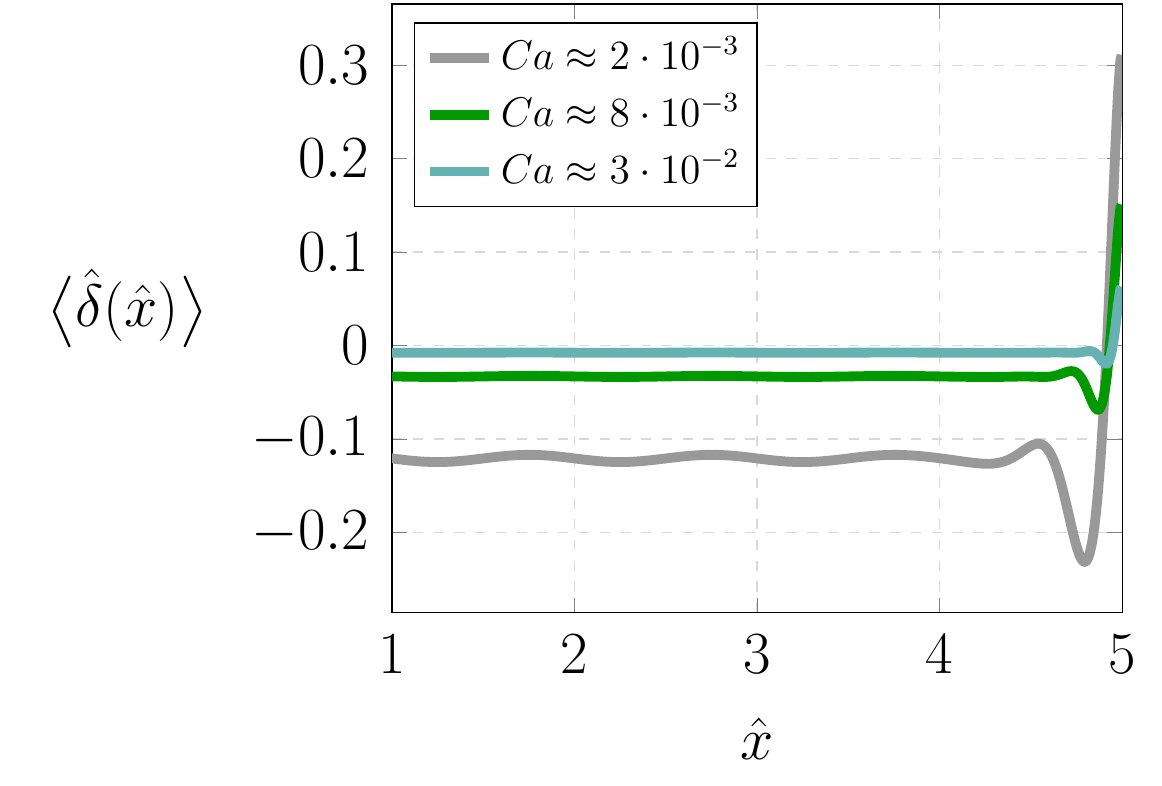}
 \caption{{Deformation of the substrate beneath the bacteria. The parameters for the simulations are $\hA=0.25$, $n=5$, $\epsilon=0.008$, $\mcal{R}=0.1$, $\ha\approx 3\times 10^{-3}$, $\eta=10^3$.}}
 \label{fig:substrate}
 \end{figure}
 
\bigskip
We also show, in Fig.~\ref{fig:substrate}, the effects of the capillary number on the deformation field, for a very soft substrate ($\eta=10^3$). We find, in accordance with Eq.~(\ref{eq:ridge_extent}), that increasing the capillary number results in a ridge that is both smaller (since $\Delta\sim Ca^{-1}$) and narrower (since $\hat{l}_c\sim Ca^{-1/3}$). This is because as $Ca$ increases, the interfacial tension $\gamma$ decreases, thereby limiting the capillary-induced ridge to a smaller and smaller extent at the leading edge of the bacteria.

\addtocounter{subsection}{1}
\subsection{Influence of slime-air interfacial thickness $\ha$}
Fig.~\ref{fig:speeda} shows that, as the softness parameter increases (i.e $1/\eta$ decreases), the thickness of the slime-air interface at the leading edge of the bacteria selects the value of $\eta$ beyond which capillarity effects start to contribute to the bacterial gliding speed. We find that the thinner the interface, the stronger its influence on the gliding speed since the jump across the interface is sharper. However, we also find that the exact value of the $\ha$ is of little importance when the gliding occurs on very soft substrates. Fig.~\ref{fig:speeda} shows that the asymptotic solution $\bl\hV_{\infty}\br$ (Eq.~(\ref{eq:gspeed})), plotted here for $\ha\approx 3\times10^{-3}$, is a good approximation in the limit $\eta\rightarrow\infty$ over a range of values for $\ha$ spanning two orders of magnitude.
\begin{figure}[h!]
\centering
\includegraphics[scale=0.7]{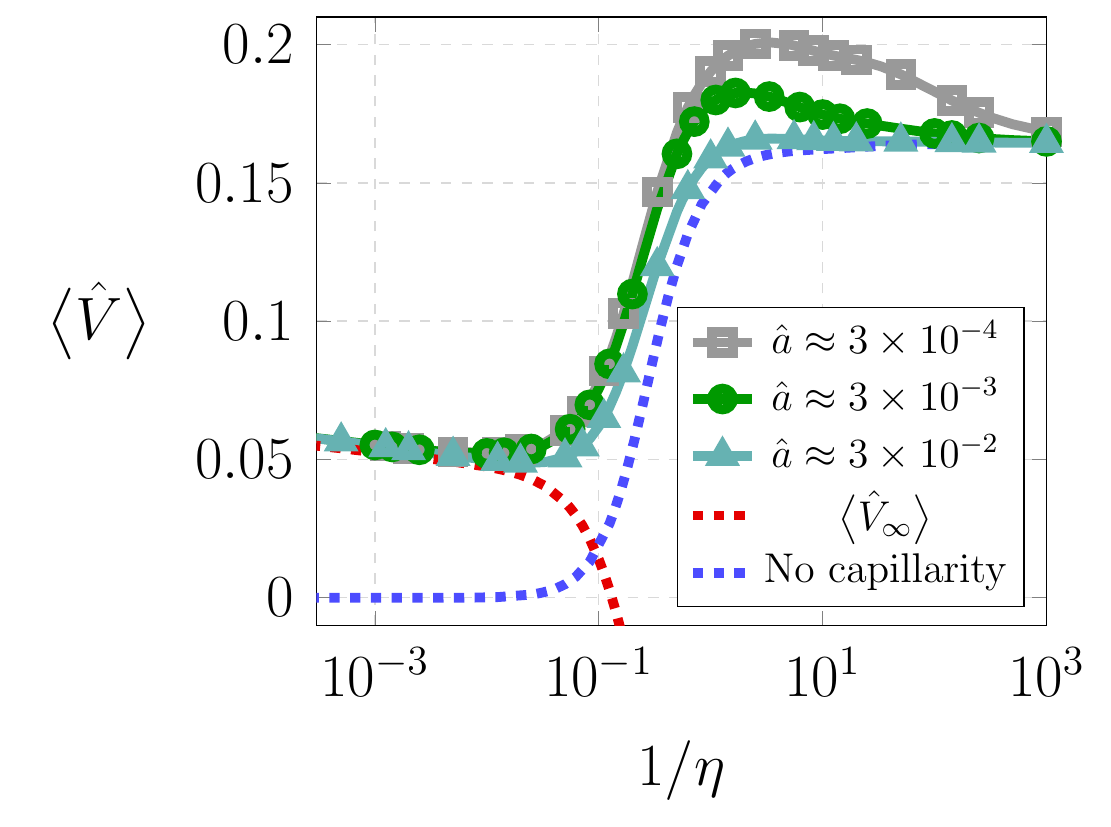}
\caption{{Time-averaged gliding speed as a function of the softness parameter for different values of the slime-air interfacial thickness $\ha$. Here $\hA=0.25$, $n=5$, $\epsilon=0.008$, $\mcal{R}=0.1$, $Ca\approx 6\times 10^{-3}$.}}
\label{fig:speeda}
\end{figure}

\bigskip



\endgroup

\end{document}